\documentclass[12pt]{article}
\usepackage{amsmath}
\usepackage{graphicx}
\usepackage{enumerate, setspace}
\usepackage{subcaption}
\usepackage[round]{natbib}
\usepackage{url} 
\usepackage{float}
\usepackage{mathtools}
\usepackage{chngcntr}
\usepackage{fontenc}  
\usepackage{lmodern}

\usepackage{algorithm,algpseudocode,float}
\usepackage{lipsum}
\usepackage{amssymb}
\usepackage{xcolor}
\usepackage{color}
 \usepackage{multirow}
\definecolor{green}{RGB}{0, 128, 0}

\makeatletter
\newenvironment{breakablealgorithm}
  {
   \begin{center}
     \refstepcounter{algorithm}
     \hrule height.8pt depth0pt \kern2pt
     \renewcommand{\caption}[2][\relax]{
       {\raggedright\textbf{\ALG@name~\thealgorithm} ##2\par}%
       \ifx\relax##1\relax 
         \addcontentsline{loa}{algorithm}{\protect\numberline{\thealgorithm}##2}%
       \else 
         \addcontentsline{loa}{algorithm}{\protect\numberline{\thealgorithm}##1}%
       \fi
       \kern2pt\hrule\kern2pt
     }
  }{
     \kern2pt\hrule\relax
   \end{center}
  }
\makeatother

\usepackage{hyperref}					
\hypersetup{colorlinks,
	linkcolor=blue,%
	citecolor=blue}

\newtheorem{theorem}{Theorem}
\newtheorem{lemma}{Lemma}
\newtheorem{remark}{Remark}
\newtheorem{proposition}{Proposition}
\newtheorem{assumption}{Assumption}
\newtheorem{corollary}{Corollary}

 \title{\bf Conformalized Multiple Testing under Unknown Null Distribution with Symmetric Errors} 
  \author{Yang Tian\thanks{
    Center for Data Science and School of Mathematical Sciences, Zhejiang University.}\hspace{.2cm}, Zinan Zhao\thanks{
    Center for Data Science and School of Mathematical Sciences, Zhejiang University.}
    \hspace{0.01cm} and Wenguang Sun\thanks{
    Center for Data Science and School of Management, Zhejiang University.}}
  \date{}

\begin{document}

\def\spacingset#1{\renewcommand{\baselinestretch}%
{#1}\small\normalsize} \spacingset{1}

\maketitle

\bigskip
\begin{abstract}

This article addresses a fundamental concern, first raised by \cite{efron2004large}, regarding the selection of null distributions in large-scale multiple testing. In modern data-intensive applications involving thousands or even millions of hypotheses, the theoretical null distribution of the test statistics often deviates from the true underlying null distribution, severely compromising the false discovery rate (FDR) analysis.  We propose a conformalized 
empirical Bayes method using self-calibrated empirical null samples (SENS) for both one-sample and two-sample multiple testing problems. The new framework not only sidesteps the use of potentially erroneous theoretical null distributions, which is common in conventional practice, but also mitigates the impact of estimation errors in the unknown null distribution on the validity of FDR control, a challenge frequently encountered in the empirical Bayes FDR literature. In contrast to the empirical Bayes approaches (cf.  \citealp{efron2004large, jin2007estimating, sun2007oracle}) that rely on Gaussian assumptions for the null models, SENS imposes only a weak condition on the symmetry of the error distribution, and leverages conformal tools to achieve FDR control in finite samples. Moreover, SENS incorporates structural insights from empirical Bayes into inference, exhibiting higher power compared to frequentist model-free methods. We conduct an in-depth analysis to establish a novel optimality theory for SENS under Efron’s two-group model and demonstrate its superiority over existing empirical Bayes FDR methods and recent model-free FDR methods through numerical experiments on both simulated and real data.

\end{abstract}

\noindent%
\vfill
\newpage
\spacingset{1} 

\section{Introduction}\label{sec:intro}

This article revisits a fundamental issue raised in \cite{efron2004large} concerning the choice of the null distribution in large-scale testing problems and its impact on subsequent false discovery rate (FDR; \citealp{benjamini1995controlling}) analysis. We start with the one-sample case (the two-sample case is detailed in Section \ref{subsec:two sample case}), where \(n_i\) repeated measurements are collected for each study unit \(i \in [m]\coloneqq \{1, \ldots, m\}\):
\begin{equation}\label{equ:model}
  X_{ij} = \mu_i + \epsilon_{ij}, \quad \mathbb{E}(\epsilon_{ij}) = 0, \quad i \in [m], \quad j \in [n_i],
\end{equation}
where \(\mu_i\) denotes the unknown effect size relative to the baseline level, \(\epsilon_{ij}\) are random errors and $n_i$ is the number of observations in unit $i$. The goal is to identify non-null effects, which can be formulated as the following multiple testing problem: 

\begin{equation}\label{MT:formulation}
H_{0,i}: \mu_i = 0 \quad \text{versus} \quad H_{1,i}: \mu_i \neq 0; \quad i \in[m]. 
\end{equation}
Next, we outline conventional practices under the problem formulations \eqref{equ:model} and \eqref{MT:formulation}, followed by a discussion of the key challenges.

\subsection{Theoretical null vs. empirical null in large-scale inference}\label{subsec:choice-null}

The standard practice in multiple testing involves computing a summary statistic, such as a $t$-statistic, for unit $i$. This statistic is then converted into a $z$-value or a $p$-value. Consider the following random mixture model for $z$-values \citep{efron2001empirical}:

\begin{equation*}\label{model:two-group}
Z_1, \ldots, Z_m \stackrel{i.i.d.}{\sim} F^*(z) = (1 - \pi)F^*_0(z) + \pi F^*_1(z),
\end{equation*} 
where \(\pi \coloneqq \mathbb{P}(\mu_i \neq 0)\) indicates the non-null proportion, and \(F^*\), \(F^*_0\), and \(F^*_1\) respectively represent the mixture, null and non-null cumulative distribution functions (CDF). The corresponding probability densities are denoted as \(f^*\), \(f^*_0\), and \(f^*_1\).

A critical assumption underpins conventional FDR analyses is that the $z$-values associated with null hypotheses follow the standard normal distribution $\Phi$, referred to as the \emph{theoretical null} distribution. This assumption lays the foundation for encoding the common pattern observed in null cases, namely the ``normal state’’, which enables the effective separation of non-null effects that deviate from this commonality. However, as highlighted by \cite{efron2004large}, observed \(z\)-values instead follow the \emph{empirical null} distribution \(F_0^*\), which can differ substantially from its theoretical counterpart $\Phi$. This discrepancy can be attributed to various factors, including the complex data structure in large-scale studies, interdependencies among test statistics, and presence of unobserved confounders. Using an incorrect theoretical null can substantially degrade statistical inference, leading to inflated error rates, reduced statistical power, and erroneous scientific conclusions. This highlights the necessity of moving away from methods based on theoretical null distributions.

\subsection{The empirical null approach: advantages and challenges}\label{subsec:empirical null approach}

We review and scrutinize a prominent line of research built upon \cite{efron2004large}, which advocates using the empirical null for large-scale inference. This approach utilizes a data-driven baseline to identify ``interesting'' or ``abnormal'' study units, providing more reliable and interpretable inferences. It is particularly useful in analyzing complex biological studies where the theoretical null  frequently does not align with the prevalent patterns observed in large datasets. The works of \cite{jin2007estimating}, \cite{cai2010optimal}, and \cite{roquain2022false} show that, under certain regularity conditions, estimators for the empirical null distribution \( F_0^* \) can be constructed with high precision.

\begin{figure}[htbp!]
    \centering
    \includegraphics[width=0.85\linewidth, height=2.2in]{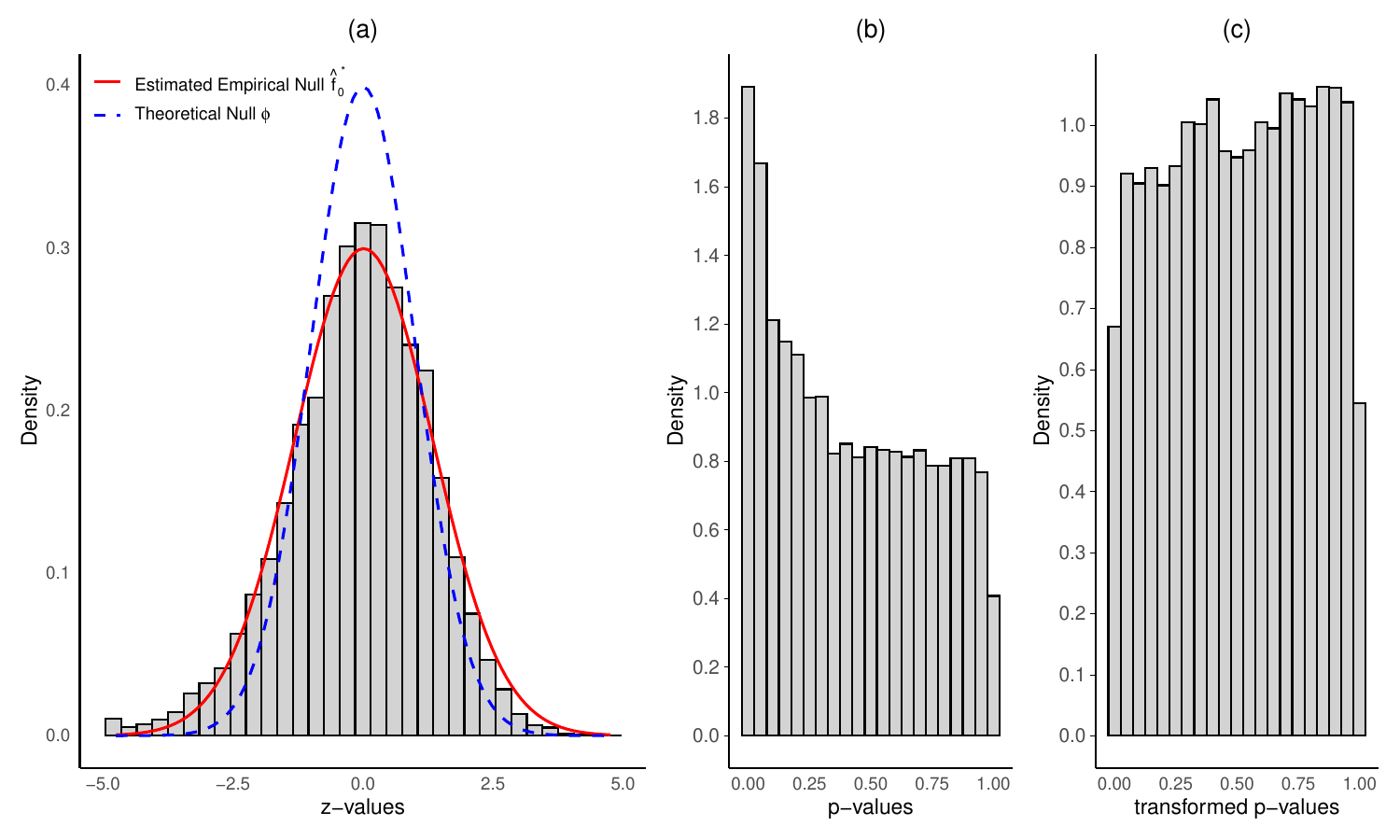}
    \caption{\small Histograms of gene expression data (a) \(z\)-values; (b) \(p\)-values obtained using the theoretical null distribution; (c) transformed \(p\)-values using the estimated empirical null $\mathcal{N}(0.01,1.33^2)$.}
    \label{fig:intro}
\end{figure}

The empirical null approach can be illustrated through a differential analysis of gene expression data collected from 20 insulin-sensitive individuals \citep{wu2007effect}; see Section \ref{sec:application} for details. Figure \ref{fig:intro}(a) presents the theoretical null (blue dashed line) and the estimated empirical null (red solid line) of \(z\)-values. We can see that the theoretical null curve is much narrower than its empirical counterpart. Moreover, the \(p\)-values derived from the theoretical null [Panel (b)] significantly deviate from the expected uniform pattern. By contrast, the pattern of transformed \(p\)-values, derived from the estimated empirical null, aligns more closely with the uniform distribution [Panel (c)], which enhances the interpretability and accuracy of subsequent FDR analyses. 

Estimating the unknown null distribution presents significant challenges. Current methodologies rely heavily on parametric models and regularity conditions that are difficult to validate in practice. For example, \cite{efron2004large}, \cite{jin2007estimating}, and \cite{cai2010optimal} assume that the empirical null distribution follows a Gaussian distribution \(\mathcal{N}(\mu_0, \sigma_0^2)\). However, the complexities inherent in real-world data, such as unobserved covariates, frequently result in null density tails that are heavier than those of a Gaussian distribution. Furthermore, even if the empirical null theoretically conforms to the assumed Gaussian family, the validity of FDR analyses involves asymptotic arguments that require additional assumptions regarding the estimation consistency of \(\mu_0\) and \(\sigma_0^2\) \citep{sun2007oracle}; see \cite{roquain2022false} for related discussions of impossibility results in this context. A critical concern is that even small errors in estimating \(\mu_0\) and \(\sigma_0^2\) can significantly degrade the quality of the FDR analysis, thus considerably restricting the applicability of the empirical null framework. Next, we outline our proposal, which leverages modern tools in conformal inference to overcome the limitations of existing empirical Bayes methods.


\subsection{A preview of our proposal and contributions}\label{subsec:preview}

The primary objective of this work is to develop a methodology that leverages a data-driven empirical null  indirectly, bypassing the challenges of direct estimation of the unknown null and avoiding reliance on a potentially incorrect theoretical null. We adopt a semi-supervised multiple testing (SSMT) strategy that utilizes self-calibrated empirical null samples (SENS). The proposed SENS algorithm involves three steps. First, we employ a sample-splitting strategy to simultaneously construct (a) test samples~\(\mathbf{T}\) that provide evidence against the null  and (b) calibrated null samples~\(\mathbf{T}^0\) that capture the configuration of the unknown null distribution. Second, we derive conformal scores from these samples, ensuring they fulfill pairwise exchangeability conditions. Finally, we utilize a mirror process to establish a thresholding rule for FDR control.

Our work makes several contributions. First, in contrast to existing frequentist model-free methods (cf. \citealp{arlot2010some, arias2017distribution, zou2020new, ge2021clipper}), SENS is fundamentally motivated by an empirical Bayes perspective. It effectively exploits structural information in the sample by employing the local false discovery rate (lfdr; \citealp{efron2001empirical}); lfdr-based methods have been shown to possess certain optimality properties and to outperform frequentist p-value-based approaches \citep{sun2007oracle}. Second, unlike existing empirical Bayes methods that utilize estimated nulls and rely on Gaussian assumptions (cf. \citealp{efron2001empirical, jin2007estimating}), SENS adaptively generates null samples and requires only the symmetry of the error distribution. Finally, we present a novel power analysis demonstrating that SENS achieves asymptotic optimality under a specific variant of Efron's two-group model. To our knowledge, this analysis provides the first optimality theory for the mirror-based FDR methods introduced in \cite{barber2015controlling}. In summary, by integrating the assumption-lean validity of conformal inference with the structural insights of empirical Bayes, SENS provides a powerful, model-free tool for large-scale multiple testing.

\subsection{Related work}\label{subsec:related-works}

The proposed SENS Algorithm is closely connected to several significant lines of research. First, our basic strategy aligns with the core principles of empirical Bayes FDR methods (\citealp{efron2004large, jin2007estimating, sun2007oracle}), which advocate for the use of empirical null distributions in large-scale inference. However, given our distinct model assumptions and inferential strategies, SENS does not address the full range of issues raised in \cite{efron2004large}. In particular, mean shifts resulting from structured dependencies remain unresolved; see Section \ref{subsec:comparisons efron} of the Supplement for further discussion. 

Second, SENS employs innovative tools from SSMT \citep{blanchard2010semi, mary2022semi} and distribution-free inference \citep{yang2021bonus, bates2023testing, marandon2024adaptive, liang2024integrative}. Unlike existing conformal methods, SENS does not require the availability of null samples, which are often difficult to acquire due to the unknown null distribution. Moreover, existing methods typically assume joint exchangeability among null samples across the calibration and test sets, which fails to hold under our setup. In contrast, SENS relies on a weaker pairwise exchangeability assumption for controlling the FDR, thereby broadening its applicability across various complex scenarios. 

Third, as a conformalized empirical Bayes method, SENS significantly differs from existing model-free FDR approaches (e.g., \citealp{arlot2010some, arias2017distribution, zou2020new, ge2021clipper}), which are inherently frequentist and do not capitalize on the advantages offered by lfdr-type statistics. We present a comprehensive comparison between SENS and its frequentist competitors in Section \ref{sec:comparisons model-free} of the Supplement.

Finally, SENS leverages the ``plus and minus'' trick to construct test samples using all available data, thereby avoiding the information loss common with FDR methods based on sample-splitting (e.g., \citealp{fithian2014optimal, lei2021general, du2023false, Dai2023splitting}). Moreover, our sample-splitting strategy is fundamentally different -- and not directly comparable -- to the unit-splitting methods employed in the conformal inference literature (e.g., \citealp{bates2023testing, marandon2024adaptive}).

\subsection{Organization}\label{subsec:organization}

The paper is structured as follows. In Section \ref{sec:ssmt}, we discuss the semi-supervised setup and outline the basic framework. Section \ref{sec:SENS} describes the SENS Algorithm, with asymptotic optimality theory established in Section \ref{sec:asymptotic optimality}. Sections \ref{sec:simulation} and \ref{sec:application} investigate the numerical performance of SENS and compare it with existing methods using both simulated and real-world data. The proofs and other auxiliary results are provided in the Supplement.

\section{Generating Null Samples from the Unknown}\label{sec:ssmt}

\subsection{Basic setting and problem formulation}\label{sec:preliminaries}

This article leverages the SSMT framework (cf. \citealp{blanchard2010semi, mary2022semi}) to address challenges posed by unknown null distributions. In high-dimensional settings characterized by complex data structures and intricate models, the SSMT formulation enables FDR analysis without requiring precise knowledge of the null distribution of test statistics, offering enhanced flexibility and robustness compared to traditional methods. We begin by reviewing the preliminaries of SSMT and then discuss constructing null samples for the one-sample (Section \ref{subsec:SENS-construction}) and two-sample (Section \ref{subsec:two sample case}) scenarios.

In SSMT, users have access to a null training sample (NTS) comprising a set of labeled null data points drawn from a common but unknown distribution. The test sample contains $m$ new observations, and the goal is to identify those that deviate from the commonality. This setup is referred to as the out-of-distribution testing or outlier detection problem in the conformal inference literature (e.g., \citealp{Gru69, bates2023testing, marandon2024adaptive, liang2024integrative}). While our original problem [defined by \eqref{equ:model} and \eqref{MT:formulation}] stems from the classical setup, the SSMT framework offers two key advantages in addressing the challenges discussed in \cite{efron2004large} and \cite{jin2007estimating}: (a) the ability to make reliable inferences without needing to estimate the unknown null, and (b) the adaptability to accommodate the complexities inherent in real-world data and intricate machine learning algorithms.

Let \(\mathcal{H}_0\) and \(\mathcal{H}_1\) denote the index sets of null and non-null hypotheses, respectively, and let \(\pmb{\delta} = (\delta_i: i \in [m]) \in \{0, 1\}^m\) represent a multiple testing rule, where \(\delta_i = 1\) indicates the rejection of \(H_{0,i}\), and \(\delta_i = 0\) otherwise. The false discovery rate \citep{benjamini1995controlling} is defined as
$
\operatorname{FDR} = \mathbb{E}\left[\{\sum_{i \in \mathcal{H}_0} \delta_i\}/{\max\left\{\sum_{i \in [m]} \delta_i, 1\right\}}\right].
$
Define the expected number of true positives as 
$
\operatorname{ETP} = \mathbb{E}\left[\sum_{i \in \mathcal{H}_1} \delta_i\right].
$
The objective is to develop \(\pmb{\delta}\) that controls the FDR at level \(\alpha\) while making the ETP as large as possible.

The SSMT formulation enables us to shift the focus from estimating the unknown null distribution to generating valid null samples that conform to that distribution. Conventionally, null samples can be provided directly \citep{blanchard2010semi}, learned from data \citep{bacon2021muse}, or obtained through black-box sampling \citep{choquet2018hd, bacon2021muse}, including via generative models \citep{ghahramani2014generative} and variational autoencoders \citep{kingma2013auto}. However, existing methods are not applicable in our setting, which involves unknown null (or incorrect theoretical null). The next two subsections introduce innovative techniques for constructing self-calibrated empirical null samples (SENS) that rely solely on a mild symmetry condition of the random errors.


\subsection{The one-sample scenario}\label{subsec:SENS-construction}


To ensure effective alignment between the classical and semi-supervised setups, we need the following mild yet essential assumption for constructing valid samples. Consider the data-generating model \eqref{equ:model}. Denote \(\mathbf{X}_i = (X_{ij}: j \in [n_i])\) for \(i \in [m]\).

\begin{assumption}\label{ass:errors}
$\forall i \in \mathcal H_0$, the errors \(\{\epsilon_{ij}: j\in[n_i]\}\) are independent and identically distributed (iid) conditional on the observations from other units, obeying a zero-symmetric density $f_{i}^{\epsilon}(\cdot\mid \mathbf{X}_1,\cdots,\mathbf{X}_{i-1},\mathbf{X}_{i+1},\cdots,\mathbf{X}_m)$ such that 

\begin{equation}\label{cond:symmetric-error}
f_{i}^{\epsilon}(-x\mid \mathbf{X}_1,\cdots,\mathbf{X}_{i-1},\mathbf{X}_{i+1},\cdots,\mathbf{X}_m) = f_{i}^{\epsilon}(x\mid \mathbf{X}_1,\cdots,\mathbf{X}_{i-1},\mathbf{X}_{i+1},\cdots,\mathbf{X}_m). 
\end{equation}
\end{assumption}

\begin{remark}\rm{

Assumption \ref{ass:errors} relaxes the Gaussian assumptions widely employed in classical methods, making our method suitable for heavy-tailed distributions. Moreover, unlike \cite{efron2004large} and \cite{jin2007estimating}, which assume a common null of \(z\)-values, Assumption \ref{ass:errors} explicitly focuses on the error density functions, which are allowed to vary across different units. This flexibility accommodates heterogeneous nulls~--~an important issue discussed in \cite{efron2008simultaneous}, \cite{cai2009simultaneous}, and \cite{SunMcLain:2012}. Finally, although the more intuitive assumption is that errors are iid within a unit and independent across units, we adopt Assumption \ref{ass:errors} instead because it permits dependence across units. } 
\end{remark}

To streamline our discussion, we assume that \( n_i \geq 4 \) for all \( i \in [m] \); cases where \( 2 \leq n_i < 4 \) are discussed in Remark \ref{rem:small-n} below. The construction of SENS comprises three steps. The first step randomly partitions the observations within each unit \( i \) as $\{X_{ij}: j\in[n_i]\}=\{X_{ij}: j\in\mathcal{N}_{i1}\}\cup \{X_{ij}: j\in\mathcal{N}_{i2}\}$, with sample sizes  \( n_{i1}\coloneqq|\mathcal{N}_{i1}|= \lceil n_i/2 \rceil \) and \( n_{i2}=n_i-n_{i1}\), where \( \lceil x \rceil \) represents the smallest integer greater than or equal to \( x \). The second step computes,  for \( i \in [m] \), {\small
\begin{equation}\label{eq:VS}
V_i = \bar{X}_{i1} + \bar{X}_{i2},  V^0_i =\bar{X}_{i1} - \bar{X}_{i2},  S_i = \sqrt{\frac{n_i}{n_{i1}n_{i2}}\cdot\frac{(n_{i1}-1) S^2_{i1} + (n_{i2}-1) S^2_{i2}}{(n_i-2)}},
\end{equation}}
where \( \bar{X}_{ik} = \frac{1}{n_{ik}} \sum_{j \in \mathcal{N}_{ik}} X_{ij} \) and \( S^2_{ik} = \frac{1}{n_{ik}-1} \sum_{j \in \mathcal{N}_{ik}} (X_{ij} - \bar{X}_{ik})^2 \) for \( k = 1, 2 \). 
The final step transforms pairs \((V_i, S_i)\) and \((V^0_i, S_i)\) into standardized statistics:
\begin{equation}\label{eq:T}
T_i = \Phi^{-1}\left\{G_{t,n_i-2}(V_i / S_i)\right\}\quad \mbox{and}\quad T^0_i = \Phi^{-1}\left\{G_{t,n_i-2}(V^0_i / S_i)\right\}, 
\end{equation}
where \( G_{t,n_i-2}(\cdot) \) is the CDF of a \(t_{n_i - 2} \)-distribution. Figure \ref{fig:1} in Section \ref{sec:diagram} of the Supplement provides a schematic illustration of the data processing steps. Let \(\mathbf{T} = (T_i)_{i=1}^m\) and \(\mathbf{T}^0 = (T^0_i)_{i=1}^m\) denote the test samples and calibration samples, respectively.

\begin{remark}\label{rem:small-n}
\rm{We briefly mention two technical issues. First, for \( n_i = 2 \), we let \( T_i = \frac{X_{i1} + X_{i2}}{\sqrt{2}} \) and \( T^0_i = \frac{X_{i1} - X_{i2}}{\sqrt{2}} \). For \( n_i = 3 \), we replace \( S_i \) with \( S_{i1} \); then $(T_i, T_i^0)$ are still computed using \eqref{eq:T}. Second, it may appear plausible to utilize the sample standard deviation \(S_i^* = \sqrt{\frac{1}{n_i - 1} \sum_{j=1}^{n_i} (X_{ij} - \bar{X}_i)^2}\) in the construction. However, as demonstrated in Section \ref{insights and counter examples} of the Supplement, the use of \(S_i^*\) undermines a crucial exchangeability condition.}
\end{remark}

 The following theorem establishes that \((T_i^0, T_i)\) are pairwise exchangeable when \(i \in \mathcal{H}_0\). This exchangeability ensures that our self-calibrated samples accurately mirror their counterparts from the empirical null distribution. 

\begin{theorem}\label{lem:SENS}
Under model \eqref{equ:model} and Assumption \ref{ass:errors}, the samples $\left\{(T_i, T_i^0): i\in[m]\right\}$ constructed via \eqref{eq:VS} and \eqref{eq:T} are pairwise exchangeable under the null, i.e. 

\begin{equation}\label{def:pw-ex}
\mbox{$\forall i\in\mathcal{H}_0$,\quad $(T_i,T^0_i\mid \mathbf{T}_{-i},\mathbf{T}^0_{-i})~\stackrel{d}=~(T^0_i,T_i\mid \mathbf{T}_{-i},\mathbf{T}^0_{-i})$},
\end{equation} 
where $\mathbf{T}_{-i}=(T_1,\cdots,T_{i-1},T_{i+1},\cdots,T_m)$ and $\mathbf{T}^0_{-i}=(T^0_1,\cdots,T^0_{i-1},T^0_{i+1},\cdots,T^0_m)$.
\end{theorem}

Our framework, which requires pairwise exchangeability, differs from a typical SSMT setup, which assumes joint exchangeability (i.e., labeled null samples follow a common distribution and are exchangeable with the null samples in the test data). The joint exchangeability fails to hold for two reasons. First, \(S_i\) induces correlation between \(T_i\) and \(T^0_i\), whereas such correlation is absent between \(T^0_i\) and \(T^0_j\) for \(i \neq j\). Second, according to Assumption \ref{ass:errors}, the random errors can be heterogeneously distributed across the units, violating the exchangeability between \(T_i\) and \(T_j\) for \(i \neq j\). Moreover, the construction of \(\mathbf{T}\) and \(\mathbf{T}^0\) utilizes an innovative ``plus and minus'' technique, which creates a test sample \(T_i\) that closely resembles the conventional \(z\)-statistic, effectively avoiding the efficiency loss typically associated with standard sample-splitting techniques.


\subsection{The two-sample scenario}\label{subsec:two sample case}

Suppose we collect \( n_{xi} \) and \( n_{yi} \) repeated measurements under two conditions:

\begin{equation}\label{equ:model-two-sample}
\begin{aligned}
  X_{ij} &= \mu_{xi} + \epsilon_{xij}, \quad \mathbb{E}(\epsilon_{xij}) = 0, \quad i \in [m], \quad j \in [n_{xi}], \\
  Y_{ij} &= \mu_{yi} + \epsilon_{yij}, \quad \mathbb{E}(\epsilon_{yij}) = 0, \quad i \in [m], \quad j \in [n_{yi}],
\end{aligned}
\end{equation}
where \( \mu_{xi} \) and \( \mu_{yi} \) represent the unknown effect sizes. To identify which units exhibit differential effects across two conditions, consider a multiple testing problem:

\begin{equation}\label{MT:formulation-two-sample}
H_{0,i}: \mu_{xi} = \mu_{yi} \quad \text{vs.} \quad H_{1,i}: \mu_{xi} \neq \mu_{yi}; \quad i \in[m]. 
\end{equation}

For the two-sample problem defined by \eqref{equ:model-two-sample} and \eqref{MT:formulation-two-sample}, we need a symmetry assumption similar to Assumption \ref{ass:errors}. Denote \(\mathbf{X}_i = (X_{ij}: j \in [n_{xi}])\), \(\mathbf{Y}_i = (Y_{ij}: j \in [n_{yi}])\) and  \(\mathbf{K_{-i} }= (\mathbf{X}_1, \ldots, \mathbf{X}_{i-1}, \mathbf{X}_{i+1}, \ldots, \mathbf{X}_m, \mathbf{Y}_1, \ldots, \mathbf{Y}_{i-1}, \mathbf{Y}_{i+1}, \ldots, \mathbf{Y}_m)\), for \(i \in [m]\).

\begin{assumption}\label{ass:errors-two-sample}
$\forall i \in \mathcal H_0$, \(\{\epsilon_{xij}: j \in [n_{xi}]\}\) are independent of \(\{\epsilon_{yij}: j \in [n_{yi}]\}\) conditional on $\mathbf{K_{-i}}$. Moreover, within unit $i$, \(\{\epsilon_{xij}: j \in [n_{xi}]\}\) are iid conditional on $\mathbf{K_{-i}}$, $\{\epsilon_{yij}: j \in [n_{yi}]\}$ are iid conditional on $\mathbf{K_{-i}}$. The errors obey zero-symmetric density functions: 

\begin{equation}\label{cond:two-symmetric-error}
  f_{xi}^{\epsilon}(-x \mid \mathbf{K_{-i}}) = f_{xi}^{\epsilon}(x \mid \mathbf{K_{-i}}), \quad f_{yi}^{\epsilon}(-y \mid \mathbf{K_{-i}}) = f_{yi}^{\epsilon}(y \mid \mathbf{K_{-i}}).
\end{equation} 
\end{assumption}

Assume that \( n_{xi} \geq 4 \) and \( n_{yi} \geq 4 \) for all \( i \in [m] \). The construction of the test and calibration samples follows a strategy analogous to the one-sample case: the terms \(\bar{X}_{i1}\) and \(\bar{X}_{i2}\) in \eqref{eq:VS} are replaced by \((\bar{X}_{i1}-\bar{Y}_{i1})\) and \((\bar{X}_{i2}-\bar{Y}_{i2})\) in \eqref{equ:VS-two-sample}, respectively. The standardization factor \(S_i\) is carefully designed to ensure the exchangeability condition. As we shall see, the proposed FDR procedure is based entirely on the pairs \(\{(T_i, T_i^0): i\in[m]\}\). Hence we deliberately use the same notation $T_i$ and $T_i^0$ for both the one-sample and two-sample cases to emphasize this unified framework.

The first step randomly partitions the observations as $\{X_{ij}: j\in[n_{xi}]\}=\{X_{ij}: j\in\mathcal{N}_{xi1}\}\cup \{X_{ij}: j\in\mathcal{N}_{xi2}\}$ and $\{Y_{ij}: j\in[n_{yi}]\}=\{Y_{ij}: j\in\mathcal{N}_{yi1}\}\cup \{Y_{ij}: j\in\mathcal{N}_{yi2}\}$. Let \( \bar{X}_{ik} = \frac{1}{n_{xik}} \sum_{j \in \mathcal{N}_{xik}} X_{ij} \), and \( \bar{Y}_{ik} = \frac{1}{n_{yik}} \sum_{j \in \mathcal{N}_{yik}} Y_{ij} \), for \( k = 1, 2 \). Further, let $S_{xik}^2 = \frac{1}{n_{xik}-1} \sum_{j \in \mathcal{N}_{xik}} (X_{ij} - \bar{X}_{ik})^2$ and $S_{yik}^2 = \frac{1}{n_{yik}-1} \sum_{j \in \mathcal{N}_{yik}} (Y_{ij} - \bar{Y}_{ik})^2$. Denote $S_{xi}^2=\frac{(n_{xi1}-1) S^2_{xi1} + (n_{xi2}-1) S^2_{xi2}}{(n_{xi}-2)}$ and $S_{yi}^2=\frac{(n_{yi1}-1) S^2_{yi1} + (n_{yi2}-1) S^2_{yi2}}{(n_{yi}-2)}$.

The second step computes,  for \( i \in [m] \), 
{\small
\begin{equation}\label{equ:VS-two-sample}
V_i = (\bar{X}_{i1}-\bar{Y}_{i1})+(\bar{X}_{i2}-\bar{Y}_{i2}), V^0_i = (\bar{X}_{i1}-\bar{Y}_{i1})-(\bar{X}_{i2}-\bar{Y}_{i2}), S_i = \sqrt{\frac{n_{xi}}{n_{xi1} n_{xi2}}S_{xi}^2+\frac{n_{yi}}{n_{yi1} n_{yi2}}S_{yi}^2}. 
\end{equation}}

The final step transforms pairs \((V_i, S_i)\) and \((V^0_i, S_i)\) into standardized statistics:

\begin{equation}\label{equ:T-two-sample}
T_i = \Phi^{-1}\left\{G_{t,n_{xi}+n_{yi}-4}(V_i / S_i)\right\}\quad \mbox{and}\quad T^0_i = \Phi^{-1}\left\{G_{t,n_{xi}+n_{yi}-4}(V^0_i / S_i)\right\}, 
\end{equation}
where \( G_{t,n_{xi}+n_{yi}-4}(\cdot) \) is the CDF of a \(t_{n_{xi}+n_{yi}-4} \)-distribution. The pairwise exchangeability property can be similarly established for the two-sample case. 
\begin{theorem}\label{lem:SENS-two-sample}
Under model \eqref{equ:model-two-sample} and Assumption \ref{ass:errors-two-sample}, $\left\{(T_i, T_i^0): i \in [m]\right\}$, constructed via \eqref{equ:VS-two-sample} and \eqref{equ:T-two-sample}, satisfy the pairwise exchangeability condition \eqref{def:pw-ex}.
\end{theorem}

\section{Conformalized Multiple Testing with SENS}\label{sec:SENS}


We first outline the basic strategy for methodological development (Section \ref{subsec:sens-pre}), then detail the construction of conformity scores (Sections \ref{subsec:working-model} and \ref{subsec:sens-score}), and finally present the SENS Algorithm (Section \ref{subsec:sens-alg}) while investigating its theoretical properties (Section \ref{subsec:finite sample validity}).

\subsection{Basic strategy and general considerations}\label{subsec:sens-pre}

Sections~\ref{subsec:SENS-construction} and~\ref{subsec:two sample case} describe strategies for converting conventional multiple testing problems into the SSMT framework. We provide a unified treatment for both the one-sample and two-sample scenarios using the common notation~\(\mathbf{T}\) and~\(\mathbf{T}^0\).

Let \(g(\cdot)\) denote a score function derived from prior knowledge or training data, where the scores tend to be small under the alternative. One plausible approach to SSMT is the conformal Benjamini–Hochberg (cfBH) procedure, which first computes, for all $i \in [m]$,

\begin{equation}\label{eq:cp}
p_i\left(T_i\right) = \frac{1+\left|\left\{T \in \mathbf{T}^{0}: g\left(T\right) < g\left(T_i\right)\right\}\right|}{1+\left|\mathbf{T}^{0}\right|}, 
\end{equation}
and then applies the classical BH procedure \citep{benjamini1995controlling} to $p_i\left(T_i\right)$'s. If the elements in \(\{T_i: i \in \mathcal{H}_0\}\) and \(\{T_j^0: j \in [m]\}\) are jointly exchangeable, then the standardized ranks in \eqref{eq:cp} are valid \(p\)-values and satisfy the PRDS condition \citep{bates2023testing}, implying that the cfBH method is valid for FDR control. 

However, in our setup, \eqref{eq:cp} is no longer a valid p-value because \(T_i\) and \(T_i^0\) are only pairwise exchangeable under the null. To bypass the conventional $p$-value framework, we consider a class of scores $\mathbf{U}=(U_i)_{i=1}^m$ and $\mathbf{U}^0=(U_i^0)_{i=1}^m$ in the form

\begin{equation}\label{g-class}
 \left\{U_i\equiv g(T_i; \mathbf{T}, \mathbf{T}^0), U^0_i\equiv g(T^0_i; \mathbf{T}, \mathbf{T}^0): i\in[m]\right\}.    
\end{equation}

Let $\mathbf{U}_{-i}=(U_1,\dots,U_{i-1},U_{i+1},\dots,U_m)$ and $\mathbf{U}^0_{-i}=(U^0_1,\dots,U^0_{i-1},U^0_{i+1},\dots,U^0_m)$.
Then, $\mathbf{U}$ and $\mathbf{U}^0$ are pairwise exchangeable under the null if

\begin{equation}\label{pwex:scores}
\mbox{$\forall i\in\mathcal{H}_0$,\quad $(U_i,U^0_i\mid \mathbf{U}_{-i},\mathbf{U}^0_{-i})~\stackrel{d}=~(U^0_i,U_i\mid \mathbf{U}_{-i},\mathbf{U}^0_{-i})$}.
\end{equation}

Constructing a powerful score function \(g(\cdot)\) that effectively integrates both \(\mathbf{T}\) and \(\mathbf{T}^0\) is a challenging task. The following proposition, inspired by \cite{barber2015controlling}, offers guiding principles on how to exploit pairwise exchangeable data points [\eqref{def:pw-ex}] to develop pairwise exchangeable scores [\eqref{pwex:scores}]. 

\begin{proposition}
\label{prop:score-class}
Suppose $\mathbf{T}$ and $\mathbf{T}^0$ satisfy \eqref{def:pw-ex}. Let $(\mathbf{T},\mathbf{T}^0)_{\operatorname{swap}(\mathcal{J})}$ denote the operation of swapping $T_i$ and $T_i^0$ for each $i\in\mathcal J\subset[m]$. Consider a generic class of scores $(U_i, U_i^0)$ defined by \eqref{g-class}. Then $U_i$ and $U_i^0$ satisfy \eqref{pwex:scores} if, $\forall \mathcal{J} \subset[m]$,

\begin{equation}\label{equ:score function}
    g(\cdot;(\mathbf{T},\mathbf{T}^0)_{\operatorname{swap}(\mathcal{J})})=g(\cdot;(\mathbf{T}, \mathbf{T}^0)).
\end{equation} 
\end{proposition}

\subsection{The two-group model and oracle scores}\label{subsec:working-model}

This section employs the following two-group mixture model \citep{Efron_2010} as a working model to illustrate the principles of score construction outlined in Proposition \ref{prop:score-class}:

\begin{equation}\label{equ:working-model}
    T_i \stackrel{i.i.d.}{\sim} f(t) = (1 - \pi) f_0(t) + \pi f_{1}(t), \quad i \in [m].
\end{equation}

\begin{remark}\rm{
While an effective working model that approximates reality is crucial for precise inference, our method remains valid even if the model is misspecified -- as long as condition \eqref{pwex:scores} holds. Prominent works promoting the use of the empirical null \citep{efron2004large, jin2007estimating} fundamentally relied on Model \eqref{equ:working-model}, which has long been a staple in large-scale inference due to its simplicity and proven effectiveness. Although more generic and complex models for $\{T_i: i\in[m]\}$ (especially for heterogeneous units) might be considered, we focus on Model \eqref{equ:working-model} to maintain consistency with the established literature and to clearly highlight the unique merits of our proposal.}
\end{remark}

The thresholding rule based on \(\text{lfdr}(t) = (1-\pi)\frac{f_0}{f}(t)\) is optimal in that it maximizes power subject to an FDR constraint \citep{sun2007oracle, HelRos19}. Next, we discuss constructing data-driven score functions that effectively emulate lfdr according to the principles outlined in Proposition \ref{prop:score-class}.

\subsection{Construction of data-driven scores}\label{subsec:sens-score}

In practice, the unknown \(f\) and \(f_0\) must be estimated from data. Proposition \ref{prop:score-class} indicates that to achieve pairwise exchangeability among scores, it is essential to construct \emph{symmetric estimators}, wherein the test samples \(\mathbf{T}\) and calibration samples \(\mathbf{T}^0\) play equal roles. This section develops innovative conformalization techniques, including \emph{mixing} and \emph{filtering}, which transform conventional estimators into symmetric estimators. We first focus on estimating \( f \), then turn to the more complex task of estimating \( f_0 \).

Conventional kernel estimators for \( f \), which rely solely on the test samples \(\mathbf{T}\), violate the principles outlined in Proposition \ref{prop:score-class}.
To ensure a balanced contribution from both \(\mathbf{T}\) and \(\mathbf{T}^0\), we adopt a mixing technique, inspired by \cite{marandon2024adaptive}, that directly estimates \( f_\text{mix} =(f_0 + f)/2 \) as a replacement for \( f \). The resulting estimator is given by

\begin{equation} \label{equ:hat_f_mix}
    \hat{f}_{mix}~\coloneqq~\hat{f}_{mix}(t;\mathbf{T}, \mathbf{T}^0)~=~\frac{\sum_{i=1}^m [K_{h_{mix}}(t - T_i) + K_{h_{mix}}(t - T^0_i)]}{2m}, 
\end{equation}
where \( K_{h}(t) = h^{-1} K(t / h) \) represents symmetric kernel functions that satisfy:
$\int K(t) \, dt = 1$,  $\int t K(t) \, dt = 0$,  and $\int t^2 K(t) \, dt < \infty.$ 
We recommend using well-established techniques for selecting the bandwidth \( h_{mix} \) (cf.  \citealp{silverman2018density, sheather1991reliable}), ensuring that \( h_{mix}((\mathbf{T}, \mathbf{T}^0)_{\Pi}) = h_{mix}(\mathbf{T}, \mathbf{T}^0) \) holds, where \(\Pi\) is any permutation of the elements in the vector \( (\mathbf{T}, \mathbf{T}^0) = (T_1, \ldots, T_m, T^0_1, \ldots, T^0_m) \). 

When estimating \( f_0 \), the mixing technique becomes ineffective as it is infeasible to extract a pure null sample from the test data. We propose to generate a new sample \(\tilde{\mathbf{T}}^0 = (\tilde{T}^0_i)_{i=1}^m\) through the following transformation: 

\begin{equation}\label{equ:camouflaged T_0}
\tilde{T}_i^0 = \left\{
\begin{array}{ll}
T_i & \text{if } |T_i| \leq |T^0_i| \\
T^0_i & \text{otherwise}
\end{array}
\right..
\end{equation}
According to Assumption \ref{ass:errors} and our strategy for constructing $\mathbf T$, the null density \( f_0 \) must be symmetric about 0. Therefore, we propose the following zero-symmetric kernel [option ``KN'' in Algorithm \ref{alg:SENS}] estimator:

\begin{equation}\label{f0-symm}
\hat{f}_0(t) \coloneqq \hat{f}_0(t; \mathbf{T}, \mathbf{T}^0) = \frac{\sum_{i=1}^m K_{h_0}(t - \tilde{T}^0_i) + \sum_{i=1}^m K_{h_0}(t + \tilde{T}^0_i)}{2m},
\end{equation}
where \( h_0 \) is a data-driven bandwidth satisfying \( h_{0}((\tilde{\mathbf{T}}^0, -\tilde{\mathbf{T}}^0)_{\Pi}) = h_{0}(\tilde{\mathbf{T}}^0, -\tilde{\mathbf{T}}^0) \). This zero-symmetric estimator, which leverages the symmetry constraint, is both more efficient and interpretable than a kernel estimator that utilizes \(\tilde{\mathbf{T}}^0\) alone. 

\begin{remark}\rm{
The strategy in \eqref{equ:camouflaged T_0} is to \emph{filter out} the true identities of \(T_i\) and \(T_i^0\), thereby enforcing a symmetrized estimator in which \(T_i\) and \(T_i^0\) are treated equally. Moreover, \eqref{equ:camouflaged T_0} retains the smaller values of \(|T_i|\) and \(|T^0_i|\), as larger values are more likely to come from the alternative. While this filtering technique shares conceptual similarities with the masking approach \citep{lei2018adapt} and the symmetrization method \citep{zhao2024false}, it is specifically designed for different objectives and utilizes distinct strategies. We formally quantify the bias in \eqref{f0-symm} and present a debiasing method in Section \ref{subsec:bias of filter} of the Supplement. However, we have chosen to adopt the current method as it is simple and the validity is unaffected by the bias. }
\end{remark}

If it is known that \( f_0 \) is Gaussian, one can directly estimate \( f_0 \) using the mixed sample \((\mathbf{T}, \mathbf{T}^0)\). The mixture density of \((\mathbf{T}, \mathbf{T}^0)\), using the notation from \eqref{equ:working-model}, is given by \( f_\text{mix} = (1 - \frac{\pi}{2}) f_0(t) + \frac{\pi}{2} f_1(t) \). The Jin-Cai estimator \citep{jin2007estimating}, briefly described in Section \ref{subsec:JC} of the Supplement, remains a viable option for constructing pairwise exchangeable scores [option ``JC'' in Algorithm \ref{alg:SENS}]. The relative strengths of the kernel method and Jin-Cai method are investigated for both Gaussian and non-Gaussian $f_0$ in Section \ref{sec:simulation}.

\begin{remark}\rm{While conventional empirical Bayes FDR methods aim to estimate the empirical null, we aim to estimate the working null $f_0$ in \eqref{equ:working-model}. In contrast to  \cite{efron2004large} and \cite{jin2007estimating}, our framework remains valid even if our working null deviates from the true empirical null. To distinguish various concepts of null distributions, we provide a detailed discussion in Section \ref{sec:null distribution} of the Supplement.
}
\end{remark}

Consider the symmetric estimators \(\hat{f}_0(t)\) and \(\hat{f}_{mix}(t)\) defined in \eqref{f0-symm} and \eqref{equ:hat_f_mix}. We propose to construct scores via 

\begin{equation}\label{equ:g(t)} 
(U_i, U^0_i) = \{ g(T_i), g(T^0_i)\}, \quad \mbox{where} \quad g(t) = {\hat{f}_0(t)}/{\hat{f}_{mix}(t)}.
\end{equation}

The next proposition shows that scores constructed via \(g(t)\) are pairwise exchangeable.
\begin{proposition}\label{pro1}
 Consider \(g(t)\) constructed via \eqref{equ:g(t)}. Then \(\mathbf{U} = \{U_i = g(T_i)\}_{i=1}^m\) and \(\mathbf{U}^0 = \{U^0_i = g(T^0_i)\}_{i=1}^m\) satisfy the pairwise exchangeability condition \eqref{pwex:scores}.
\end{proposition}

Our numerical studies demonstrate that the ``conformalized'' function \(g(t)\) remains highly effective.  Furthermore, Section \ref{sec:asymptotic optimality} demonstrates that \(g(t)\) is asymptotically optimal under certain regularity conditions.

\subsection{The SENS Algorithm}\label{subsec:sens-alg}

We discuss how the Selective SeqStep+ algorithm \citep{barber2015controlling}, henceforth referred to as the BC algorithm, can be adapted to our SSMT framework. The BC algorithm was originally employed by the knockoff filter for variable selection in regression. The knockoff filter differs from SENS in several key aspects; see Remark~\ref{rem:knockoff} for related discussion. 

The rejection rule can be determined in three steps. First, a test statistic \(G_i\) is constructed for each testing unit \(i\) using an anti-symmetric function $\gamma(x,y)$:

\begin{equation}\label{eq:anti-sym}
G_i = \gamma\left(U_i, U^0_i\right) = \operatorname{sign}\left(U^0_i - U_i\right) \cdot \left[\exp\left(-U_i\right) \vee \exp(-U^0_i)\right], \quad \forall i \in [m].
\end{equation}

\begin{remark}\rm{
The anti-symmetric function \(\gamma(x, y)\) is carefully selected to emulate a thresholding rule based on the ranking of \(U_i\), a strategy shown to be optimal for FDR analysis in Section~\ref{sec:asymptotic optimality}. Detailed explanations of the rationale for this choice, alternative anti-symmetric functions, and supporting simulation comparisons are provided in Supplement Section~\ref{aux-sim:anti-func}.
}
\end{remark}
Next, the threshold is chosen via a mirror process: 

\begin{equation}\label{equ:tau}
\tau = \inf \left\{\lambda \in \{|G_i|\}_{i=1}^m : \frac{1 + \sum_{i=1}^{m} \mathbb{I}(G_i \leq -\lambda)}{\sum_{i=1}^{m} \mathbb{I}(G_i \geq \lambda)} \leq \alpha \right\}.
\end{equation}
Finally, we reject \(H_{0,i}\) if \(G_i \geq \tau\). By mathematical conventions, $\tau=\infty$ if the infimum is taken over an empty set, whence the algorithm makes no rejections. We summarize the SENS Algorithm below. 

\singlespacing

\begin{breakablealgorithm}
\caption{The SENS Algorithm}  
\label{alg:SENS}
 \begin{algorithmic}[1]
    \Require Observations $\{X_{ij}:j\in[n_i]\}_{i=1}^m$, target FDR level $ \alpha $, option from \{``JC'', ``KN''\}. 
    \Ensure The rejection set $ \mathcal{R} $.
    
    \State Randomly partition the observations and compute \( (T_i, T^0_i) \), \( i \in [m] \).         
    \State Estimate \( \hat{f}_{mix} \) using Equation \eqref{equ:hat_f_mix}.
     \If{option==``JC''}
       \State Estimate $\hat{f}_0$ using the Jin-Cai method [\eqref{equ:JC}]. 
    \ElsIf{option==``KN''}
        \State Estimate $\hat{f}_0$ using the kernel method [\eqref{f0-symm}].
    \EndIf 
    \State Compute \( g(t) \) via \eqref{equ:g(t)}. Let \(\mathbf{U} = \{U_i = g(T_i)\}_{i=1}^m\) and \(\mathbf{U}^0 = \{U^0_i = g(T^0_i)\}_{i=1}^m\). 
    \State Compute test statistics \(\{G_i: i\in[m]\}\) via \eqref{eq:anti-sym}. 
    \State Determine the threshold \( \tau \) via \eqref{equ:tau} and let 
    \(\mathcal{R} = \{i \in [m]: G_i \geq \tau\}.\) 
 \end{algorithmic}
\end{breakablealgorithm}

\medskip
\spacingset{1} 

SENS employs an empirical Bayes approach: it extracts structural  information from the data and utilizes scores that emulate the lfdr. This distinguishes it from frequentist BC-type algorithms; detailed comparisons are provided in Section~\ref{sec:comparisons model-free} of the Supplement. To reduce uncertainties associated with sample-splitting, we introduce a derandomized version of the SENS algorithm in Section~\ref{subsec:derandomized SENS} of the Supplement.

\subsection{Impacts of distribution shifts and finite sample validity}\label{subsec:finite sample validity}

This section presents a unified theory based on $e$-values that (a) characterizes the limitations of conventional methods under distribution shifts, and (b) establishes the finite-sample validity of the SENS Algorithm for FDR control. Denote

\begin{equation}\label{equ:e-value}
e_i = \frac{m \mathbb{I}(G_i \geq \tau)}{1 + \sum_{j=1}^{m} \mathbb{I}(G_j \leq -\tau)},\quad i \in [m],
\end{equation}
where \(G_i\) and \(\tau\) are defined in \eqref{eq:anti-sym} and \eqref{equ:tau}, respectively. Let \(e_{(1)} \geq e_{(2)} \geq \cdots \geq e_{(m)}\) denote the order statistics of  \(\{e_i: i \in [m]\}\). The rejection set of the $e$-BH procedure \citep{wang2022false, ren2024derandomised} is given by $\mathcal{R}_{\text{ebh}} = \{i \in [m] : e_i \geq e_{(\hat{k})}\}$, where $\hat{k} = \max \left\{i : \frac{i e_{(i)}}{m} \geq \frac{1}{\alpha}\right\}$. The following proposition demonstrates that the BC-based Algorithm \ref{alg:SENS} is equivalent to this $e$-BH-based algorithm.

\begin{proposition}\label{pro:e-SENS}
    If we implement the $e$-BH procedure with $e$-values in \eqref{equ:e-value}, then the rejection set $\mathcal{R}_{\text {ebh }}=\mathcal{R}$, where $\mathcal{R}$ is the set of rejected hypotheses by Algorithm \ref{alg:SENS}.
\end{proposition}

Leveraging the $e$-BH perspective of SENS outlined in Proposition \ref{pro:e-SENS}, we demonstrate that distribution shifts from the empirical null to the theoretical null render the $e$-values in \eqref{equ:e-value} invalid. Specifically, within the framework of SSMT, \(\mathbf{T}^0\) and \(\mathbf{T}\) may be viewed as samples drawn from the theoretical null and empirical null distributions, respectively. Suppose $m$ pairs of scores \(\{(U_i^0, U_i),\, i\in[m]\}\) are constructed for deployment within a BC algorithm. In the presence of distribution shifts, \(U_i^0\) and \(U_i\) become non-exchangeable. Our subsequent theory quantifies how the breakdown of exchangeability affects the validity of $e$-values and further characterizes the impact on FDR analyses.

Let \( f \) be a generic notation representing the probability density or mass function. 
Consider the following two conditional distributions:  
$p_i^{U, U^0}(u, v) \coloneqq f(U_i = u, U_i^0 = v \mid \mathbf{U}_{-i}, \mathbf{U}^0_{-i})$ and $ p_i^{U^0, U}(u, v) \coloneqq f(U_i^0 = u, U_i = v \mid \mathbf{U}_{-i}, \mathbf{U}^0_{-i})$. The amount of shift can be captured by the degree of discrepancy between the two distributions \(p_i^{U, U^0}\) and \(p_i^{U^0, U}\). We quantify this discrepancy using the following observed Kullback-Leibler (KL) divergence:

\begin{equation}\label{equ:KL}
\widehat{\mathrm{KL}}_i\coloneqq \log \left\{ \frac{p_i^{U, U^0}(U_i, U_i^0)}{p_i^{U^0, U}(U_i, U_i^0)} \right\}.
\end{equation}
It follows that  
$
\mathbb{E}\left[\widehat{\mathrm{KL}}_i\right] = \mathrm{d}_{\mathrm{KL}}\left(p_i^{U, U^0} \| p_i^{U^0, U}\right),
$
where \(\mathrm{d}_{\mathrm{KL}}(p_1 \| p_2)\) denotes the KL divergence between two distributions \(p_1\) and \(p_2\).

The following theorem, adapted from the robust knockoff theory in \cite{barber2020robust}, formalizes the impact of exchangeability -- specifically, the swapping of \(U_i\) and \(U_i^0\) in the conditional distribution \(p_i^{U, U^0}\) -- on the validity of $e$-values and subsequent FDR analyses.

\begin{theorem}\label{the:robust knockoff}
Consider the $e$-values defined in \eqref{equ:e-value} and the observed KL divergence defined in \eqref{equ:KL}. Then we have (a) for any $\epsilon \geq 0$

\begin{equation*}\label{equ:approximate $e$-values}
    \mathbb{E}\left[\textstyle\sum_{i\in\mathcal{H}_0}e_i\right] \leq \inf _{\epsilon \geq 0}\left\{m \left[\text{\normalfont e}^\epsilon + \textstyle\sum_{i\in\mathcal{H}_0}\mathbb{P}\left(\widehat{\mathrm{KL}}_i>\epsilon\right)\right]\right\}.
\end{equation*}
(b) If we apply the $e$-BH algorithm with these $e$-values, the corresponding FDR level satisfies 

\[
\mathrm{FDR} \leq \inf _{\epsilon \geq 0}\left\{\alpha \left[\text{\normalfont e}^\epsilon + \sum_{i\in\mathcal{H}_0}\mathbb{P}\left( \widehat{\mathrm{KL}}_i>\epsilon\right)\right]\right\}.
\]
\end{theorem}

\begin{remark}\label{rem:knockoff}
\rm{
We highlight several key distinctions from the theory presented in \cite{barber2020robust}. Firstly, our framework focuses on the SSMT setup, whereas the robust knockoff approach considers variable selection in regression contexts. Unlike the knockoff method, our approach does not involve a response variable and relies exclusively on calibration and test data. Secondly, our method requires pairwise exchangeability only for null cases, while Model-X knockoff assumes this property for \emph{all X-variables}. Finally, although the two proofs share overarching techniques, the theory within our SSMT framework diverges from that in \cite{barber2020robust}, conveying quite different interpretations.}
\end{remark}

Theorem \ref{the:robust knockoff} highlights that, within the SSMT framework, the FDR analysis can be undermined by distribution shifts, which often result from conventional practices that directly utilize the theoretical null (or biased null training samples) in analysis. In contrast, the SENS framework eliminates the distribution shift by employing self-calibrated samples that accurately represent the true null distribution of test samples. In conjunction with carefully constructed conformal scores, we have \(\widehat{\mathrm{KL}}_i = 0\) for all \(i \in \mathcal{H}_0\), which guarantees the validity of Algorithm \ref{alg:SENS} (as a corollary of Proposition \ref{pro:e-SENS} and Theorem \ref{the:robust knockoff}). 

\begin{corollary}\label{the:SENS}
(Finite-sample validity of SENS). Consider model \eqref{equ:model}. Suppose that (a) Assumption \ref{ass:errors} holds; (b) $(U_i,U^0_i)$ are constructed via Algorithm \ref{alg:SENS}, and there is no tie between $ U_i $ and $U^0_i $ almost surely. Then the $e$-values defined in \eqref{equ:e-value} are exact generalized $e$-values: $\mathbb{E}\left(\sum_{i\in\mathcal{H}_0}e_i\right) \leq m$. Consequently, Algorithm \ref{alg:SENS} controls the FDR at level $\alpha$.
\end{corollary}

\section{Power analysis and asymptotic optimality}\label{sec:asymptotic optimality}
We present a power analysis of BC-type algorithms, upon which the SENS method is built, within the SSMT framework. Although the analysis is conducted under idealized conditions and relies on strong assumptions, our new theory is significant, as they delineate the conditions under which BC-type algorithms achieve the optimality benchmark. 

Our analysis extends beyond the specific \(T_i\) and \(T^0_i\) presented in previous sections. Consider pairs \(\{(T_i, T_i^0) : i \in [m]\}\), which may exhibit dependence within each pair but are mutually independent across pairs, and which obey the following model: 

\begin{equation}  
\begin{aligned}  
    T_i \stackrel{i.i.d.}{\sim} f_m(t) = (1-\pi)f_0(t) + \pi f_{1m}(t), \quad T^0_i \stackrel{i.i.d.}{\sim} f_0(t),\quad i \in [m].
\end{aligned}  
\label{equ:theoretical-working-model}  
\end{equation}  
The subscript \( m \) in \( f_m \) and \( f_{1m} \) reflects the asymptotic regime in our theoretical analysis, where \( f_{1m} \) varies with the dimension \( m \). Given that \( T_i \) and \( T_i^0 \) are typically standardized, the null density \( f_0 \) remains the same for all \( m \).

\begin{remark}\label{rem:asy}\rm{
The SENS Algorithm, which utilizes the working model \eqref{equ:working-model}, is a conformal method that is \emph{model-free}: the validity of FDR control is guaranteed even when the working model is mis-specified. In contrast, our power analysis is not model-free, and the asymptotic optimality of SENS is achieved only when \eqref{equ:theoretical-working-model} aligns with the underlying true model. }
\end{remark}

Following the framework outlined in \cite{sun2007oracle}, our power analysis is organized into two parts. The first part (Proposition \ref{pro3}) derives an oracle procedure that aims to maximize the ETP while satisfying a constraint on the marginal FDR, defined as \(\operatorname{mFDR}=\mathbb{E}[\sum_{i \in \mathcal{H}_0} \delta_i]/\mathbb{E}[\sum_{i \in [m]}\delta_i]\). The mFDR is asymptotically equivalent to the FDR under some regularity conditions \citep{genovese2002operating, tony2019covariate} and has been employed to facilitate the theoretical analysis. The second part investigates the conditions under which SENS achieves the benchmark ETP level of the oracle procedure asymptotically (Theorem \ref{the:Asymptotic optimality}).

Consider an oracle setting where $f_0$, $\pi$ and $f_{1m}$ in model \eqref{equ:theoretical-working-model} are known. Define the (oracle) score function $r_m(t)={2 f_0(t)}/\{f_0(t)+f_m(t)\}$.
In Section \ref{sec:compatible} of the Supplement, we show that $r_m(T_i)-\hat{f}_0(T_i)/{\hat{f}_{mix}(T_i)}=o_p(1)$. The following proposition establishes \( r_m(t) \) as the optimal score function, justifying our choice of data-driven score function \eqref{equ:g(t)}. 

\begin{proposition}\label{pro3}
  Consider $\{T_i: i\in[m]\}$ generated from \eqref{equ:theoretical-working-model}. Let $\pmb\delta$ denote a generic decision rule based on $\{T_i:i\in[m]\}$ and $\mathcal{D}_\alpha$ be the collection of $\pmb\delta$ such that $\text{mFDR}_{\boldsymbol{\delta}} \leq \alpha$. Consider a class of decision rules $\boldsymbol{\delta}^R(\lambda)=\{\mathbb{I}(R_i\leq \lambda): i\in[m]\}$. Define $\lambda^R={\sup}\{\lambda: \text{mFDR}_{\boldsymbol{\delta}^R(\lambda)}\leq\alpha\}$ and $\boldsymbol{\delta}^\text{OR}=(\mathbb{I}\{R_i\leq \lambda^R\}: i\in[m])$. Then $\pmb\delta^\text{OR}$ is optimal in the sense that $\forall \boldsymbol{\delta} \in \mathcal{D}_{\alpha}$, $\text{ETP}_{\boldsymbol{\delta}}\leq \text{ETP}_{\boldsymbol{\delta}^\text{OR}}$. 
\end{proposition}

Proposition \ref{pro3} offers a benchmark for characterizing the optimal performance of any \( \alpha \)-level mFDR procedure. We will next examine whether SENS asymptotically achieves this optimal power benchmark. Our analysis requires the following additional assumption.

\begin{assumption}\label{ass:optimality} 
Let $\phi_{\sigma}(t-\mu)$ denote the density function of a $\mathcal N(\mu,\sigma^2)$ variable. In the random mixture model \eqref{equ:theoretical-working-model}, the null and non-null densities are given by

\begin{equation}\label{model:nmix}
f_0(t)=\phi_{\sigma_0}(t-\mu_0)\; \mbox{ and }\; f_{1m}(t)=\phi_{\sigma_0}(t-\mu_m), \; \mbox{with \(\mu_m \to \infty \) \;as\; \( m \to \infty \)}. 
\end{equation}
\end{assumption}

The two-point Gaussian mixture model \eqref{model:nmix} has been widely used in high-dimensional sparse inference \citep{donoho2004higher, meinshausen2006estimating, cai2007estimation, TonyCaioptimal2017, arias2017distribution}. Under the Gaussian assumption, we apply the Jin--Cai method to estimate \( \hat{f}_0 \). Moreover, the $\alpha$-mixing condition in \cite{jin2007estimating} is automatically satisfied by Model~\eqref{equ:theoretical-working-model}; therefore, the Jin--Cai method yields strongly consistent estimates of \( \mu_0 \) and \( \sigma_0 \). Corollary~\ref{the:SENS} and the next theorem together establish the asymptotic optimality of the SENS algorithm.

\begin{theorem}\label{the:Asymptotic optimality}
    (Asymptotic optimality of SENS). Suppose $\{T_i,T^0_i:i\in[m]\}$ are generated from  model \eqref{equ:theoretical-working-model} and Assumption \ref{ass:optimality} holds. Then SENS  (Algorithm \ref{alg:SENS}) with the option set to ``JC'' satisfies $\text{ETP}_{SENS}/\text{ETP}_{\boldsymbol{\delta}^\text{OR}}=1+o(1)$. 
\end{theorem}

In contrast to existing power analyses (cf.~\citealp{arias2017distribution, Dai2023splitting}), we focus on a practically relevant scenario characterized by the discovery boundary \citep{TonyCaioptimal2017}; see Supplement Section~\ref{subsec:power analysis} for further discussion. Our power analysis provides novel and valuable insights. Specifically, \cite{sun2007oracle} demonstrates that the adaptive $z$-value (AZ) procedure asymptotically achieves oracle performance under the two-group model~\eqref{equ:working-model}. However, BC-type algorithms (such as SENS) and the AZ procedure operate in fundamentally different ways: the former is based on mirror processes, while the latter employs a moving average of the estimated lfdrs. Within the challenging SSMT framework, we have developed innovative techniques to establish the asymptotic optimality of SENS (and consequently, the broader class of BC-type algorithms). A detailed comparison of SENS with AZ is provided in  Section~\ref{subsec:comparison} of the Supplement.

\section{Simulation}\label{sec:simulation}

We present simulation results investigating the performance of SENS for both the one-sample case (Section \ref{subsec:sim-one sample}) and  two-sample case (Section \ref{subsec:sim-two sample}). Additional simulation results are provided in the Supplement, detailing the following: (a) the choice of anti-symmetric functions (Appendix \ref{aux-sim:anti-func}); (b) additional comparisons under the SSMT setup (Appendix \ref{simu:SSMT}); (c) the derandomized SENS Algorithm (Appendix \ref{simu:dsens}); and (d) further comparisons with recent model-free methods (Appendix \ref{sec:add-simu-res}). Given the numerous methods being compared and our use of central trends (FDR and AP) to illustrate interesting patterns, the variance comparisons for FDP and true positive proportion (TPP), which are also of great interest, are provided in Section \ref{subsec:variance} of the Supplement.

Unless otherwise specified, the nominal FDR level is \(\alpha = 0.05\) and the number of tests is \(m = 2000\). The reported FDR level is obtained as the average of the false discovery proportion (FDP), defined as \((\sum_{i \in \mathcal{H}_0} \delta_i) / \max\left\{\sum_{i \in [m]} \delta_i, 1\right\}\), and the average power (AP) is reported as the average of \((\sum_{i \in \mathcal{H}_1} \delta_i) / |\mathcal{H}_1|\), both averaged across 200 independent datasets. 

Below we summarize the notation for the different methods included in our comparison:
\begin{description}
\item (a) SENS\_JC: Algorithm \ref{alg:SENS} with the option set to ``JC'';  
\item (b) SENS\_KN: Algorithm \ref{alg:SENS} with the option set to ``KN'';  
\item (c) BH\_TN: BH method applied to the \(p\)-values converted from $z$-values based on the theoretical null distribution \(\mathcal{N}(0,1)\);  
\item (d) BH\_EEN: BH method applied to the \(p\)-values converted from $z$-values based on the estimated empirical null distribution (via the Jin-Cai method); 
\item (e) sfBH:  The BH procedure applied to two-sided valid $p$-values for testing symmetry, specifically using the sign-flipping method as outlined in \cite{arlot2010some}, with 1000 bootstrap iterations; see Appendix \ref{subsec:sfBH} for a detailed description;
\item (f) stBC:  The vanilla BC method in \cite{arias2017distribution} applied to the $t$-statistics \( T^*_i \) based on all repeated measurements for study unit \( i \);
\item (g) RESS: A model-free procedure based on sample-splitting \citep{zou2020new};
\item (h) CLIPPER: a model-free procedure for two-sample multiple testing \citep{ge2021clipper}.
\end{description}

\subsection{The one-sample case}\label{subsec:sim-one sample}

The following data generating model is considered:

\begin{equation}\label{model:simu-var}
\begin{aligned}
X_{ij} & = \mu_i + \epsilon_{ij}, j\in[n],\quad \mu_i \stackrel{i.i.d.}{\sim}(1 - \pi) \delta_0 + \pi \mathcal{N}(-\mu,\mu^2),\quad \sigma_i\stackrel{i.i.d.}{\sim} \mathcal{U}(0.05,\sigma_{\text{max}}),\\
\epsilon_{ij}\mid \sigma_i &\stackrel{i.i.d.}{\sim} (1-\beta)\mathcal{N}(0, \sigma_i^2) + \frac{3\beta}{4}\mathcal{U}(-\sqrt{3}\sigma_i, \sqrt{3}\sigma_i)+\frac{\beta}{4}\text{Laplace}(0, \sigma_i/\sqrt{2}).
\end{aligned}
\end{equation}

\textbf{Simulation 1.} To investigate the impacts of distribution shifts and the Gaussian assumption on the performance of classical methods, we compare SENS\_JC, SENS\_KN, BH\_TN, and BH\_EEN across the following three settings:
\begin{description}
\item (a) \(\mu=3\), \(n = 4\), \(\beta = 1\), $\sigma_{\text{max}}=0.1$, varying \(\pi\);
\item (b) \(\pi=0.1\), \(\mu=3\), \(\beta = 1\), $\sigma_{\text{max}}=0.2$, varying \(n\); 
\item (c) \(\pi=0.05\), \(\mu=3\), \(n=4\), $\sigma_{\text{max}}=0.06$, varying \(\beta\).
\end{description}

\begin{figure}[htbp!]
    \centering
    \includegraphics[width=6in]{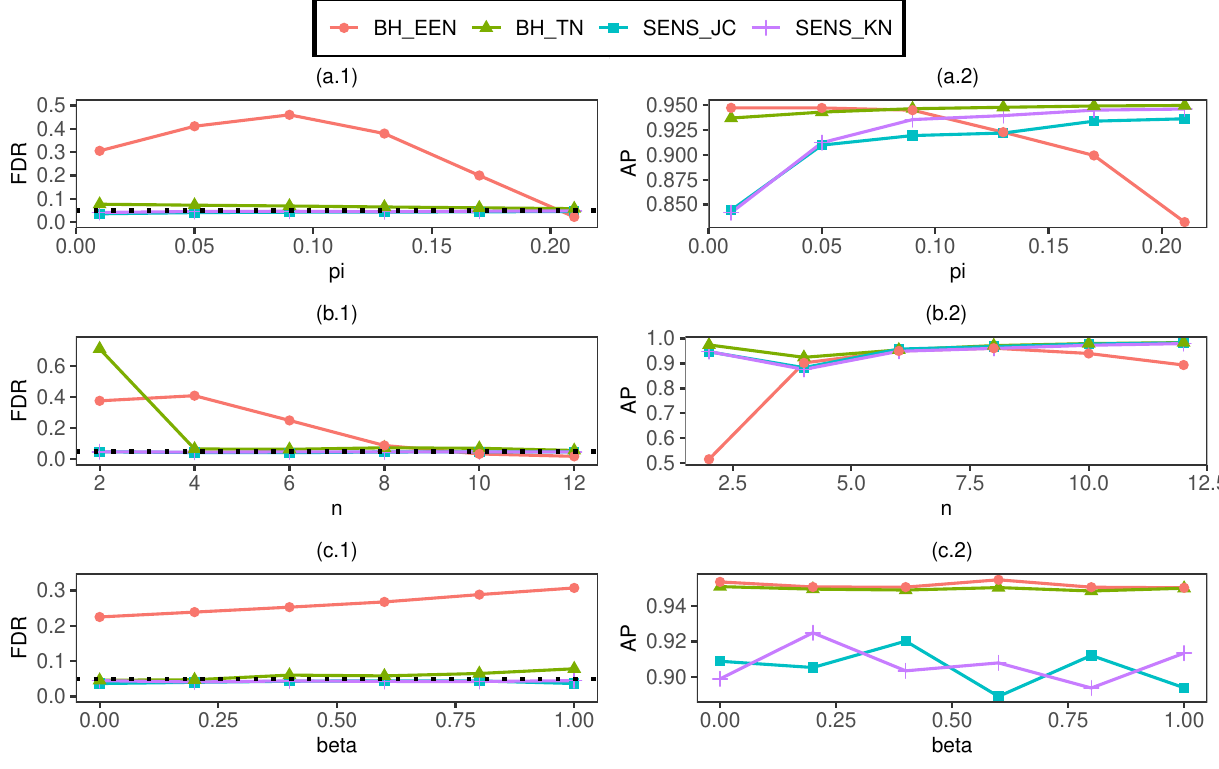}
    \caption{FDR (left column) and AP (right column) comparison: the top, middle and bottom rows correspond to settings (a), (b) and (c), respectively.}
    \label{fig:one-sample-EB}
\end{figure}

The simulation results are presented in Figure~\ref{fig:one-sample-EB}, revealing several notable patterns. First, the left column [Panels (a.1)--(c.1)] demonstrates that both BH\_TN and BH\_EEN exhibit inflated FDR levels, whereas SENS\_JC and SENS\_KN effectively control the FDR at the nominal level. The failure of BH\_TN arises from a mismatch between the theoretical and empirical null distributions, while BH\_EEN fails due to inaccuracies in estimating the empirical null -- specifically, the Jin--Cai method assumes a Gaussian null distribution, which diverges from the actual data-generating process.

Second, Panel (b.1) reveals that as the sample size $n$ increases, the empirical null converges toward the theoretical null; consequently, the FDR level of BH\_TN approaches the nominal level. Moreover, with larger $n$, the deviation from the Gaussian assumption decreases, and the Jin--Cai estimate becomes more accurate, leading the FDR level of BH\_EEN to approach the nominal level.

Third, Panel (c.1) demonstrates that as \(\beta\) increases, that is, the deviation of the empirical null from the Gaussian distribution becomes more pronounced, the FDR levels for both BH\_TN and BH\_EEN are consequently elevated. 

Finally, Panel (a.2) indicates that the average power of SENS\_KN exceeds that of SENS\_JC in most scenarios, suggesting that ``KN'' is preferable in Algorithm~\ref{alg:SENS} when the empirical null substantially deviates from Gaussian. In contrast, ``JC'' remains a suitable option when the Gaussian assumption holds approximately, as seen in Panel (b.2). In any case, an inaccurately estimated model affects only the power, not the validity, of SENS\_JC and SENS\_KN.

\textbf{Simulation 2.}  Next, we compare SENS with recently proposed model-free methods, including sfBH, stBC, and RESS. CLIPPER is not included in the comparison, as it is applicable only to the two-sample case. The comparison is conducted across the following two settings: 
\begin{description}
\item (a) \(\pi=0.1\), \(\sigma_{\text{max}}=0.3\), \(\beta = 1\), \(n=4\), varying $\mu$;
\item (b) \(\pi=0.1\), \(n=4\), \(\beta = 1\), $\mu=3$, varying \(\sigma_{\text{max}}\).
\end{description}
\begin{figure}[htbp!]
    \centering
    \includegraphics[width=6in]{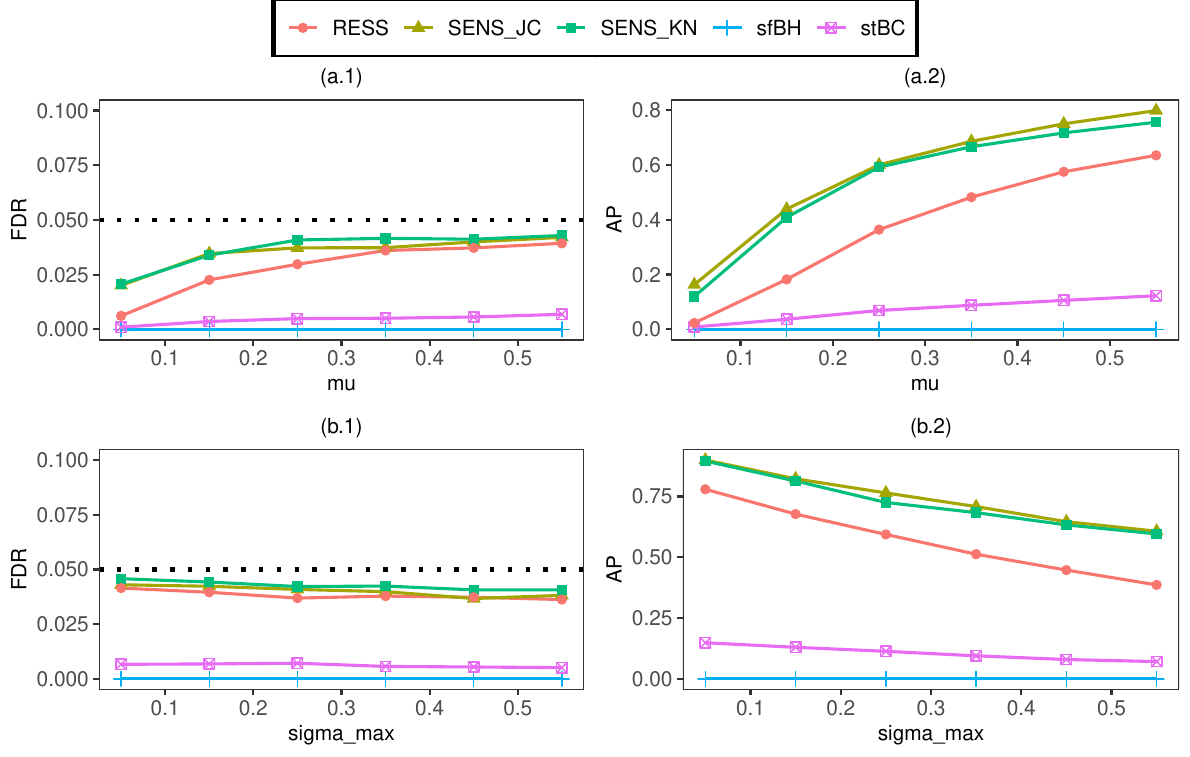}
    \caption{FDR and AP comparison for SENS (one sample) with three other model-free methods. The top and bottom rows correspond to (a) and (b), respectively.}
    \label{fig:one-sample-model-free}
\end{figure}

The simulation results are summarized in Figure \ref{fig:one-sample-model-free}. From the left column, we can see that all five methods  control the FDR at the nominal level. Furthermore, the right column indicates that the average powers of SENS\_KN and SENS\_JC are almost identical, with both methods outperforming RESS, stBC, and sfBH. The superiority of the SENS methods is attributed to the utilization of lfdr-type scores. Additional details regarding the advantages of lfdr are provided in Appendix \ref{subsec:benefits of the Lfdr-Type score function g}.

\subsection{The two-sample case}\label{subsec:sim-two sample}
The data are generated according to the following model:

$$
\begin{aligned}
X_{ij} & = \mu_{xi} + \epsilon_{xij}, j \in [n_x], \quad \mu_{xi} \stackrel{i.i.d.}{\sim} (1 - \pi_x) \delta_0 + \pi_x \delta_{\mu_x},\quad \sigma_x \stackrel{i.i.d.}{\sim} \mathcal{U}(0.05,\sigma_{x,\text{max}}),
\\
\quad Y_{ij} & = \mu_{yi} + \epsilon_{yij}, j \in [n_y], \quad \mu_{yi} \stackrel{i.i.d.}{\sim} (1 - \pi_y) \delta_0 + \pi_y \delta_{\mu_y},\quad \sigma_y \stackrel{i.i.d.}{\sim} \mathcal{U}(0.05,\sigma_{y,\text{max}}),\\
\epsilon_{xij}\mid \sigma_x &\stackrel{i.i.d.}{\sim} \mathcal{N}(0,\sigma_x^2),\quad \epsilon_{yij}\mid \sigma_y \stackrel{i.i.d.}{\sim}  (1-\beta)\mathcal{N}(0, \sigma_y^2) + \frac{3\beta}{4}\mathcal{U}(-\sqrt{3}\sigma_y, \sqrt{3}\sigma_y)+\frac{\beta}{4}\text{Laplace}(0, \sigma_y/\sqrt{2}).
\end{aligned}  
$$

\begin{figure}[htbp!]
    \centering
    \includegraphics[width=5.6in]{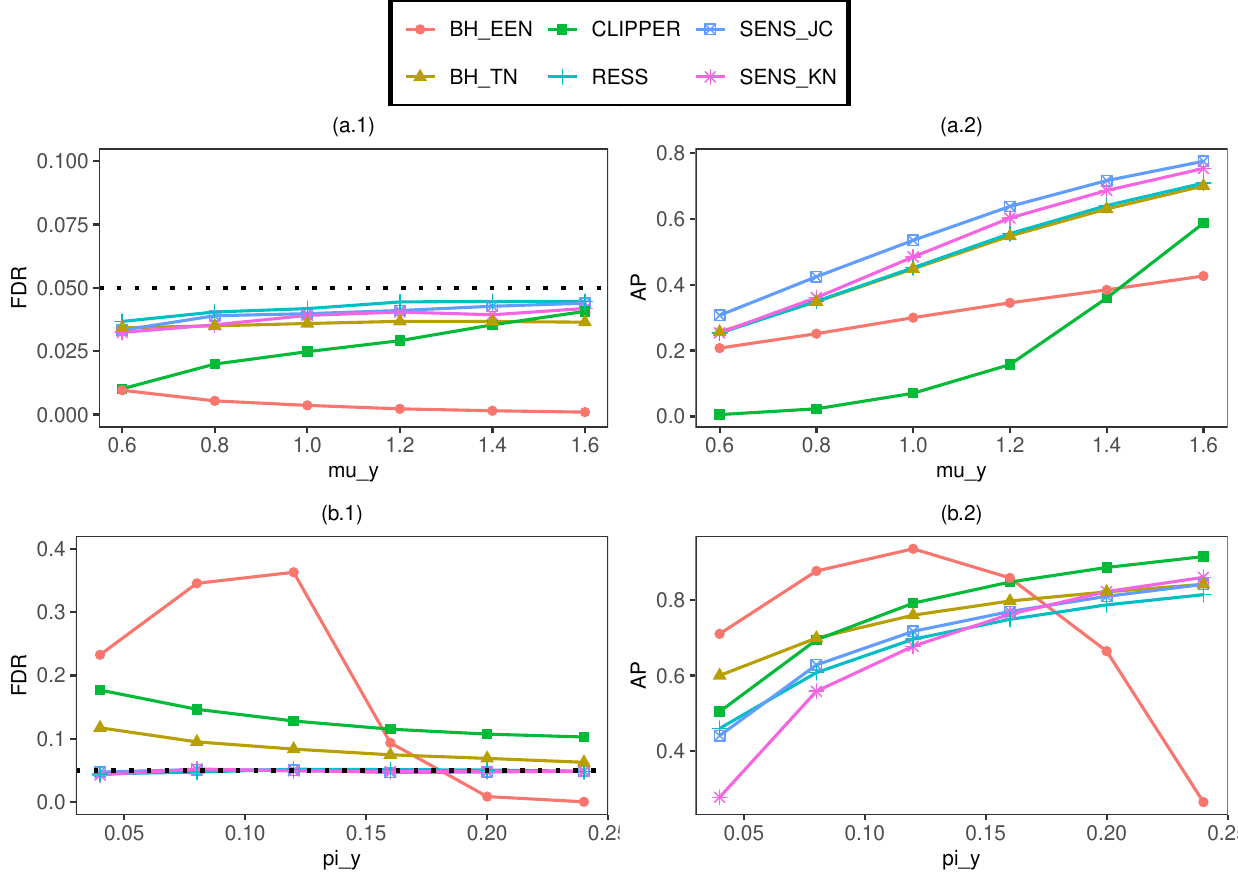}
    \caption{FDR and AP comparison: SENS (two sample) vs. BH (theoretical null \(\mathcal{N}(0,1)\)) vs. BH (estimated empirical null via the Jin-Cai method) vs. CLIPPER. vs. RESS.}
    \label{fig:Clipper}
\end{figure}

For the two-sample case, the stBC method is no longer applicable. Thus we compare SENS with BH\_TN, BH\_EEN,  CLIPPER and RESS. 
We focus on the following two settings: (a) $\pi_x=0.1$, $\pi_y=0.2$, $n_x=n_y=50$, \(\mu_x=-1\), $\sigma_{x,\text{max}}=\sigma_{y,\text{max}} = 4$, $\beta=0$, varying \(\mu_y\);
 (b) $\pi_x=0.05$, $n_x=8$, $n_y=15$, \(\mu_x=1\), $\mu_y=-2$, $\sigma_{x,\text{max}}=2$ \(\sigma_{y,\text{max}} = 1\), $\beta=1$, varying \(\pi_y\). The simulation results are presented in Figure \ref{fig:Clipper}.

In setting (a), where the error distributions of the two samples are identical, all methods successfully control the FDR. The approximate power ranking is as follows: 

$$
\mbox{SENS\_JC} > \mbox{SENS\_KN} > \mbox{RESS} > \mbox{BH\_TN} > \mbox{BH\_EEN} > \mbox{CLIPPER}.
$$
The superior power of SENS methods can be attributed to their ability to construct scores that emulate the lfdr.

In setting (b), where the two samples have different error distributions, CLIPPER, BH\_TN, and BH\_EEN fail to control the FDR. The failure of the two BH methods is due to deviations from the true null distribution, either because of an incorrect theoretical null or a poorly estimated empirical null, while CLIPPER's failure results from its requirement for equal variances under the two conditions. In contrast, RESS and the two SENS methods are valid for FDR control. Among these valid methods, SENS\_JC demonstrates the strongest overall performance, with RESS also showing competitive performance.

\section{Real Data Experiments}\label{sec:application}
To gain insights into the strengths and limitations of each algorithm, we compare SENS with its competitors using two microarray datasets: one representing a one-sample paired design and the other a two-sample design. For both cases, we apply the proposed methods, SENS\_JC and SENS\_KN, which are implemented using the datasets $\mathbf{T}$ and $\mathbf{T}^0$, constructed according to the methods outlined in Section \ref{sec:SENS}. In our analysis, the SENS\_JC, SENS\_KN, RESS, sfBH, and CLIPPER algorithms -- each utilizing random data splitting or sign flipping -- were implemented 50 times, with the average results reported. Comparisons are conducted at various FDR levels: $\alpha = 0.025$, 0.05, and 0.075. 

\subsection{One-sample paired case}\label{subsec:application one-sample}
The first dataset is a one-sample paired case that examines the impact of insulin on gene expression and biochemical pathways in human skeletal muscle. Muscle biopsies were collected from $n^{(1)} = 20$ insulin-sensitive individuals before and after euglycemic hyperinsulinemic clamps \citep{wu2007effect}. The goal of the analysis is to identify which of the $m^{(1)} = 12,626$ genes show differential expression in response to the clamp procedure.

\noindent\textbf{1. Data preprocessing.} For each insulin-sensitive individual, we measure gene expression levels under two conditions: before and after the clamps. We use the following steps to preprocess the data. 
\begin{description}
\item Step (a). Calculate \( X^{(1)*}_{ij} \), the pairwise difference in gene expression levels for gene \( i \in [m^{(1)}] \) in individual \( j \in [n^{(1)}] \).

\item Step (b). Compute the sample mean \( \bar{X}^{(1)*}_i \) for each gene \( i \in [m^{(1)}] \), and define \( M^{(1)} = \texttt{median}(\bar{X}^{(1)*}_i : i \in [m^{(1)}]) \).

\item Step (c). Obtain the ``centralized" data points \( X^{(1)}_{ij} = X^{(1)*}_{ij} - M^{(1)} \). This preprocessing step, also used in \cite{strimmer2008unified} and \cite{hu2022locom}, addresses structural biases in random errors and potential global shifts, allowing the analysis to focus on gene expression variability rather than systematic shifts.
\end{description}

\noindent\textbf{2. Construction of test and calibration samples.} We have obtained datasets in the unified format, as presented in Equations \eqref{equ:model} and \eqref{MT:formulation}, after following the prescribed data preprocessing steps. Consequently, the test and calibration samples, denoted as \( \mathbf T \) and \( \mathbf T^0 \), can be constructed according to the steps outlined in Section \ref{subsec:SENS-construction}.

\begin{figure}[htbp!]
    \center
\includegraphics[width=\textwidth,height=0.25\textheight]{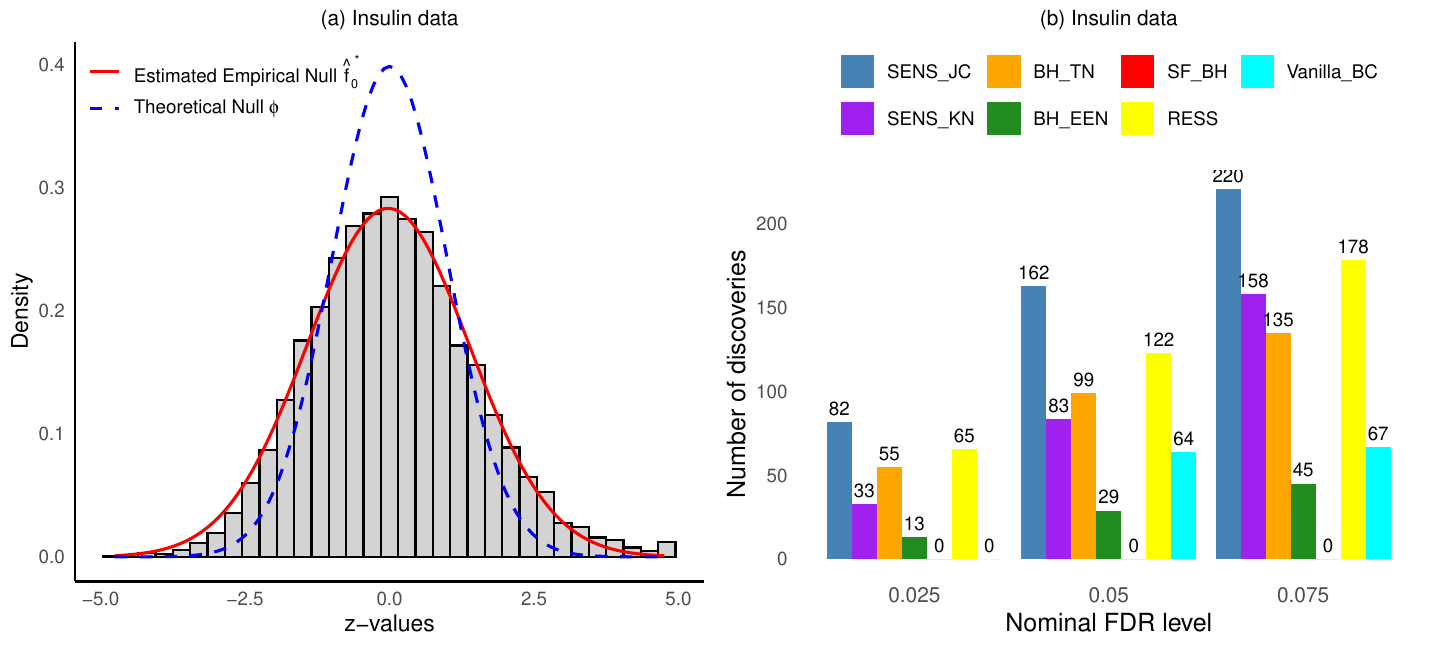}
    \caption{\small Analysis of the insulin data: The left panel displays the histogram of $z$-values alongside the theoretical null $\mathcal{N}(0, 1)$ and the estimated empirical null $\mathcal{N}(0.01, 1.33^2)$. The right plot shows the average number of discoveries for each method at various nominal FDR levels.}
     \label{fig:DE analysis_one sample}
\end{figure}

\noindent\textbf{3. Methods and results.} For the insulin data, we perform paired t-tests, converting the $t$-statistics to $z$-values, denoted as \( \mathbf Z^{(1)} = \{ Z_i^{(1)} : i \in [m^{(1)}] \}\). We estimate the empirical null distribution using the Jin-Cai method and the estimated null distributions for the dataset is  \(\mathcal{N}(0.01, 1.33^2)\). The histogram of the $z$-values is presented in the left column of Figure \ref{fig:DE analysis_one sample}, where we also overlay the theoretical null and estimated empirical null densities. We observe significant distribution shifts between the theoretical null and estimated null. 

 The number of discoveries made by each method at various FDR levels is summarized in the right column of Figure \ref{fig:DE analysis_one sample}. Among the methods, SENS\_JC yields the largest number of discoveries, followed by RESS, BH, SENS\_KN, BH\_EEN, and sfBH.

\subsection{Two-sample case}\label{subsec:application two-sample}
The second dataset is a two-sample case focused on identifying clinically relevant subclasses of bladder carcinoma. This dataset consists of expression microarray data from $n^{(2)}=21$ well-characterized bladder tumors, with $n^{(2)}_x=10$ samples from the T2-T4 tumor stage and $n^{(2)}_y=11$ from the T1 stage \citep{0Identifying}. The objective here is to determine which of the $m^{(2)} = 5724$ genes are differentially expressed between the two stages.

\noindent\textbf{1. Construction of test and calibration samples.}
The data is already in the unified format for the two-sample case, as presented in Equations \eqref{equ:model-two-sample} and \eqref{MT:formulation-two-sample}. The test and calibration samples, denoted as $\mathbf{T}$ and $\mathbf{T}^0$, can then be constructed following the steps for the two-sample case outlined in Section \ref{sec:SENS}.

\noindent\textbf{2. Methods and results.}
For the bladder tumor data, we perform two-sample t-tests and convert the $t$-statistics to $z$-values, denoted as $\mathbf{Z}^{(2)} = \{ Z_i^{(2)} : i \in [m^{(2)}] \}$. We estimate the empirical null distribution using the Jin-Cai method, with the estimated null distribution being $\mathcal{N}(-0.07, 1.34^2)$. The histogram of the $z$-values and the number of discoveries made by each method at various FDR levels exhibit patterns similar to those observed in the one-sample case, and are presented in Figure \ref{fig:DE analysis_two sample} of the Supplement.
 The figure shows that the densities of both the theoretical null and the estimated empirical null are significantly shifted, with SENS\_JC yielding the largest number of discoveries.


\medskip

\spacingset{0}
\footnotesize 
\bibliographystyle{plainnat}
\bibliography{reference}

\spacingset{1} 

\normalsize
\newpage
\begin{center}
\appendix
{\Large\bf Online Supplementary Material for \\ \vskip 0.5cm \LARGE ``Conformalized Multiple Testing under Unknown Null Distribution with Symmetric Errors''}
\end{center}

This supplement contains supplementary methodological details (Section \ref{sec:supple-mater}), the proofs for primary theory (Section \ref{sec:supple-proof}), and auxiliary numerical results (Section \ref{simu:aux}).

\setcounter{equation}{0}
\renewcommand{\theequation}{A.\arabic{equation}}
\setcounter{figure}{0}
\setcounter{page}{1}

\renewcommand{\thefigure}{A.\arabic{figure}}

\section{Supplementary Methodological Details}\label{sec:supple-mater}

\subsection{Comparisons with Efron’s paradigm}\label{subsec:comparisons efron}
\textbf{Philosophical alignment with Efron's paradigm.} \cite{efron2004large} emphasizes that systematic discrepancies may exist between the empirical null and theoretical null distributions in FDR analysis. Such discrepancies -- manifesting as shifted means and inflated/reduced variances relative to the classical theoretical null model $\mathcal{N}(0,1)$ -- often stem from unobserved confounders, structured dependencies, or model mis-specifications in large-scale studies. Rather than rigidly adhering to a pre-specified theoretical baseline, the empirical null is estimated directly from the observed data, thereby redefining the ``normal state'' and mitigating potential distortions in the null distribution.

Efron’s empirical null approach is a cornerstone of empirical Bayes multiple testing, providing an intuitive and effective framework for separating abnormal signals from baseline observations. This data-driven methodology not only improves the precision in pinpointing significant cases but also yields more meaningful and reliable FDR analyses.

Our approach is philosophically aligned with the core principles of Efron’s paradigm in that (a) both works aim to learn the ``normal state'' from the data, and (b) both works adopt an empirical Bayes perspective and employ the lfdr-type statistics as the basic building blocks for large-scale multiple testing. 

\noindent\textbf{Departures from Efron’s Paradigm.} While our proposal shares similar philosophical principles with Efron’s framework, SENS introduces a refined strategy for addressing the issue of empirical null. In particular, our method diverges from Efron’s approach both in its model assumptions and its methodological focus. In the following discussion, we provide detailed justifications for our assumptions and elaborate on key distinctions.

\begin{description}
    \item \textit{Model Assumptions.} Within the empirical Bayes framework, it is necessary to posit certain beliefs regarding the null distribution; without such assumptions, the problem becomes unidentifiable, rendering the separation of ``interesting'' cases from baseline  infeasible. \cite{efron2004large} and \cite{jin2007estimating} assume that the empirical null follows a Gaussian distribution. In contrast, our method only requires the random errors to have distributions symmetric about zero -- a mild and natural condition that implies a symmetric null distribution, with the centered Gaussian being a special case.

    \item \textit{Methodological Focus.} Under the assumption of a Gaussian empirical null, \cite{efron2004large} proposes that deviations from the theoretical null $\mathcal{N}(0,1)$ can be effectively summarized by a shifted mean $\mu_0$ and an altered variance $\sigma_0^2$. A typical strategy is then to directly estimate these parameters and use the resulting $\mathcal{N}(\hat \mu_0, \hat\sigma_0^2)$ in subsequent FDR analyses. In contrast, our method addresses the issue of shifted means by subtracting a baseline estimate. This centralization, also employed by \cite{efron2004large} and \cite{efron2007correlation}, assumes that, following adjustment, the residual errors are symmetrically distributed about zero. Our primary methodological contribution lies in devising techniques to flexibly estimate this zero-symmetric distribution without imposing further assumptions. Our formulation naturally leads to multiple testing procedures that detect deviations from zero, which is applicable to both one-sample and two-sample testing problems (see the next item for further explanations). 

    \item \textit{General Considerations and Intuitive Justifications.} Both SENS and Efron’s proposal fundamentally assume symmetry about a central value in the null distribution. In Efron’s framework, permitting a non-zero mean introduces additional flexibility, albeit at the cost of remaining confined to the Gaussian family. Conversely, while our method imposes the restriction of a centered (zero-mean) null, it gains flexibility by avoiding strict Gaussian assumptions and accommodating a wider range of distribution families, including those with heavier tails. This zero-mean restriction is both practical and contextually justified. For instance, in the one-sample setting (see Section \ref{subsec:application one-sample}), subtracting an appropriate baseline from the data naturally leads to a null hypothesis centered at zero. In the two-sample scenario (see Section \ref{subsec:two sample case}), the inherent structure of the problem suggests that any departures from the ``normal state'' should be interpreted relative to a zero-centered difference. In both contexts, enforcing a zero-mean null seems to align well with the underlying scientific rationale.

    \item \textit{The Dependence Issue.} SENS is restricted to scenarios in which the shifted means are not driven by structured dependence \citep{schwartzman2008empirical}. Addressing dependence in large-scale multiple testing remains a formidable  challenge. Although Efron’s estimation of the empirical null might capture structured dependence, subsequent computation of $p$-values based on this estimated null, followed by application of the BH procedure, can lead to substantial confusions and complications. Specifically, estimation approaches (e.g., Jin-Cai's method) are typically derived under independence or specific dependence structures, leaving it unclear whether these methods would remain valid under arbitrary dependence. Furthermore, even if $p$-values can be computed from the estimated empirical null, their complex dependence structure may undermine the validity of the BH procedure. One potential remedy is to address dependence effects by employing methods such as those proposed by \cite{sun2009large}, \cite{fan2012estimating}, and \cite{du2023false}. However, these approaches fundamentally diverge from the empirical null framework. Consequently, while the empirical null remains a powerful concept, its application under dependence warrants further careful investigation. 
        
\end{description} 

In summary, our approach aligns with the core principles of Efron’s paradigm while introducing substantive methodological distinctions.

\subsection{Comparisons with Existing Model-free Methods}\label{sec:comparisons model-free}

In this section, we provide a detailed comparison of SENS with existing model-free methods [cf. \citep{arlot2010some, arias2017distribution, zou2020new, ge2021clipper, wang2024conformalized}] that have been briefly mentioned in Section \ref{subsec:related-works} of the main text. The key distinction is that these methods are frequentist in nature and therefore unable to leverage the advantages of lfdr-type statistics. In contrast, SENS is built upon an empirical Bayes framework and constructs score functions that effectively emulate the lfdr, thereby achieving higher power and asymptotic optimality. Table \ref{table:comparison} summarizes the key assumptions and properties of SENS and existing methods, with each method's advantages highlighted in bold. We then provide detailed discussions comparing SENS with each method.

\begin{table}[htpb!]
\centering
\footnotesize
\begin{tabular}{|p{2.2cm}|p{1.8cm}|p{1.8cm}|p{1.8cm}|p{1.7cm}|p{2.2cm}|p{2.5cm}|}
\hline
Methods & Zero-mean assumption & Symmetry assumption & Gaussian assumption & Optimality property & Finite-sample theory & Applicability \\
\hline
Efron; Jin-Cai & \textbf{no} & yes & yes & \textbf{yes} & no & \textbf{both}\\
\hline
sfBH & yes & yes & \textbf{no} & no & \textbf{yes} & one-sample only \\
\hline
stBC & yes & yes & \textbf{no} & no & \textbf{yes} & \textbf{both}\\
\hline
RESS(I\&S) & yes & yes & \textbf{no} & no & \textbf{yes} & \textbf{both}\\
\hline
RESS(D/A) & yes & \textbf{no} & \textbf{no} & no & no & \textbf{both}\\
\hline
CLIPPER& yes & \textbf{no} & \textbf{no} & no & \textbf{yes} & two-sample only\\
\hline
SENS & yes & yes & \textbf{no} & \textbf{yes} & \textbf{yes} & \textbf{both}\\
\hline
\end{tabular}
\caption{Comparisons of various related methods based on key assumptions and properties}
\label{table:comparison}
\end{table}

\begin{enumerate}
    \item The sfBH procedure by \cite{arlot2010some}. The sfBH method, which applies the traditional BH procedure to \(p\)-values constructed via sign-flipping techniques, represents pioneering work in model-free inference. We highlight several potential advantages of SENS over sfBH. First, sfBH requires a large sample size for each testing unit, whereas SENS only requires \(n_i \geq 2\). Second, SENS is computationally efficient, while the effective implementation of sfBH—often necessitating a large number of replications—can be very time-consuming. Finally, the sign-flipping \(p\)-values employed by sfBH are less efficient than the lfdr-based statistics used by SENS, and adjusting these p-values from sfBH within the EB framework appears infeasible. A detailed description of sfBH and supporting simulations are presented in Appendix \ref{subsec:sfBH}.

    \item The stBC method by \cite{arias2017distribution}. This method involves applying the BC algorithm with symmetric t-statistics. There are three important limitations of the stBC method: First, the stBC procedure is not directly applicable to two-sided alternatives. Specifically, for the case where \( H_{1,i}: \mu_i < 0 \), the power is zero. Second, even for the one-sided alternative, the rejection rule based on the absolute value of the test statistic, which corresponds to the \texttt{minP} method in \cite{TonyCaioptimal2017}, may be suboptimal [cf. Theorem 5 in \cite{TonyCaioptimal2017}]. The numerical superiority of SENS over stBC is illustrated Figure \ref{fig:special SSMT setting} in Appendix \ref{subsec:stBC}. Third, we wish to emphasize that the asymptotic theory presented in \cite{arias2017distribution} exclusively addresses the attainment of the classification boundary. However, from our perspective, a more relevant notion of optimality in the context of FDR analysis is the \emph{discovery boundary}, as discussed in \cite{TonyCaioptimal2017}. A detailed discussion of this issue can be found in Section \ref{subsec:power analysis}.

    \item The RESS method by \cite{zou2020new}.     
    RESS employs summary statistics to conduct FDR analysis under a frequentist paradigm, relying on the independence between the absolute value and sign of the statistic under the null hypothesis. We outline several important distinctions between SENS and RESS. First, different from RESS, SENS is a conformalized empirical Bayes method that creates self-calibrated empirical null samples, ensuring their pairwise exchangeability with the test samples under the null. Second, RESS splits the data into two parts, computes a t-statistic on each split, and then multiplies the two t-statistics to form the final summary statistic. This sequence of operations leads to some information loss, as the product of two t-statistics provides less information than a t-statistic computed using the entire dataset. By contrast, the sample-splitting strategy in SENS employs the plus-minus trick that effectively avoids information loss. Third, SENS is specifically designed to calibrate scores to emulate the lfdr statistics, which is demonstrated to be optimal \citep{sun2007oracle}.

    The RESS method exhibits different properties under varying scenarios. For independent and symmetric errors, RESS(I\&S) ensures finite-sample FDR control. However, for dependent or asymmetric errors, RESS(D/A) only provides asymptotic FDR control. Additionally, neither RESS(I\&S) nor RESS(D/A) includes a power analysis. In contrast, SENS offers finite-sample FDR control and possesses certain optimality properties. The numerical comparison of SENS with RESS is provided in Appendix \ref{subsec:RESS}.
    
    \item The CLIPPER method by \cite{ge2021clipper}. This method assumes that the measurement errors for all replicates across the two conditions are independent and identically distributed. The homogeneity assumption, which underpins both the methodological and theoretical developments in CLIPPER, is arguably its most significant limitation. Our simulation studies in Section \ref{subsec:sim-two sample} demonstrate that CLIPPER fails to control the FDR when the variances of measurements differ between the two conditions. In contrast, SENS accommodates heterogeneous variances across conditions and does not require independence across all features.

Several additional distinctions between SENS and CLIPPER are worth noting. First, the two methods are grounded in different objectives: CLIPPER is designed to develop model-free FDR methods under the frequentist paradigm, whereas SENS is formulated to handle an unknown null distribution within the empirical Bayes framework. Second, CLIPPER computes summary statistics, ranks them, and derives symmetric statistics through this ranking process. In contrast, SENS begins by generating calibrated null samples and subsequently constructs scores that mimic the lfdr, thereby leveraging the full strength of the empirical Bayes approach. Third, although both methods have been shown to control the FDR in finite samples, SENS additionally provides rigorous theoretical guarantees for asymptotic optimality. As shown in Figure \ref{fig:Clipper} in Section \ref{subsec:sim-two sample}, when both CLIPPER and SENS are valid for FDR control, SENS is more powerful. Fourth, the SENS framework is versatile and can be applied to both one-sample and two-sample settings, while CLIPPER has been developed solely for the two-sample context.

    \item The SCPV method by \cite{wang2024conformalized}. This work offers an interesting perspective and was brought to our attention by an insightful referee. We note that both SCPV and SENS address the issue of distorted null distributions; however, the two methods originate from fundamentally different research lines and are not directly comparable. Consequently, SCPV is not included in Table \ref{table:comparison} and is not numerically compared with SENS in the simulations. Instead, we provide a high-level comparison below to highlight the key distinctions.
      
    First, SENS addresses simultaneous inference through a one-stage selection procedure with FDR control, whereas SCPV handles post-selection inference via a two-stage approach involving selection followed by inference. Second, although both SENS and SCPV aim to correct distorted null distributions, they tackle fundamentally distinct challenges. SENS addresses the problem of an unknown null distribution within an empirical Bayes framework [a central issue highlighted by \citet{efron2004large}] while SCPV focuses on the distortion induced by selection [a serious concern emphasized by \citet{benjamini2005false}]. Third, SCPV relies on external calibration samples from the null distribution and assumes joint exchangeability. In contrast, SENS constructs self-calibrated null samples directly from the observed data and only requires pairwise exchangeability—a strictly weaker assumption that better accommodates heterogeneity among testing units. Fourth, SENS employs the BC procedure (also known as the Selective SeqStep+ algorithm; see \citet{barber2015controlling}), whereas SCPV builds upon the classical BH procedure. Finally, SCPV establishes the FDR control theory for the BH procedure with conformal $p$-values via a leave-one-out technique, while SENS develops a martingale-based theory for BC-type procedures and further establishes a novel theory of asymptotic optimality.
\end{enumerate}

\subsection{The Standardization Step: A Counter Example}\label{insights and counter examples}

We will now resume our discussion from Remark \ref{rem:small-n} in Section \ref{subsec:SENS-construction}. In the construction of SENS, we assert that the appropriate formula for \(S_i\) is given by \(S_i = \sqrt{\frac{(n_{i1}-1) S^2_{i1} + (n_{i2}-1) S^2_{i2}}{(n_i-2)}}\), which is essential in the standardization step. A seemingly plausible and natural alternative is \(S_i^* = \sqrt{\frac{1}{n_i - 1} \sum_{j=1}^{n_i} (X_{ij} - \bar{X}_i)^2}\). Note that both \(S_i\) and \(S_i^*\) are consistent estimators. Remark \ref{rem:small-n} contends that \(S_i^*\) is unsuitable for constructing \(T_i\) and \(T_i^0\) because it undermines the pairwise exchangeability. Denote $T^*_i = \Phi^{-1}\left\{G_{t,n_i-2}(V_i / S^*_i)\right\}$ and $T^{0,*}_i = \Phi^{-1}\left\{G_{t,n_i-2}(V^0_i / S_i)\right\}$ for $i\in[m]$. We will now elaborate on this point with Proposition \ref{pro:counter example}, the proof of which will be provided in Section \ref{proof:pro counter example}.

\begin{proposition}\label{pro:counter example}
   (Counter Example). Suppose \( X_{ij} \stackrel{\text{i.i.d.}}{\sim} \mathcal{N}(0,1) \), \( i \in \mathcal{H}_0 \),  \( j \in [n_i] \), where we assume, for simplicity, that \( n_i \) is an even integer. Let \( \mathbf{T}^*_{-i} = (T^*_1, \cdots, T^*_{i-1}, T^*_{i+1}, \cdots, T^*_m) \) and \( \mathbf{T}^{0,*}_{-i} = (T^{0,*}_1, \cdots, T^{0,*}_{i-1}, T^{0,*}_{i+1}, \cdots, T^{0,*}_m) \). Then we have \( T^*_i \not\stackrel{d}{=} T^{0,*}_i \). It follows that the pairwise exchangeability condition fails to hold.
\end{proposition}

\subsection{Schematic Diagram for Constructing SENS}
\label{sec:diagram}

\begin{figure}[htbp!]
\centering
\includegraphics[width=3.8in, height=4in]{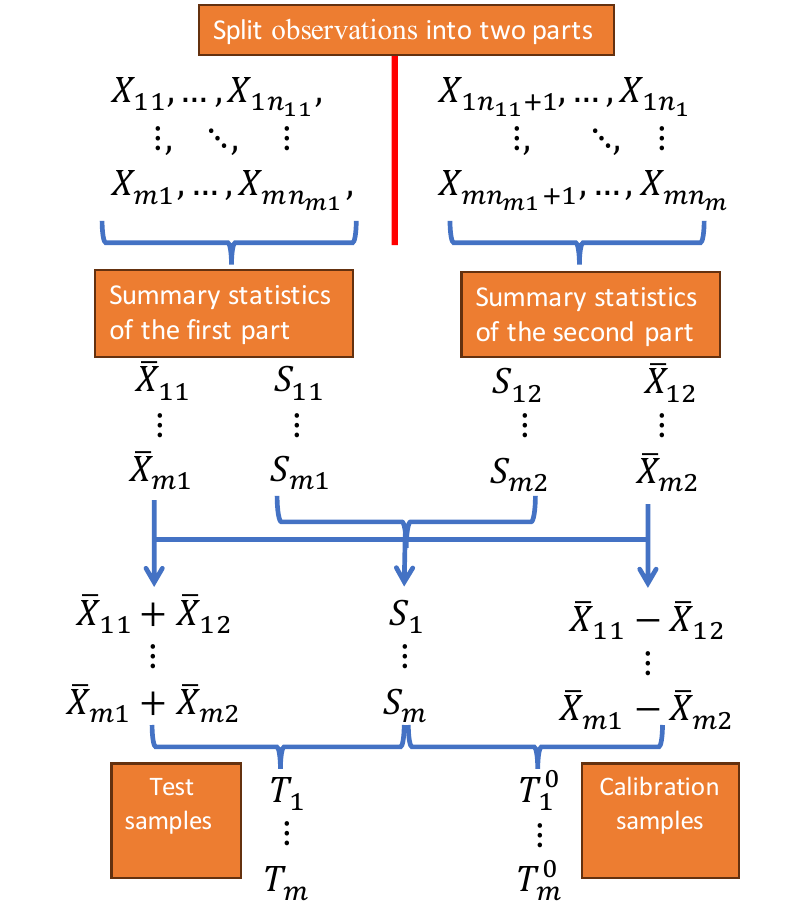}
\caption{A schematic illustration of the construction of $(\mathbf{T},\mathbf{T}^0)=\{(T_i, T_i^0): i\in[m]\}$ in one-sample case.}
\label{fig:1}
\end{figure}

We present a schematic diagram for constructing SENS. The one-sample case is illustrated in Figure \ref{fig:1}; the two-sample case is similar and hence omitted.

\subsection{\texorpdfstring{\cite{jin2007estimating}'s estimator for null distribution}{jin2007estimating's estimator for null distribution}}\label{subsec:JC}

We provide a brief description of Jin-Cai's method (discussed in Section \ref{subsec:sens-score} and employed as the ``JC'' option in the SENS algorithm) for estimating the empirical null distribution. The code for implementing this method, kindly provided by Professor Jiashun Jin from Carnegie Mellon University, is available at \url{https://github.com/Tiany5/SENS}.

Denote \((X_1, \dots, X_{2m}) = (\mathbf{T}, \mathbf{T}^0)\) the mixed sample. The Jin-Cai estimator is built upon the following empirical characteristic function:
$$
\varphi_{2m}(t) = \varphi_{2m}(t; X_1, \dots, X_{2m}) = \frac{1}{2m} \sum_{j=1}^{2m} e^{i t X_j}.
$$
Let \( \operatorname{Re}(z) \) and \( \operatorname{Im}(z) \) denote the real and imaginary parts of the complex number \( z \), respectively. Define 
$$
\sigma_0^2(f; t) = -\frac{\frac{d}{d t}|f(t)|}{t \cdot |f(t)|}, \quad \mu_0(f; t) = \frac{\operatorname{Re}(f(t)) \cdot \operatorname{Im}(f'(t)) - \operatorname{Re}(f'(t)) \cdot \operatorname{Im}(f(t))}{|f(t)|^2}.
$$
For a given \( \gamma \in (0, 1/2) \), let:
\begin{equation}\label{equ:t,sig,mu}
\begin{aligned}
\hat{t}_{2m}(\gamma) &= \hat{t}_{2m}(\gamma; \varphi_{2m}) = \inf \left\{ t : |\varphi_{2m}(t)| = (2m)^{-\gamma}, 0 \leq t \leq \log (2m) \right\}, \\
\hat{\sigma}_0^2 &= \sigma_0^2(\varphi_{2m}; \hat{t}_{2m}(\gamma)) \quad \text{and} \quad \hat{\mu}_0 = \mu_0(\varphi_{2m}; \hat{t}_{2m}(\gamma)),
\end{aligned}
\end{equation}
where \( \hat{\sigma}_0^2 \) and \( \hat{\mu}_0 \) are estimates for \( \sigma_0^2 \) and \( \mu_0 \), respectively. Let $\phi_\sigma(x-\mu)$ denote the density function of a $\mathcal N(\mu, \sigma^2)$-variable. Then the Jin-Cai estimator for $f_0(x)$ is given by 
\begin{equation}\label{equ:JC}
    \hat{f}_0(x)=\phi_{\hat{\sigma}_0}(x-\hat{\mu}_0).
\end{equation}

\subsection{The filtering method: bias quantification and correction}\label{subsec:bias of filter}
In this section, we formally quantify the bias introduced by the filtering strategy in Section \ref{subsec:sens-score}. The filtering strategy is an important and innovative method for data preprocessing, designed with three main objectives: (a) ensuring the symmetry of the data, so that $T_i$ and $T_i^0$ are treated equally, (b) utilizing only assumptions about zero symmetry for the null units, and (c) generating fake samples that closely resemble calibrated null samples. It is important to note that while the bias introduced by this strategy affects statistical efficacy, it does not undermine the validity of FDR control. For this reason, we initially did not focus on quantifying the bias. In this context, due to the complexity involved in the construction of SENS, we focus on a simplified and idealized SSMT setup, where $T_i$ and $T^0_i$ are assumed to be mutually independent. Consequently, while the analysis provides valuable insights, it only serves as an initial exploration of the problem and does not claim to offer an ultimate solution. 

\begin{enumerate}
        \item {\bf Restating the problem:} The bias comes from the fact that $T_i\sim f_\text{mix}\coloneqq (1-\pi)f_0+\pi f_1$, whereas $T_i^0\sim f_0$. Our filtering rule replaces \( T_i^0 \) by \( T_i \) when \( |T_i| \leq |T_i^0| \) and vice versa otherwise. Since the kernel estimator in Equation (19) of the manuscript is an asymptotically unbiased estimator for $f_\text{filter}(x)=\frac{1}{2}f_{\tilde{T}^0}(x)+\frac{1}{2}f_{\tilde{T}^0}(-x)$, the symmetrized density function of $\tilde{T}_i^0$. We only need to investigate the difference between $f_\text{filter}$ and $f_0$. 
        
    \item {\bf Formulae in the general case.} The density function can be computed by conditioning on two events: \( |T_i| \leq |T_i^0| \) and \( |T_i| > |T_i^0| \). Let \( F_0 \) and \( F_1 \) denote the cumulative distribution functions (CDFs) of \( f_0 \) and \( f_1 \), respectively. Furthermore, define
    $$
    \xi(x)=\pi \big\{1 - F_1(x) + F_1(-x)\big\} + 2(1-\pi)\big\{1 - F_0(x)\big\}. 
    $$
By summing the contributions from both events and treating the positive and negative parts separately, we obtain:
\begin{equation*}
f_{\tilde{T}_i^0}(x) = 
\begin{cases} 
2\big\{1 - F_0(x)\big\} f_\text{mix}(t) + f_0(x) \xi(x) & x \geq 0 \\\
2F_0(x) f_\text{mix}(t) + f_0(x)\xi(-x) & x < 0
\end{cases}.
\end{equation*}
It follows that  $f_\text{filter}(x)  =  2\big\{1 - F_0(|x|)\big\} f_\text{mix}(|x|) + f_0(|x|)\xi(|x|)$.

 \item {\bf Special Case for Normal Distributions:} For illustration, we consider a special case where $f_0(x) = \phi(x)$ and $f_1(x) = \phi(x - \mu)$. The density function simplifies to:
\begin{equation*}
   f_{\text{filter},\Phi}(x)  = 2\big\{1 - \Phi(|x|)\big\} \big\{\pi \phi(|x| - \mu) + (1-\pi)\phi(|x|)\big\} + \phi(x) \xi_{\Phi}(|x|),
\end{equation*}
where $\xi_{\Phi}(|x|)=\pi \big\{2 - \Phi(|x| - \mu) - \Phi(|x| + \mu)\big\} + 2(1-\pi)\big\{1 - \Phi(|x|)\big\}$ and \( \Phi \) is the CDF of a \( \mathcal{N}(0, 1)\)-variable. Panel (a) in Figure \ref{fig:filter} displays \( f_{\text{filter},\Phi} \) overlaid with the true null \( \phi \). The plot reveals a clear discrepancy: \( f_{\text{filter},\Phi} \) exhibits a sharp, cone-like shape that reflects the impact of the filtering step.

\begin{figure}[htbp!]
    \centering
    \includegraphics[width=0.95\linewidth, height=2.8in]{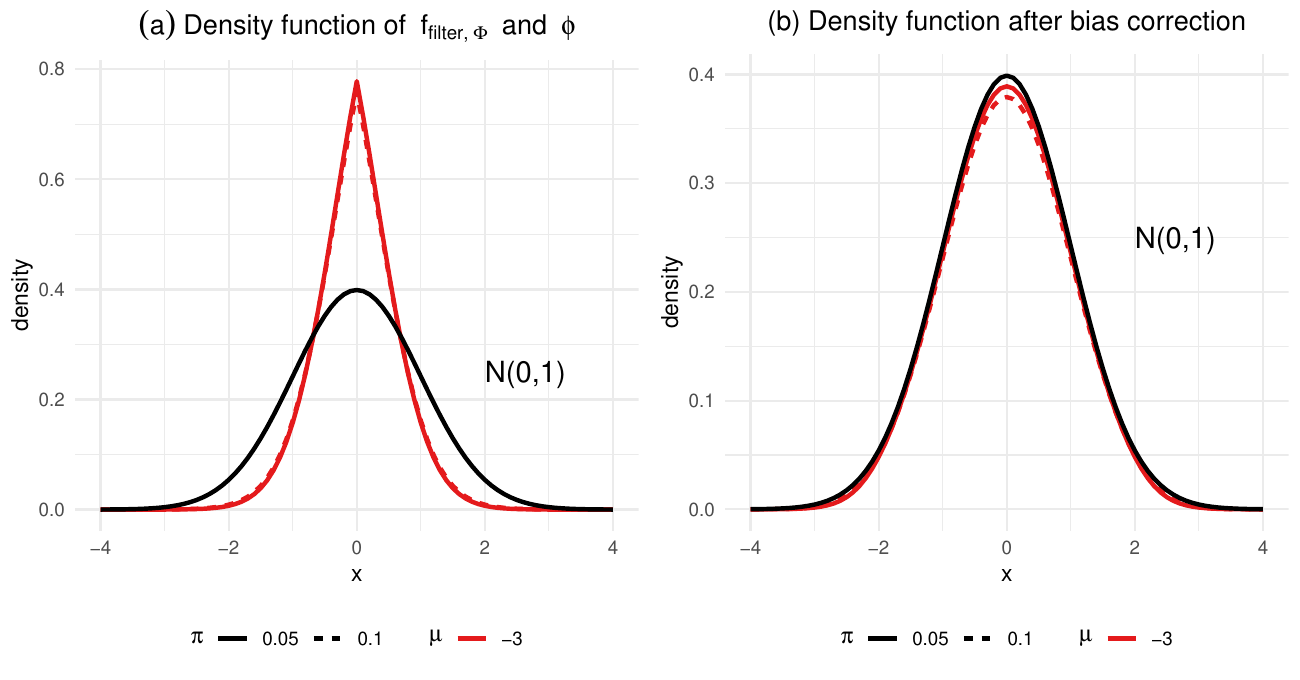}
    \caption{From left to right, Panel (a) shows $f_{\text{filter},\Phi}$ under different $\pi$, and Panel (b) shows the density function after bias correction.}
    \label{fig:filter}
\end{figure}

  \item {\bf Bias correction.} We introduce a correction method to try to minimize this deviation. Since a thorough analysis with an unknown mixture density function $f_\text{mix}$ is intractable, we consider a simplified case where both $T_i$ and $T^0_i$ follow a symmetric distribution $f_0$ centered around 0 and are independent. In this case, $f_0$ and $\tilde{f}_0$ -- the density of $\tilde{T}_i^0$ -- satisfy the following relationships: 
\begin{equation}\label{equ:f0}
    \tilde{f}_0(x) = 4f_0(x)(1-\int_{-\infty}^{|x|}f_0(u)du); \quad  f_0(x) = \frac{\tilde{f}_0(x)}{2 \sqrt{1 - 2\int_{0}^{|x|} \tilde{f}_0(t) \, dt}}. 
\end{equation}
Under the sparsity assumption in large-scale inference, $\pi$ is small. This implies that $f_{\text{mix}}$ and $f_0$, and consequently, $f_{\text{filter}}$ and $\tilde{f}_0$, are very similar. Therefore, the relationship in \eqref{equ:f0} can be exploited to develop a bias-correction method. Specifically, we first use $f_{\text{filter}}$ to approximate $\tilde{f}_0$, and then substitute $\hat{f}_{\text{filter}}$ for $f_{\text{filter}}$,  yielding the bias-corrected estimator for $f_0$:
\begin{equation}\label{bias-correct}
    \hat{f}_0(x) = \frac{\hat{f} _{\text{filter}}(x)}{2 \sqrt{1 - 2\int_{0}^{|x|} \hat{f} _{\text{filter}}(t) \, dt}}.
\end{equation} 
We illustrate the effectiveness of \eqref{bias-correct} in Figure \ref{fig:filter}. 
As shown in Panel (b), the density function after bias correction closely resembles that of $\mathcal{N}(0, 1)$. 

While this bias reduction method appears effective for normal distributions, we choose to present the original approach in the main text and leave discussion of the bias-correction option to the Appendix. There are several reasons for this decision. First, the correction step is complex and may obscure the main idea of our algorithm. Second, the formula’s accuracy is questionable, as it is derived under the assumption that $T_i$ and $T_i^0$ are independent -- a condition that does not hold in practice. Finally, a comprehensive investigation of the bias correction for other families of distributions is beyond the scope of this work, and our methodology remains valid regardless of whether this step is applied.

 \end{enumerate}

\subsection{Choice of anti-symmetric functions}\label{aux-sim:anti-func}
Below we highlight two principles for choosing an appropriate anti-symmetric function \(\gamma\):
 
\begin{enumerate}

\item Our anti-symmetric function \(\gamma\) is specifically designed for the class of scores \((\mathbf{U}, \mathbf{U}^0)\) that satisfy both of the following conditions: (i) both scores are positive, and (ii) smaller scores indicate stronger evidence against the null hypothesis. This is a natural and rich class of scores that includes the widely used conformal p-values and the lfdr-type scores utilized by SENS. A common pitfall is applying this \(\gamma\) to scores outside this class (e.g. absolute value based t-statistics [violating condition (ii)], negative correlation coefficient [violating condition (i)]), resulting in valid but highly inefficient BC-type procedures.

\item  The anti-symmetric function should be designed to preserve the ranking of \(U_i\) as much as possible. The reasoning is that in the SENS algorithm, the oracle version of \(U_i\) is \(r_m(T_i)\), which is demonstrated to be optimal for FDR control. 

The SENS algorithm is based on the following decision rule \(\mathbb{I}(U_i \leq U_i^0 \wedge \tau^\prime)\), where 
\begin{equation}
\tau^\prime=\sup \left\{\lambda\in \mathbf{U} \cup \mathbf{U}^0: \frac{1+\sum_{j=1}^m \mathbb{I}\left\{U^0_i \leq  U_i\wedge \lambda\right\}}{\left[\sum_{j=1}^m \mathbb{I}\left\{U_i \leq U^0_i \wedge \lambda\right\}\right] \vee 1} \leq \alpha\right\}.
\label{equ:tau^prime}
\end{equation}

The following lemma is proved later in Section \ref{sec:G=U}.

\begin{lemma}\label{lem:G=U}
    The decision rule \(\mathbb{I}(G_i \geq \tau)\), where \(\tau\) is defined in Equation \eqref{equ:tau}, is equivalent to \(\mathbb{I}(U_i \leq U_i^0 \wedge \tau^\prime)\), with \(\tau^\prime\) defined in Equation \eqref{equ:tau^prime}.
\end{lemma}

Lemma \ref{lem:G=U} shows that our chosen \(\gamma(\cdot, \cdot)\) ensures that \(\mathbb{I}(\gamma(U_i, U^0_i) \geq \tau)\) is equivalent to \(\mathbb{I}(U_i \leq U_i^0 \wedge \tau^\prime)=\mathbb{I}(U_i \leq U^0_i) \mathbb{I}(U_i \leq \tau')\). This establishes that the thresholding rule adopted by SENS  effectively operates based on the ranking of \(U_i\), apart from the eliminated terms due to \(\mathbb{I}(U_i \leq U^0_i)\). In contrast, alternative anti-symmetric functions (examples of which, \(\gamma_1\) and \(\gamma_2\), are provided in the next subsection) may lead to decision rules that either do not directly threshold \(U_i\) (as with \(\gamma_1\)) or apply the threshold in the wrong direction (as with \(\gamma_2\)), both of which can be highly inefficient.
\end{enumerate}


\subsubsection{Numerical results for comparison}

We present numerical results to compare the effectiveness of various anti-symmetric functions. It is important to clarify that if an anti-symmetric function does not perform well in this context, it simply indicates that it is not suitable for our specific problem, rather than suggesting that it is inherently flawed. With careful modifications based on the principles we previously outlined, many of these functions may be significantly improved.

\begin{description}
 \item (a) \( \gamma(x, y) = \text{sign}(y - x)\cdot[\exp(-x) \vee \exp(-y)] \) [SENS's anti-symmetric function];

 \item (b) \( \gamma_1(x, y) = \text{sign}(y - x)\cdot |x - y| \);

 \item (c) \( \gamma_2(x, y) = \text{sign}(y - x) \cdot (x\wedge y)\) [the anti-symmetric function in \cite{barber2015controlling} tailored to our setup].
\end{description}

We generate \(n\) observations for each unit \(i\in[2000]\): $X_{ij} = \mu_i + \epsilon_{ij},$ $j\in[n]$, where 
\begin{equation*}
\begin{aligned}
\mu_i &\stackrel{i.i.d.}{\sim}(1 - \pi) \delta_0 + \pi \mathcal{N}(-\mu,\mu^2),\quad \sigma_i\stackrel{i.i.d.}{\sim} \mathcal{U}(0.05,\sigma_{\text{max}}),\\
\epsilon_{ij}\mid \sigma_i &\stackrel{i.i.d.}{\sim} (1-\beta)\mathcal{N}(0, \sigma_i^2) + \frac{\beta}{2}\mathcal{U}(-\sqrt{3}\sigma_i, \sqrt{3}\sigma_i)+\frac{\beta}{2}\text{Laplace}(0, \sigma_i/\sqrt{2}).
\end{aligned}
\end{equation*}
The simulation examines the following settings with \(\alpha = 0.05\):  
\begin{description}
\item (a) \(\pi=0.2\), \(n = 10\), \(\beta = 1\), $\sigma_{\text{max}}=0.5$, varying \(\mu\);
\item (b) \(\mu=2\), \(\beta = 1\), \(n=10\), $\sigma_{\text{max}}=0.5$, varing $\pi$;
\item (c) \(\pi=0.2\), \(n=10\) \(\mu=2\), \(\beta = 0.5\), varying $\sigma_{\text{max}}$.
\end{description}

\begin{figure}[htbp!]
    \centering
    \includegraphics[width=1\linewidth]{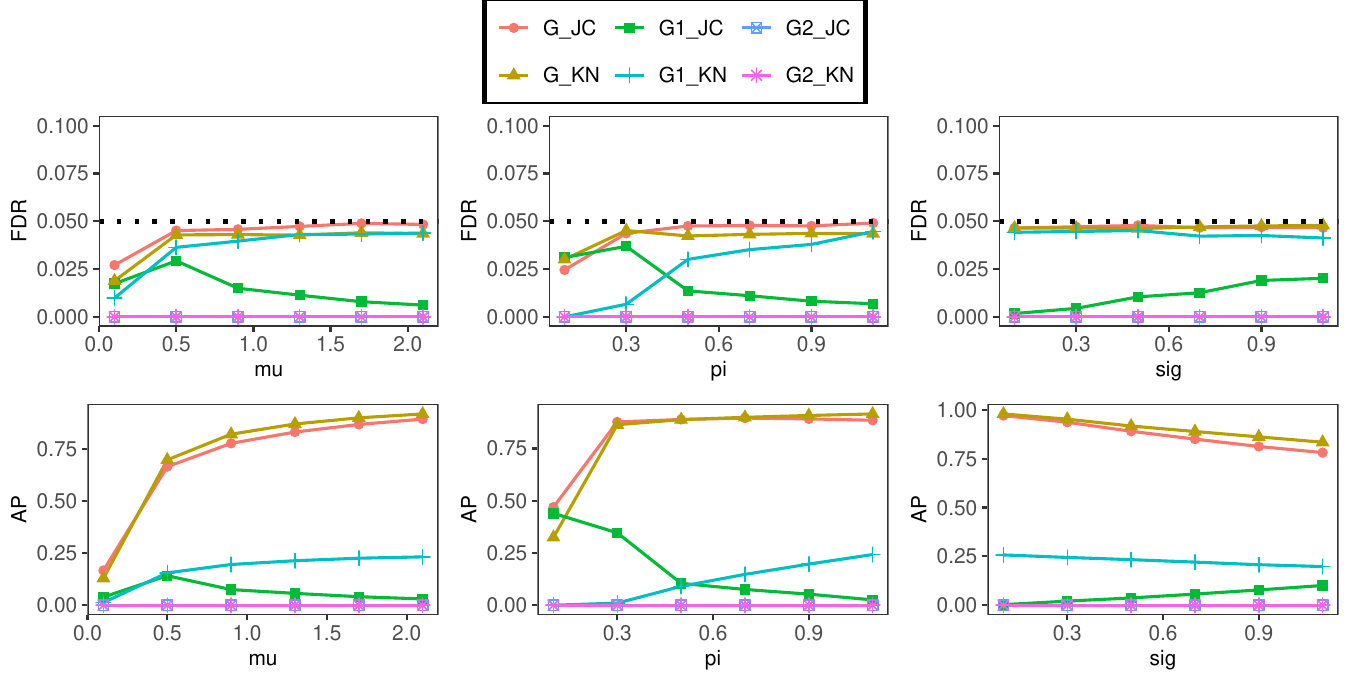}
    \caption{FDR and AP comparison for SENS with different choices of anti-symmetric functions. The left, middle, and right columns correspond to (a), (b), and (c), respectively.}
    \label{fig:choice-anti}
\end{figure}

We apply two SENS methods, SENS\_KN and SENS\_JC, with three different choices of anti-symmetric functions, $\gamma$, $\gamma_1$ and $\gamma_2$. The six combinations are denoted as G\_{KN}, G\_{JC}, G1\_{KN}, G1\_{JC}, G2\_{KN}, and G2\_{JC}, respectively. We summarize the simulation results in Figure \ref{fig:choice-anti} by presenting the average outcomes across 100 independent datasets. We can see that in all three scenarios, the FDR is effectively controlled by all methods. The anti-symmetric function \( \gamma \) achieves the best performance. In contrast, the other two functions, \( \gamma_1 \) and \( \gamma_2 \), display notably lower power across all cases.

\subsubsection{Analysis of performance factors}

Next, we present an analysis that carefully explains why \(\gamma_1(x, y)\) and \(\gamma_2(x, y)\) are unsuitable as anti-symmetric functions for our algorithm, while \(\gamma(x, y)\) is a more effective choice.

According to Lemma \ref{pro3}, the optimal decision rule is to reject smaller values of \(r_m(T_i)\), which serve as the oracle counterpart of the estimated score \(U_i\). This suggests that directly thresholding \(U_i\) is the preferred approach. Lemma \ref{lem:G=U} further demonstrates that the rule \(\mathbb{I}(\gamma(U_i, U^0_i) \geq \tau)\) is equivalent to \(\mathbb{I}(U_i \leq U^0_i)\mathbb{I}(U_i \leq \tau^\prime)\), where \(\tau^\prime\) is defined in Equation \eqref{equ:tau^prime}. This establishes a thresholding principle that effectively ranks \(U_i\), aside from the terms eliminated by \(\mathbb{I}(U_i \leq U^0_i)\), thus justifying our proposed anti-symmetric function \(\gamma\).

In contrast, following similar arguments in the proof of Lemma \ref{lem:G=U}, the rule \(\mathbb{I}(\gamma_1(U_i, U^0_i) \geq \tau_1)\) simplifies to \(\mathbb{I}(U_i \leq U^0_i)\mathbb{I}(U^0_i - U_i \leq \tau_1^\prime)\), and the rule \(\mathbb{I}(\gamma_2(U_i, U^0_i) \geq \tau_2)\) reduces to \(\mathbb{I}(U_i \leq U^0_i)\mathbb{I}(U_i \geq \tau_2^\prime)\), where \(\tau_1\), \(\tau_1^\prime\), \(\tau_2\), and \(\tau_2^\prime\) are the respective thresholds ensuring FDR control at level \(\alpha\). Notably, these alternative anti-symmetric functions either fail to directly impose thresholds on \(U_i\) (as with \(\gamma_1\)) or impose them in the opposite direction (as with \(\gamma_2\)), rendering them significantly less efficient.

\subsection{Derandomized SENS Algorithm}\label{subsec:derandomized SENS}
SENS employs sample-splitting, which can be undesirable in practice due to the additional uncertainties it introduces. To address this issue, we propose a derandomization method that draws upon the works of \cite{vovk2021values}, \cite{wang2022false}, and \cite{ren2024derandomised}. 

We summarize the derandomized SENS Algorithm in the table below (Algorithm \ref{alg:derandomized SENS}). This algorithm essentially involves running Algorithm \ref{alg:SENS} (or its equivalent \(e\)-BH) \(N\) times and then averaging the outputs. Specifically, for each \( k \in [N] \), we construct \( T^{(k)}_i \) and \( T^{0,(k)}_i \) as defined in \eqref{eq:T}, compute the scores \( \{ U^{(k)}_i, U^{0,(k)}_i \} \) using \eqref{equ:g(t)}, and calculate \( G^{(k)}_i \) via \eqref{eq:anti-sym}. The threshold \( \tau^{(k)} \) is then determined from \eqref{equ:tau}, and the generalized $e$-values \( e^{(k)}_i \) are computed as described in \eqref{equ:e-value}. Afterward, we apply the $e$-BH procedure to the summary $e$-values \( \{\bar{e}_i = \frac{1}{N} \sum_{k=1}^N e_i^{(k)}: i\in [m]\} \). 

\singlespacing
\begin{breakablealgorithm}
\label{alg:derandomized SENS}
    \caption{The derandomized SENS Algorithm}  
    \begin{algorithmic}[1]
       \Require Observations $\{X_{ij}:j\in[n_i]\}_{i=1}^m$, number of replications $N$, weights $(\alpha_k)_{k=1}^N$, target FDR level $\alpha$, option from \{``JC'', ``KN''\}.
       \Ensure The rejection set $\mathcal{R}$.
    \For{$k=1,2,\ldots,N$}
    \State Randomly partition the observations and compute \( (T_i, T^0_i) \), \( i \in [m] \).
    \State Estimate \( \hat{f}_{mix} \) using Equation \eqref{equ:hat_f_mix}.
     \If{option==``JC"}
       \State Estimate $\hat{f}_0$ using the Jin-Cai method [\eqref{equ:JC}]. 
    \ElsIf{option==``KN''}
        \State Estimate $\hat{f}_0$ using the kernel method [\eqref{f0-symm}].
    \EndIf 
    \State Compute \( g(t) \) via \eqref{equ:g(t)}. Let \(\mathbf{U} = \{U_i = g(T_i)\}_{i=1}^m \quad \text{and} \quad \mathbf{U}^0 = \{U^0_i = g(T^0_i)\}_{i=1}^m.\)
    \State Compute test statistics \( \{G_i:i\in[m]\}\) via \eqref{eq:anti-sym}.
    \State Determine the threshold $\tau$ via \eqref{equ:tau} by replacing $\alpha$ with $\alpha_k$ and construct (generalized) $e$-values $ e^{(k)}_i $ for $ i \in [m] $ via \eqref{equ:e-value}.
    \EndFor
    \State Let $\bar{e}_i=\frac{1}{N}\sum_{k=1}^N e_i^{(k)}$ for $i\in[m]$. Denote the order statistics by $\bar{e}_{(1)} \geq \bar{e}_{(2)} \geq \cdots \geq \bar{e}_{(m)}$. Let $\hat{k} = \max \left\{i : \frac{i \bar{e}_{(i)}}{m} \geq \frac{1}{\alpha}\right\}$.
    \State Let $\mathcal{R}= \left\{i\in[m] : \bar{e}_i \geq \bar{e}_{(\hat{k})}\right\}$.
    \end{algorithmic}
\end{breakablealgorithm}

\spacingset{1} 

Since the summary $e$-values \( \{\bar{e}_i: i\in [m]\} \) form a set of generalized $e$-values, Algorithm \ref{alg:derandomized SENS} guarantees valid FDR control; we state this formally in the following corollary.  

\begin{corollary}\label{the:DSENS} (Validity of derandomized SENS). Consider model \eqref{equ:model}. Suppose that (a) Assumption \ref{ass:errors} holds; (b) $(U_i,U^0_i)$ are constructed via Algorithm \ref{alg:derandomized SENS}, and there is no tie between $ U_i $ and $U^0_i $ almost surely for each $k\in[N]$. Then Algorithm \ref{alg:derandomized SENS} controls the FDR at level $\alpha$.
\end{corollary}

The problem of derandomization is inherently complex, often involving a trade-off between stability and efficiency. As a result, we do not have an optimality theory for Algorithm \ref{alg:derandomized SENS} that aligns with Theorem \ref{the:Asymptotic optimality}.

Implementing Algorithm \ref{alg:derandomized SENS} requires the selection of hyperparameters such as \(N\) and \((\alpha_k)_{k=1}^N\). We have conducted extensive numerical studies to examine the effects of these hyperparameters on the performance of the derandomized SENS (see Appendix \ref{simu:dsens} for details). The following points have been noted in our studies:
\begin{itemize}
\item Increasing $N$ reduces algorithmic variability but at the cost of higher computational burden. Beyond $N = 10$, further increases yield diminishing returns in terms of variability reduction. Moreover, $N$ has little effect on the power of the Derandomized SENS. Hence, we recommend using a moderate value for $N$ (e.g., $N = 10$); 

\item The hyperparameter $\alpha^*$ significantly influences power. Specifically, our results show that the highest power and lowest variance are achieved when $\alpha^* \leq 0.7\alpha$, whereas $\alpha^* \geq \alpha$ leads to fewer discoveries and higher variance. Therefore, we recommend using a smaller value for $\alpha^*$ (e.g., $\alpha^* = 0.5\alpha$).
\end{itemize}

Furthermore, we conduct a simulation study to compare SENS and Derandomized SENS with the appropriate hyperparameters and identify the following two key points:
(a) Derandomized SENS effectively reduces randomness by achieving lower variance in FDP and lower variability in the rejection decisions. These reductions suggest that the additional uncertainties introduced by sample splitting in SENS have been mitigated through the operation of e-value averaging;
(b) While derandomized SENS exhibits a reduction in average power (AP), it occasionally enhances the average ranking (AR). The AR metric can be more relevant than AP for assessing the effectiveness of a multiple comparison method in some scenarios (e.g., a biologist might prioritize selecting the top 10 genes over controlling the FDR). More details are provided in Appendix \ref{simu:dsens}.

\subsection{Choice of the null distribution in large-scale inference: further clarifications and illustrations}\label{sec:null distribution}

This section aims to provide further clarification on various concepts related to null distributions, which should be carefully distinguished and scrutinized in practice. The relevant concepts include: (a) theoretical null distribution, (b) empirical null distribution, (c) estimated empirical null distribution, (d) oracle null distribution of the working model, and (e) estimated null distribution of the working model. Below, we outline several important points that we would like to emphasize.

\begin{enumerate}[(i)]

\item It is essential to clarify that the concepts regarding the null distribution pertain to the distribution of summary statistics rather than the original observations. In Section \ref{subsec:choice-null}, we have followed conventional practice by selecting $z$-values as the summary statistics. Consequently, the discussions surrounding the various null distributions exclusively focus on $z$-values. In contrast, our SENS Algorithm utilizes \(\mathbf{T}\) [cf. \eqref{eq:T} in Section \ref{subsec:SENS-construction}] as the summary statistics, where \(\mathbf{T} = \{T_i: i \in [m]\}\). In this context, the interpretations of the null distributions differ from that of $z$-values.


\item When \( z \)-values are used, it’s typically assumed that the theoretical null for standardized summary statistics follows a \(\mathcal{N}(0, 1)\) distribution across all testing units. In practice, when the application of the theoretical null is deemed problematic, a viable alternative is to utilize an estimated empirical null derived directly from the summary statistics. Existing methods typically rely on the Gaussian assumption \citep{efron2004large, jin2007estimating}, which has several limitations: (a) the method may perform poorly if the Gaussian assumption is violated; for instance, it can become invalid in cases where the null distribution are heterogeneous or heavy-tailed; (b) poor estimation of model parameters may compromise subsequent inference; and (c) the method only guarantees \emph{asymptotic} control of the FDR and requires strong regularity conditions that can be challenging to validate in practical applications.

\item In the context of the SSMT setup, rather than estimating the empirical null, we propose to directly generate calibrated null samples that accurately reflect the characteristics of the empirical null distribution. This approach effectively circumvents the issues encountered by conventional methods that rely either on theoretical null or estimated empirical null distributions.

\item To construct efficient scores, we require a working model that serves as a simplified representation of the true data-generating process. Since the (theoretical) working null is unknown, we employ an estimated working null as its approximation. Importantly, the accuracy of these approximations -- both the working model's representation of reality and the alignment of the estimated working model with the (theoretical) working model -- does not compromise the validity of our approach, which remains effective for FDR control in finite samples, provided that the constructed test and calibration samples are pairwise exchangeable [cf.\eqref{def:pw-ex}] and therefore the scores are pairwise exchangeable [cf. \eqref{pwex:scores}]. 

\end{enumerate}

We employ the simulation setup outlined in Section \ref{sec:simulation} to illustrate and contrast  various concepts of null distributions. In our analysis, we utilize two types of summary statistics: \(\mathbf{Z}\) for BH\_TN and BH\_EEN, and \((\mathbf{T}^0, \mathbf{T})\) for SENS\_JC and SENS\_KN. It is notable that, despite their similarities, the summary statistics \(\mathbf{T}\) and \(\mathbf{Z}\) rely on different standardization approaches, using \(S_i\) and \(S_i^*\), respectively. For more details, refer to Remark \ref{rem:small-n} in Section \ref{subsec:SENS-construction} and Appendix \ref{insights and counter examples}.

The following concepts of null distributions should be clearly distinguished: 
\begin{itemize}
 
 \item \emph{Theoretical null distribution of \(\mathbf{Z}\)}: In conventional practice, this is assumed to be \(\mathcal{N}(0, 1)\) and is used to implement BH\_TN.
 
 \item \emph{Estimated empirical null distribution of \(\mathbf{Z}\)}: The empirical null is unknown and is estimated using the Jin-Cai method, which is applied in the implementation of BH\_EEN. 
 
 \item \emph{Theoretical and empirical null distributions of \(\mathbf{T}\)}: These concepts are not required in the SENS Algorithm, which is developed under the SSMT framework and employs \(\mathbf{T}^0\) to derive the decision rule directly. 
 
 \item \emph{``Oracle'' null of the working model (for \(\mathbf{T}\))}: This corresponds to \(f_0\) in Equation \eqref{equ:working-model}. This quantity is unknown and its deviation from the true empirical null does not affect the validity of the FDR analysis. 
 
 \item \emph{Estimated null of the working model (for \(\mathbf{T}\))}: SENS\_JC utilizes an estimated working null based on the Jin-Cai method, while SENS\_KN uses an estimated working null constructed with symmetric estimators, as described in Equation \eqref{f0-symm}. Its deviation from the true empirical null does not affect the validity of the FDR analysis. 

\end{itemize}

\subsection{Comparisons of existing techniques in power analysis}\label{subsec:power analysis}

This section presents a detailed discussion that clarifies how our power analysis differs from traditional approaches found in the literature. The discussion aims to elucidate the unique feature of our analytical setup and emphasize the novelty of our theory. As we will explain shortly, our power analysis incorporates both theoretical and practical considerations, making it particularly relevant for large-scale inference. The methodologies, techniques, and key insights of our analysis are markedly distinct from existing research.  

\begin{figure}[htbp!]
    \center
    \includegraphics[width=5.5in]{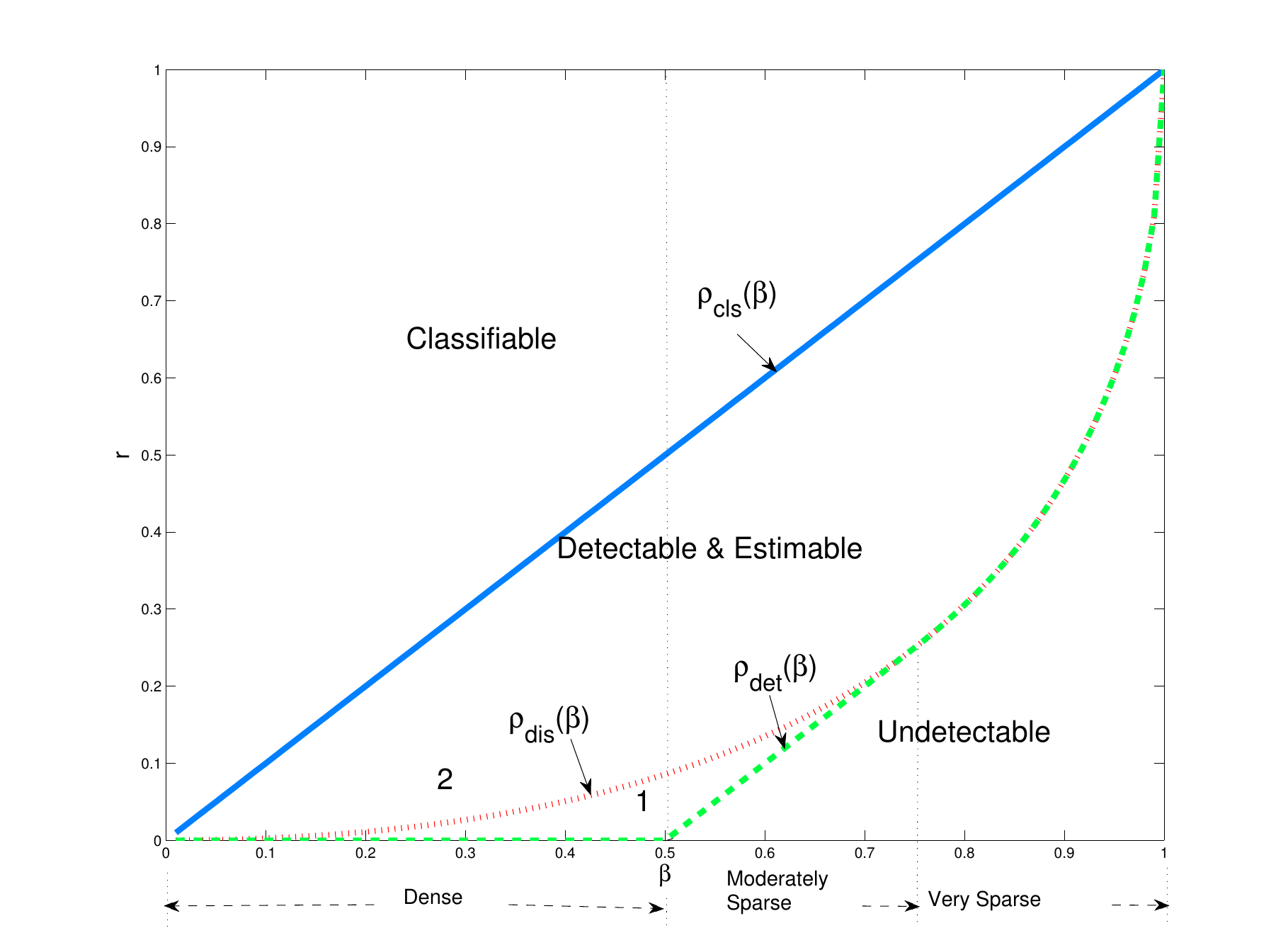}
    \caption{An illustration of detection, discovery and classification boundaries created based on Figure 1 of \cite{TonyCaioptimal2017}. }
    \label{fig:boundaries}
\end{figure}

Our power analysis is centered on the two-point normal mixture model described in \eqref{model:nmix}, a model that has gained significant interest in high-dimensional sparse inference. To frame our discussion in a concise manner, define \(\pi = m^{-\beta}\) and \(\mu_m = \sqrt{2r \log m}\), where \(\beta > 0\) and \(r > 0\) are scaling constants that indicate sparsity and signal strength, respectively. High-dimensional sparse inference encompasses a sequence of increasingly complex tasks: signal detection, signal discovery, and classification \citep{TonyCaioptimal2017}. The difference between signal detection and signal discovery is that signal detection focuses on global testing to determine whether there is any signal, while signal discovery is concerned with multiple testing to identify which are the signals. This growing complexity can be illustrated through the concept of boundaries in Figure \ref{fig:boundaries}. 
 
Specifically, the detection boundary \(\rho_{det}(\beta)\), which defines the minimum conditions necessary for identifying the presence of signals in the data, serves as an optimality benchmark, characterizing the fundamental limits in global inference. The classification boundary \(\rho_{cls}(\beta)\) delineates the precise conditions under which observations can be effectively classified into signals and noise with a negligible misclassification rate. Additionally, the discovery boundary \(\rho_{dis}(\beta)\) partitions the area between the classification and detection boundaries into two regions: the discoverable region and the undiscoverable region.

Traditional power analyses, such as those presented in \cite{arias2017distribution}, \cite{Dai2023splitting}, and \cite{marandon2024adaptive}, focus solely on the region above the classification boundary. In this domain, FDR analysis is not necessary, as signal and noise can be effectively distinguished, resulting in negligible type I and type II error rates. However, in the more challenging region that lies below the classification boundary and above the discovery boundary, it becomes possible to reliably differentiate some -- but not all -- signals from null cases. In this context, it is essential to consider the trade-offs between type I and type II errors, making FDR analysis relevant. By contrast, our power analysis addresses these significant scenarios relevant to practical applications and differs markedly from existing analyses. In these justified and practically relevant cases, our approach begins with the identification of a suitable oracle, followed by the development of a data-driven algorithm that emulates the oracle. Proposition \ref{pro3} and Theorem \ref{the:Asymptotic optimality} together establish the optimality theory of SENS, by providing conditions under which the performance of the oracle procedure can be attained by the data-driven algorithm. 

In summary, existing analyses have not accounted for such an oracle and have primarily explored an overly idealized setting above the classification boundary, where the FDR analysis may not be applicable and even irrelevant. By contrast, our power analysis provides greater relevance and utility, offering valuable insights into high-dimensional sparse inference, particularly regarding the operation of BC-type algorithms. 

\subsection{Comparisons of BH, AZ and SENS}\label{subsec:comparison}

This supplement offers an expanded discussion that continues from the end of Section \ref{sec:asymptotic optimality}. The goal is to contrast the BC algorithm, upon which SENS is built upon, with other widely used baseline algorithms in the literature, including the adaptive $z$-value procedure (AZ, \citealp{sun2007oracle}) and the Benjamini-Hochberg procedure (BH, \citealp{benjamini1995controlling}). We present both theoretical insights and simulation results to elucidate their respective strengths and weaknesses.

\subsubsection{Comparison of SENS and AZ}  

We begin by providing a brief overview of the AZ algorithm. Let \(T_i\) be i.i.d. summary statistics drawn from the mixture model in \eqref{equ:working-model}, and define the lfdr as \(\text{lfdr}(\cdot) = (1 - \pi)\frac{f_0}{f}(\cdot)\). We denote the ordered lfdr values \(\{\text{lfdr}(T_i): i \in [m]\}\) by \(\text{lfdr}_{(1)} \leq \cdots \leq \text{lfdr}_{(m)}\), with \(H_{(1)}, \ldots, H_{(m)}\) corresponding to the respective hypotheses. The AZ algorithm operates as follows:  
\begin{equation}\label{AZ-proc}
\mbox{Let} \;\; k = \max\{j: j^{-1} \sum_{i=1}^j \text{lfdr}_{(i)} \leq \alpha\}. \text{ Reject } H_{(1)}, \cdots, H_{(k)}. 
\end{equation}
This stepwise procedure asymptotically controls the FDR at the level \(\alpha\) and has been shown to be asymptotically optimal under specific regularity conditions \citep{sun2007oracle}. In contrast, the SENS Algorithm, which operates as a BC-type algorithm, is explicitly designed to control the FDR in finite samples in a model-free manner. Furthermore, under the model specified in \eqref{equ:theoretical-working-model} and under strong conditions (Assumption \ref{ass:optimality}), our power analysis (Theorem \ref{the:Asymptotic optimality}) demonstrates that SENS attains the optimality benchmark established by the AZ algorithm in an asymptotic sense. This finding is noteworthy, given that the operational mechanisms of the two algorithms differ fundamentally.

\subsubsection{Comparison of SENS and BH} 

The SENS Algorithm, which can be classified as a BC-type method, offers several advantages over the BH procedure. One of the primary limitations of the BH approach is its inherent conservativeness. Specifically, BH controls the FDR at \((1 - \pi)\alpha\), where \(\pi\) represents the proportion of non-null hypotheses. To fully capitalize on the FDR budget \(\alpha\), Storey’s adjustment is often necessary. However, in the presence of distribution shifts, accurately estimating the non-null proportion becomes challenging, as Storey’s estimator relies on prior knowledge of the null distribution.  

In contrast, the BC-type algorithms, including SENS, do not require this knowledge. BC is adaptive in the sense that as the effect size increases, the FDR level converges to the nominal level \(\alpha\). This property of adaptivity has been demonstrated in the work of \cite{barber2015controlling} and further quantified by \cite{barber2020robust}. Under the restricted model presented in \eqref{equ:theoretical-working-model} and assuming strong conditions (cf. Assumption \ref{ass:optimality}), Theorem \ref{the:Asymptotic optimality} has established the SENS Algorithm is asymptotically optimal. In the following section, we provide simulation results that illustrate how, as \(m\) increases, the SENS Algorithm becomes less conservative compared to BH; as a result, the power of SENS also surpasses that of BH.

\subsubsection{Numerical results and further discussions}
We conduct a small simulation study to compare various algorithms, including SENS (BC), BH, and AZ. The test statistics \(T_i\) and their calibration points \(T^0_i\) for \(i \in [m]\) are generated according to the following model:
\begin{equation}\label{equ:compare BH AZ SENS}
\begin{aligned}  
&T_i \stackrel{i.i.d.}{\sim} (1 - 0.1)\phi(t) + 0.1 \phi(t - 0.4 \log(m)), \\
&T^0_i \stackrel{i.i.d.}{\sim} \phi(t), \quad \mbox{Cor}(T_i,T^0_i)=0.5,\quad i \in [m], \\  
&T_i \text{ and } T_i^0 \text{ are mutually independent for } i \in [m].
\end{aligned}    
\end{equation}

To ensure the comparison of different baseline algorithms on an equal footing, we consider a setup where the theoretical null distribution of the test statistics is known, enabling BH and AZ to utilize the correctly specified theoretical null distribution. In this context, SENS is implemented using the SSMT setup, with \(\mathbf{T}^0\) already available. This ideal setting provides substantial insights into optimality issues. Notably, it has been demonstrated that AZ serves as the optimality benchmark for among all valid FDR procedures \citep{sun2007oracle}. We examine a scenario in which \(m\) is increased from small to large, which not only highlights the strengths and weaknesses of the various baseline algorithms but also reveals the critical role of Assumption \ref{ass:optimality} in our power analysis. 

\begin{figure}[htbp!]
    \center
    \includegraphics[scale=0.8]{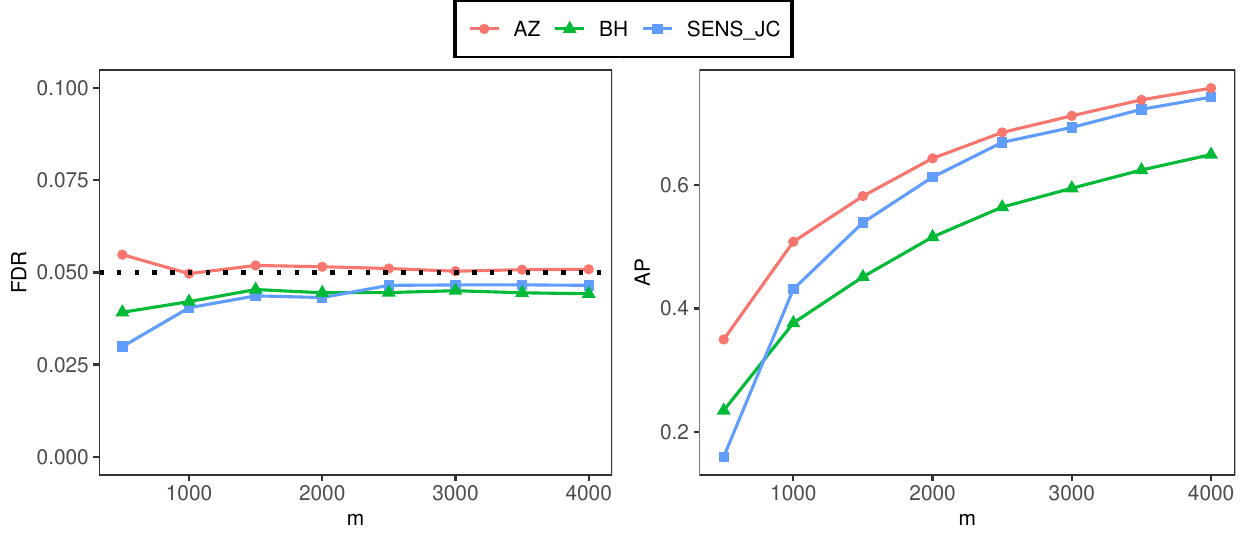}
    \caption{FDR and AP comparison for BH, AZ and SENS under model \eqref{equ:compare BH AZ SENS}.}
    \label{fig:compare BH AZ SENS}
\end{figure}

The following methods are compared in our study: (a) BH procedure implemented using $p$-values calculated under the theoretical null \( \mathcal{N}(0,1) \); (b) AZ implemented based on the $z$-values \( \mathbf{T} \), using the correct null \( \mathcal{N}(0,1) \). The SENS Algorithm is implemented with the option ``JC'' using given pairs of samples \((\mathbf{T}, \mathbf{T}^0)\). 
The simulation results are summarized in Figure \ref{fig:compare BH AZ SENS}. The following patterns can be observed. 

\begin{enumerate}

\item The FDR level of SENS\_JC is quite conservative when signals are weak, but it converges to the nominal level \(\alpha\) as \(m\) increases. In contrast, the FDR level of AZ remains at the nominal level in most scenarios, while the FDR level of BH is conservative across all settings.

\item As \(m\) increases, the power of SENS\_JC initially approaches and subsequently exceeds that of BH. When \(m\) is very large, the power of SENS\_JC converges to the optimality benchmark set by AZ.

\item AZ only controls the FDR asymptotically. We can see from the left panel that there is a slight inflation of the error rate in a range of $m$ values. In contrast, both SENS\_JC and BH control the FDR below the nominal level \(\alpha\) for all values of $m$.

\end{enumerate}

\subsection{Advantages of the lfdr-type scores}\label{subsec:benefits of the Lfdr-Type score function g}

Our score function \( g \) is designed to emulate the lfdr for two primary reasons. First, as demonstrated in \cite{sun2007oracle}, a decision rule based on lfdr statistics is optimal; hence, using a score function that approximates the lfdr enhances statistical power. Second, an lfdr-type score function aligns well with Efron's empirical Bayes framework, which explicitly leverages a data-driven empirical null to characterize the ``normal state'' in large-scale multiple testing. We offer the following two supporting explanations.

\subsubsection{Comparison with the absolute value score function}
A natural choice for the non-learned score function is \(g(\cdot) = -|\cdot|\). Suppose the anti-symmetric function is \(\gamma(x, y)\) as specified in Equation \eqref{eq:anti-sym}, and we employ 
$$
G_i = \{\gamma(-|T_i|, -|T^0_i|)\},\quad \mbox{for $i \in [m]$},
$$
to implement Algorithm \ref{alg:SENS}. However, the choice \(g(\cdot) = -|\cdot|\) may lead to efficiency loss. 

To illustrate, we generate observations according to the following model: 
{\small\begin{equation}\label{equ:setting1}
\begin{aligned}
X_{ij} & = \mu_i + \epsilon_{ij}, \quad\mu_i \stackrel{i.i.d.}{\sim}(1 - \pi) \delta_0 + \pi \mathcal{N}(-\mu,\mu^2),\quad \sigma_i\stackrel{i.i.d.}{\sim} \mathcal{U}(0.05,\sigma_{\text{max}}),\\
\epsilon_{ij}\mid \sigma_i &\stackrel{i.i.d.}{\sim} (1-\beta)\mathcal{N}(0, \sigma_i^2) + \frac{3\beta}{4}\mathcal{U}(-\sqrt{3}\sigma_i, \sqrt{3}\sigma_i)+\frac{\beta}{4}\text{Laplace}(0, \sigma_i/\sqrt{2}), 
\end{aligned}
\end{equation}
}
for $i\in[m], j\in[n]$. In our simulation studies, we fix \(m=2000\) and \(\alpha = 0.05\), and examine the following settings:
\begin{description}  
\item (a) \(\mu=3\), \(n = 4\), \(\beta = 1\), $\sigma_{\text{max}}=0.3$, varying \(\pi\);
\item (b) \(\pi=0.2\), \(\mu=3\), \(\beta = 1\), \(n=4\), varying $\sigma_{\text{max}}$. 
\end{description}

\begin{figure}[htbp!]
    \centering
    \includegraphics[scale=0.75]{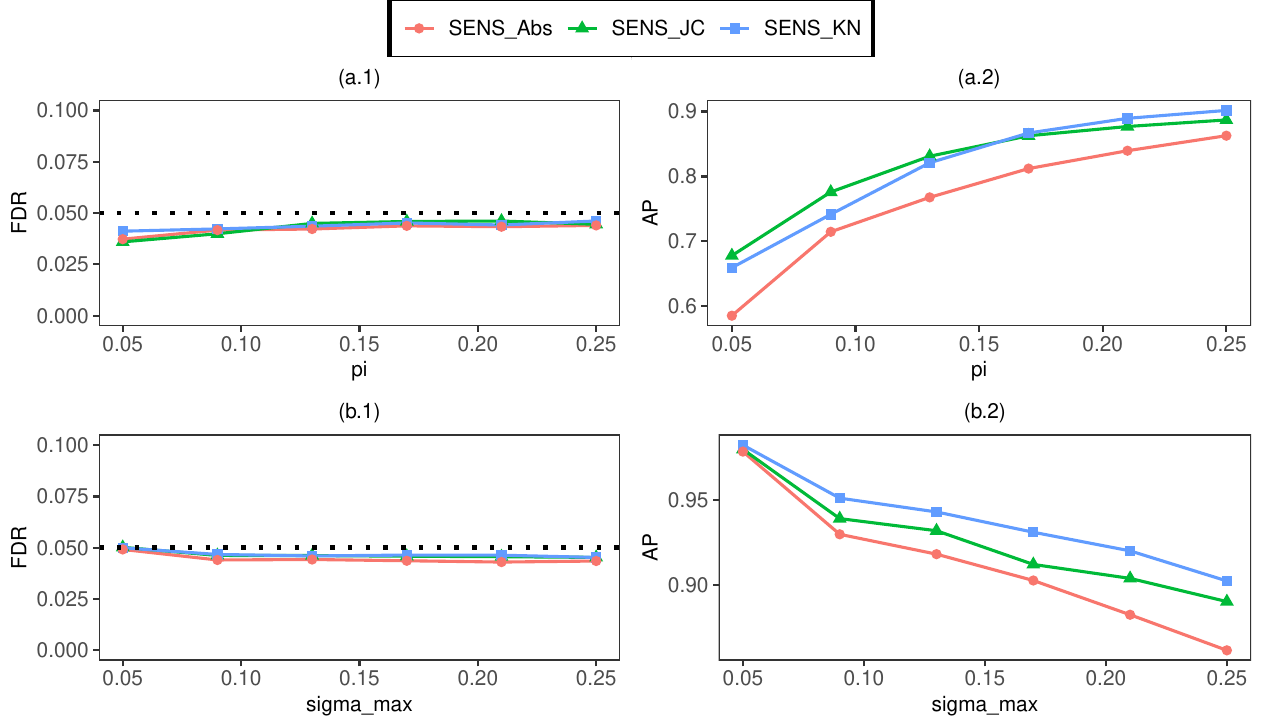}
    \caption{FDR and AP comparison between SENS and SENS\_Abs. The top and bottom rows correspond to Settings (a) and (b), respectively.}
    \label{fig:setting_g}
\end{figure}

We apply SENS\_JC, SENS\_KN, and SENS\_Abs (i.e., SENS with the score function \(g(\cdot)=-|\cdot|\)) to the simulated data. The FDR and AP levels for the different methods, calculated by averaging results from 100 independent datasets, are shown in Figure \ref{fig:setting_g}. We can see that all three methods control the FDR at the nominal level; moreover, SENS\_JC and SENS\_KN outperform SENS\_Abs, demonstrating the power gain achieved by the learned score function that emulates the lfdr.

\subsubsection{The ranking issue}

\begin{figure}[htbp!]
\centering
\includegraphics[scale=0.7]{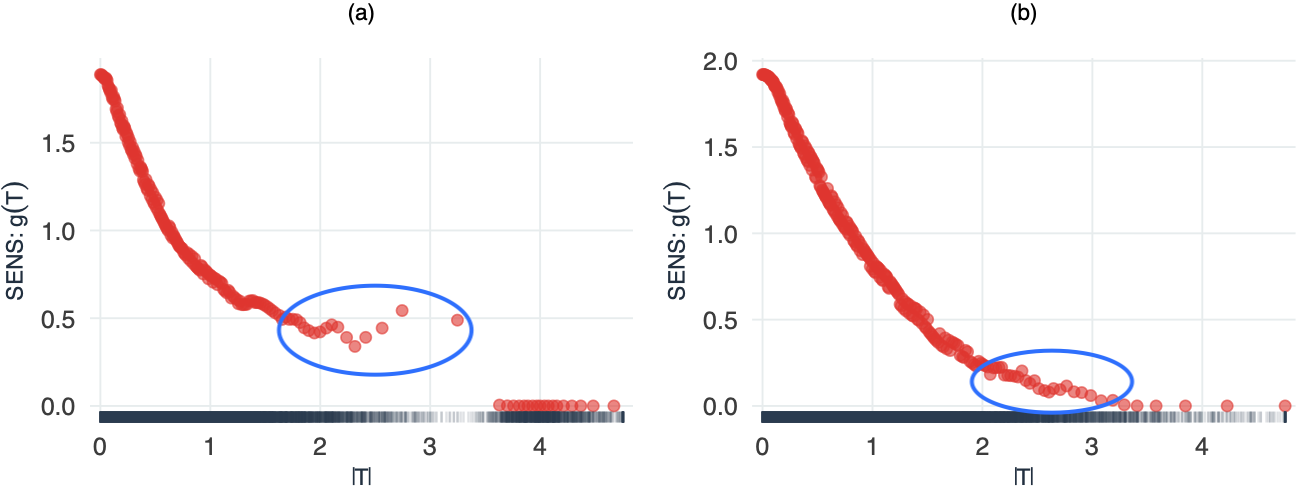}
\caption{The scatter plot of $g(T)$ (constructed via the option ``KN'') as a function of $|T|$: the left and right panels illustrate the results from simulated and real data, respectively.}
\label{fig:g absolute value}
\end{figure}

The lfdr statistics produce a more efficient ranking of hypotheses than do p-values, enabling a beneficial reordering of \(|T_i| \). To illustrate this, we consider the setting in \eqref{equ:setting1} with $m = 10^5$, $\mu = 3$, $n = 4$, $\beta = 0.5$, $\sigma_{\text{max}} = 0.1$, and $\pi = 0.1$. Panel (a) in Figure 4 shows a scatter plot of the pairs $(g(T_i), |T_i|)_{i=1}^m$, while Panel (b) presents the analogous scatter plot from the real data analysis (both using option ``KN'' in Algorithm \ref{alg:SENS}). The blue-circled regions in both Figure \ref{fig:g absolute value}(a) and Figure \ref{fig:g absolute value}(b) indicate a clear reordering trend that underlies the efficiency gain of the lfdr-type function. This reordering explains the patterns we observed in Figure \ref{fig:setting_g} above, where the SENS\_KN method achieves higher power at the same FDR level compared to the SENS\_Abs method due to its superior ranking. Thus, these analyses confirm that the lfdr-type function offers a more accurate ranking relative to the absolute value-based method.

\setcounter{equation}{0}
\renewcommand{\theequation}{B.\arabic{equation}}
\setcounter{figure}{0}
\renewcommand{\thefigure}{B.\arabic{figure}}

\section{Proofs for Primary Theory}\label{sec:supple-proof}

\subsection{Proof of Theorem \ref{lem:SENS}}

Let $\mathbf{K}=(\mathbf{X}_1, \cdots, \mathbf{X}_{i-1}, \mathbf{X}_{i+1}, \cdots, \mathbf{X}_m)$. According to Assumption \ref{ass:errors}, we have $\forall i\in\mathcal H_0$,
\begin{equation}\label{equ:X|X}
\begin{aligned}
    &(X_{ij}: j \in \mathcal{N}_{i2}\mid \mathbf{K}) \stackrel{d}{=} (-X_{ij}: j \in \mathcal{N}_{i2}\mid \mathbf{K}) \quad \mbox{and}\\
    &(X_{ij}: j \in \mathcal{N}_{i1}\mid\mathbf{K}) \text{ is independent of } (X_{ij}: j \in \mathcal{N}_{i2}\mid\mathbf{K}).
\end{aligned}
\end{equation}
Note that \(\bar{X}_{i1}\) and \(\bar{X}_{i2}\) are independent, we conclude that
\[
(\bar{X}_{i1}, \bar{X}_{i2}\mid\mathbf{K}) \stackrel{d}{=} (\bar{X}_{i1}, -\bar{X}_{i2}\mid\mathbf{K}), \quad \forall i\in\mathcal H_0.
\]
Recalling the definitions $V_i = \sqrt{\frac{n_{i1} n_{i2}}{n_i}} (\bar{X}_{i1} + \bar{X}_{i2})$ and $V^0_i = \sqrt{\frac{n_{i1} n_{i2}}{n_i}} (\bar{X}_{i1} - \bar{X}_{i2}),$
we have 
$$
(V_i,V^0_i\mid\mathbf{K}) \stackrel{d}{=} (V^0_i,V_i\mid\mathbf{K}), \quad \forall i\in\mathcal H_0.
$$

When \( n_i = 2 \), we have \( S_i = 1 \). It is evident that:
\begin{equation}\label{V|S}
    (V_i, V^0_i \mid \mathbf{K}, S_i) \stackrel{d}{=} (V^0_i, V_i \mid \mathbf{K}, S_i), \quad \forall i\in\mathcal H_0.
\end{equation}

In the scenario where \( n_i = 3 \), we have \( S_i = S_{i1} \). According to the second equation in \eqref{equ:X|X}, it can be seen that \((\bar{X}_{i2} \mid \mathbf{K})\) is independent of \((S_{i1} \mid \mathbf{K})\). We conclude that  
\[
(\bar{X}_{i1}, \bar{X}_{i2}, S_{i1} \mid \mathbf{K}) \stackrel{d}{=} (\bar{X}_{i1}, -\bar{X}_{i2}, S_{i1} \mid \mathbf{K}), \quad \forall i\in\mathcal H_0.
\]
By substituting \(S_{i1}\) with \(S_{i}\) and utilizing the definitions of \(V_i\) and \(V_i^0\), we establish that 
$(V_i, V^0_i, S_{i} \mid \mathbf{K}) \stackrel{d}{=} (V_i^0, V_i, S_{i} \mid \mathbf{K}),
$
which indicates that Equation \eqref{V|S} remains valid for the case of $n_i=3$.

For \( n_i \geq 4 \), recall that \( S_i \) is defined as 
$
S_i = \sqrt{\frac{(n_{i1}-1) S^2_{i1} + (n_{i2}-1) S^2_{i2}}{(n_i-2)}}
$
[cf. Equation \eqref{eq:VS}]. We start by considering the first equation in \eqref{equ:X|X}. Applying the sample mean function and the sample standard deviation function to both datasets, \( \{ X_{ij} : j \in \mathcal{N}_{i2} \} \) and \( \{-X_{ij} : j \in \mathcal{N}_{i2} \} \), we conclude that 
\begin{equation}\label{equ:X,S|K}
    (\bar{X}_{i2}, S_{i2} \mid \mathbf{K}) \stackrel{d}{=} (-\bar{X}_{i2}, S_{i2} \mid \mathbf{K}),
\quad \forall i\in\mathcal H_0,
\end{equation}
where \( \bar{X}_{i2} \) and \( -\bar{X}_{i2} \) represent the sample means of the two datasets, with \( S_{i2} \) being the sample standard deviation for both.

By Assumption \ref{ass:errors}, \((\bar{X}_{i1}, S_{i1} \mid \mathbf{K})\) and \((\bar{X}_{i2}, S_{i2} \mid \mathbf{K})\) are mutually independent. Moreover, \((\bar{X}_{i1}, S_{i1} \mid \mathbf{K})\) and \((-\bar{X}_{i2}, S_{i2} \mid \mathbf{K})\) must also be independent. We conclude that 
\[
(\bar{X}_{i1}, S_{i1}, \bar{X}_{i2}, S_{i2} \mid \mathbf{K}) \stackrel{d}{=} (\bar{X}_{i1}, S_{i1}, -\bar{X}_{i2}, S_{i2} \mid \mathbf{K}), \quad \forall i\in\mathcal H_0.
\]
The above equation can also be expressed as 
$$
(\bar{X}_{i1}, \bar{X}_{i2} \mid \mathbf{K}, S_{i1}, S_{i2}) \stackrel{d}{=} (\bar{X}_{i1}, -\bar{X}_{i2} \mid \mathbf{K}, S_{i1}, S_{i2}),  \forall i\in\mathcal H_0.
$$
Using the definition of \( S_i \), we have $\forall i\in\mathcal H_0$,
\[
(\bar{X}_{i1}, \bar{X}_{i2} \mid \mathbf{K}, S_i, S_{i1}, S_{i2}) \stackrel{d}{=} (\bar{X}_{i1}, -\bar{X}_{i2} \mid \mathbf{K}, S_i, S_{i1}, S_{i2}).
\]
From the definitions of \( V_i \) and \( V^0_i \), it follows that 
\[
(V_i, V^0_i \mid \mathbf{K}, S_i, S_{i1}, S_{i2}) \stackrel{d}{=} (V^0_i, V_i \mid \mathbf{K}, S_i, S_{i1}, S_{i2}), \quad \forall i\in\mathcal H_0.
\]
By integrating out \( (S_{i1}, S_{i2}) \), we conclude that Equation \eqref{V|S} holds for \( n_i \geq 4 \). Therefore, we have successfully established Equation \eqref{V|S} for all \( n_i \geq 2 \). 

The definitions of \(\mathbf{T}\) and \(\mathbf{T}^0\) in \eqref{eq:T} imply that
\begin{equation}\label{T|S}
(T_i,T^0_i\mid\mathbf{K},S_i) \stackrel{d}{=} (T^0_i,T_i\mid\mathbf{K},S_i), \quad \forall i\in\mathcal H_0.
\end{equation}
Both $T_j$ and $T^0_j$ are constructed exclusively based on $\mathbf{X}_j$, which is independent of $\mathbf{K}$. Therefore, it follows from \eqref{T|S} that
$$
(T_i,T^0_i\mid\mathbf{T}_{-i},\mathbf{T}^0_{-i},\mathbf{K},S_{i}) \stackrel{d}{=} (T^0_i,T_i\mid\mathbf{T}_{-i},\mathbf{T}^0_{-i},\mathbf{K},S_{i}), \quad \forall i\in\mathcal H_0.
$$
The proof is completed by integrating out \((\mathbf{K},S_{i})\). 

\subsection{Proof for Theorem \ref{lem:SENS-two-sample}}
According to Assumption \ref{ass:errors-two-sample}, we have $\forall i\in\mathcal H_0$,
\begin{equation}\label{equ:X|X-two-sample}
\begin{aligned}
    &(X_{ij}-\mu_{xi}: j \in \mathcal{N}_{xi2}\mid \mathbf{K_{-i}}) \stackrel{d}{=} (-(X_{ij}-\mu_{xi}): j \in \mathcal{N}_{xi2}\mid \mathbf{K_{-i}}) \text{ and }\\
    &(X_{ij}-\mu_{xi}: j \in \mathcal{N}_{xi1}) \text{ is independent of } (X_{ij}-\mu_{xi}: j \in \mathcal{N}_{xi2})\text{ conditional on }\mathbf{K_{-i}},\\
    &(Y_{ij}-\mu_{yi}: j \in \mathcal{N}_{yi2}\mid \mathbf{K_{-i}}) \stackrel{d}{=} (-(Y_{ij}-\mu_{yi}): j \in \mathcal{N}_{yi2}\mid \mathbf{K_{-i}}) \text{ and }\\
    &(Y_{ij}-\mu_{yi}: j \in \mathcal{N}_{yi1}) \text{ is independent of } (Y_{ij}-\mu_{yi}: j \in \mathcal{N}_{yi2})\text{ conditional on }\mathbf{K_{-i}}.
\end{aligned}
\end{equation}

Note that $\bar{X}_{i1} - \mu_{xi}$ and $\bar{X}_{i2} - \mu_{xi}$ are conditionally independent given $\mathbf{K_{-i}}$, and similarly, $\bar{Y}_{i1} - \mu_{yi}$ and $\bar{Y}_{i2} - \mu_{yi}$ are conditionally independent given $\mathbf{K_{-i}}$, we conclude that
\begin{equation*}
\begin{aligned}
&(\bar{X}_{i1} - \mu_{xi}, \bar{X}_{i2} - \mu_{xi} \mid \mathbf{K_{-i}}) \stackrel{d}{=} (\bar{X}_{i1} - \mu_{xi}, -(\bar{X}_{i2} - \mu_{xi}) \mid \mathbf{K_{-i}}) \mbox{ and }\\
&(\bar{Y}_{i1} - \mu_{yi}, \bar{Y}_{i2} - \mu_{yi} \mid \mathbf{K_{-i}}) \stackrel{d}{=} (\bar{Y}_{i1} - \mu_{yi}, -(\bar{Y}_{i2} - \mu_{yi}) \mid \mathbf{K_{-i}}), \quad \forall i \in \mathcal{H}_0.
\end{aligned}
\end{equation*}

Since for all $i \in \mathcal{H}_0$, the sets $\{\epsilon_{xij} : j \in [n_{xi}]\}$ and $\{\epsilon_{yij} : j \in [n_{yi}]\}$ are conditionally independent given $\mathbf{K_{-i}}$, we can combine this with the previous equation to obtain:
{\small
\begin{equation*}
\begin{aligned}
&(\bar{X}_{i1}+\bar{X}_{i2}-2\mu_{xi} \mid \mathbf{K_{-i}}) \stackrel{d}{=} (\bar{X}_{i1}-\bar{X}_{i2} \mid \mathbf{K_{-i}}) \mbox{ and } 
(\bar{Y}_{i1}+\bar{Y}_{i2}-2\mu_{yi} \mid \mathbf{K_{-i}}) \stackrel{d}{=} (\bar{Y}_{i1}-\bar{Y}_{i2} \mid \mathbf{K_{-i}})\\
&(\bar{X}_{i1}+\bar{X}_{i2}-2\mu_{xi},\bar{X}_{i1}-\bar{X}_{i2}) \mbox{ and } 
(\bar{Y}_{i1}+\bar{Y}_{i2}-2\mu_{yi} ,\bar{Y}_{i1}-\bar{Y}_{i2})\mbox{ are mutually independent conditional on }\mathbf{K_{-i}}.
\end{aligned}
\end{equation*}
}
Recalling the definitions \( V_i = (\bar{X}_{i1} + \bar{X}_{i2}) - (\bar{Y}_{i1} + \bar{Y}_{i2}) \) and \( V^0_i = (\bar{X}_{i1} - \bar{X}_{i2}) - (\bar{Y}_{i1} - \bar{Y}_{i2}) \), and given that for \( i \in \mathcal{H}_0 \), \( \mu_{xi} = \mu_{yi} \), we obtain the following distributional equality:
\[
(V_i, V^0_i \mid \mathbf{K_{-i}}) \stackrel{d}{=} (V^0_i, V_i \mid \mathbf{K_{-i}}), \quad \forall i \in \mathcal{H}_0.
\]

For \( n_{xi} \geq 4 \) and \(n_{yi}\geq4\),  recall that \( S_i \) is defined as the function of $S_{xi1}$, $S_{xi2}$, $S_{yi1}$ and $S_{yi2}$, [cf. Equation \eqref{equ:VS-two-sample}]. We start by considering the first equation in \eqref{equ:X|X-two-sample}. By applying the sample mean and sample standard deviation functions to the two datasets for \( X_i \) and \( Y_i \), we examine 
the following sets:
\begin{itemize}
    \item \( \{ X_{ij}-\mu_{xi} : j \in \mathcal{N}_{xi2} \} \) and \( \{ -(X_{ij}-\mu_{xi}) : j \in \mathcal{N}_{xi2} \} \),
    \item \( \{ Y_{ij}-\mu_{yi} : j \in \mathcal{N}_{yi2} \} \) and \( \{ -(Y_{ij}-\mu_{yi}) : j \in \mathcal{N}_{yi2} \} \).
\end{itemize} 

From this, we conclude that $\forall i \in \mathcal{H}_0$,
\[
(\bar{X}_{i2} - \mu_{xi}, S_{xi2} \mid \mathbf{K_{-i}}) \stackrel{d}{=} (-(\bar{X}_{i2} - \mu_{xi}), S_{xi2} \mid \mathbf{K_{-i}}), (\bar{Y}_{i2} - \mu_{yi}, S_{yi2} \mid \mathbf{K_{-i}}) \stackrel{d}{=} (-(\bar{Y}_{i2} - \mu_{yi}), S_{yi2} \mid \mathbf{K_{-i}}),
\]
where \( \bar{X}_{i2} - \mu_{xi} \) and \( -(\bar{X}_{i2} - \mu_{xi}) \) represent the sample means of the two datasets for \( X_i \), with \( S_{xi2} \) as the sample standard deviation for both. Similarly, \( \bar{Y}_{i2} - \mu_{yi} \) and \( -(\bar{Y}_{i2} - \mu_{yi}) \) represent the sample means for \( Y_i \), with \( S_{yi2} \) as the corresponding sample standard deviation.

By Assumption \ref{ass:errors-two-sample}, \((\bar{X}_{i1}-\mu_{xi}, S_{xi1}, \bar{Y}_{i1}-\mu_{yi}, S_{yi1} \mid \mathbf{K_{-i}})\) and \((\bar{X}_{i2}-\mu_{xi}, S_{xi2}, \bar{Y}_{i2}-\mu_{yi}, S_{yi2} \mid \mathbf{K_{-i}})\) are mutually independent. Moreover, \((\bar{X}_{i1}-\mu_{xi}, S_{xi1}, \bar{Y}_{i1}-\mu_{yi}, S_{yi1} \mid \mathbf{K_{-i}})\) and \((-(\bar{X}_{i2}-\mu_{xi}), S_{xi2}, -(\bar{Y}_{i2}-\mu_{yi}), S_{yi2} \mid \mathbf{K_{-i}})\) must also be independent. We conclude that $\forall i\in\mathcal H_0$,
\[
\begin{aligned}
    &(\bar{X}_{i1}-\mu_{xi}, S_{xi1}, \bar{X}_{i2}-\mu_{xi}, S_{xi2},\bar{Y}_{i1}-\mu_{yi}, S_{yi1}, \bar{Y}_{i2}-\mu_{yi}, S_{yi2} \mid \mathbf{K_{-i}}) \\
    &\stackrel{d}{=} (\bar{X}_{i1}-\mu_{xi}, S_{xi1}, -(\bar{X}_{i2}-\mu_{xi}), S_{xi2},\bar{Y}_{i1}-\mu_{yi}, S_{yi1}, -(\bar{Y}_{i2}-\mu_{yi}), S_{yi2} \mid \mathbf{K_{-i}}).
\end{aligned}
\]
The above equation can also be expressed as $\forall i\in\mathcal H_0$,
\[
\begin{aligned}
    &(\bar{X}_{i1}-\mu_{xi}, \bar{X}_{i2}-\mu_{xi},\bar{Y}_{i1}-\mu_{yi}, \bar{Y}_{i2}-\mu_{yi} \mid \mathbf{K_{-i}}, S_{xi1}, S_{xi2}, S_{yi1}, S_{yi2}) \\
    &\stackrel{d}{=} (\bar{X}_{i1}-\mu_{xi}, -(\bar{X}_{i2}-\mu_{xi}),\bar{Y}_{i1}-\mu_{yi}, -(\bar{Y}_{i2}-\mu_{yi}) \mid \mathbf{K_{-i}}, S_{xi1}, S_{xi2}, S_{yi1}, S_{yi2}).
\end{aligned}
\]
Using the definition of \( S_i \) and integrating out \((S_{xi1}, S_{xi2}, S_{yi1}, S_{yi2}) \), we have $\forall i\in\mathcal H_0$,
{\small\[
(\bar{X}_{i1}-\mu_{xi}, \bar{X}_{i2}-\mu_{xi},\bar{Y}_{i1}-\mu_{yi}, \bar{Y}_{i2}-\mu_{yi} \mid \mathbf{K_{-i}}, S_i) \stackrel{d}{=} (\bar{X}_{i1}-\mu_{xi}, -(\bar{X}_{i2}-\mu_{xi}),\bar{Y}_{i1}-\mu_{yi}, -(\bar{Y}_{i2}-\mu_{yi}) \mid \mathbf{K_{-i}},S_i).
\]}
From the definitions of \( V_i \) and \( V^0_i \), it follows that 
\[
(V_i, V^0_i \mid \mathbf{K_{-i}}, S_i) \stackrel{d}{=} (V^0_i, V_i \mid \mathbf{K_{-i}}, S_i), \quad \forall i\in\mathcal H_0.
\]

The definitions of \(\mathbf{T}\) and \(\mathbf{T}^0\) in \eqref{equ:T-two-sample} imply that
\begin{equation}\label{T|S-two-sample}
(T_i,T^0_i\mid\mathbf{K_{-i}},S_i) \stackrel{d}{=} (T^0_i,T_i\mid\mathbf{K_{-i}},S_i), \quad \forall i\in\mathcal H_0.
\end{equation}
Both $T_j$ and $T^0_j$ are constructed exclusively based on $\mathbf{X}_j$, which is independent of $\mathbf{K_{-i}}$. Therefore, it follows from \eqref{T|S-two-sample} that
$$
(T_i,T^0_i\mid\mathbf{T}_{-i},\mathbf{T}^0_{-i},\mathbf{K_{-i}},S_{i}) \stackrel{d}{=} (T^0_i,T_i\mid\mathbf{T}_{-i},\mathbf{T}^0_{-i},\mathbf{K_{-i}},S_{i}), \quad \forall i\in\mathcal H_0.
$$
The proof is completed by integrating out \((\mathbf{K_{-i}},S_{i})\). 
   
\subsection{Proof of Proposition \ref{prop:score-class}}

Let $\psi(x, y)$ be a vector-valued symmetric function satisfying $\psi(x, y)=$ $\psi(y, x)$. Consider two random elements $X$ and $Y$ that are pairwise exchangeable, i.e. $(X, Y) \stackrel{d}{=}(Y, X)$. Then we have
\begin{equation}\label{equ:lem2.1}
  (X, Y, \psi(X, Y)) \stackrel{d}{=}(Y, X, \psi(Y, X)) =(Y, X,\psi(X, Y)) .  
\end{equation}

Suppose we are interested in utilizing function $g\left(t;(\mathbf{T}, \mathbf{T}^0) \right)$
swapping-invariance property \eqref{equ:score function} to construct conformity scores. This implies that $g$ is fully determined by the unordered pairs $\left\{T_1, T^0_1\right\}, \cdots,\left\{T_m, T^0_m\right\}$. Let $\left\{T_i, T^0_i\right\}$ represent the unordered set of $T_i$ and $T^0_i$. Further denote $\mathbf{T}_{-i}=(T_1,\cdots,T_{i-1},T_{i+1},\cdots,T_m)$ and $\mathbf{T}^0_{-i}=(T^0_1,\cdots,T^0_{i-1},T^0_{i+1},\cdots,T^0_m)$. We can rewrite the scores defined in \eqref{g-class} as:
\begin{equation}\label{equ:lem2.2}
    U_i=g\left(T_i ;\left\{T_i, T^0_i\right\},\left(\mathbf{T}_{-i}, \mathbf{T}^0_{-i}\right)\right), \quad U^0_i=g\left(T^0_i ;\left\{T_i, T^0_i\right\},\left(\mathbf{T}_{-i}, \mathbf{T}^0_{-i}\right)\right).
\end{equation}
 
Let $\mathbf{W}_i \equiv\left(\mathbf{U}_{-i}, \mathbf{U}^0_{-i}, T_i \vee T^0_i, T_i \wedge T^0_i\right)$. The vector $\mathbf{W}_i$ comprises two components. The first part encompasses scores from units excluding $i$: $\left(\mathbf{U}_{-i}, \mathbf{U}^0_{-i}\right)$, while the second part $\left(T_i \vee T^0_i, T_i \wedge T^0_i\right)$ provides the values of the unordered set $\left\{T_i, T^0_i\right\}$.
Note that $\left(T_i \vee T^0_i, T_i \wedge T^0_i\right)=\left(T^0_i \vee T_i, T^0_i \wedge T_i\right)$, and the scores $U_i$ and $U^0_i$ are swapping invariant [cf. \eqref{equ:lem2.2}]. Given $(\mathbf{T}_{-i}, \mathbf{T}^0_{-i})$, the following mapping
$$
\left(T_i, T^0_i\right) \mapsto \mathbf{W}_i \equiv\left(\mathbf{U}_{-i}, \mathbf{U}^0_{-i}, T_i \vee T^0_i, T_i \wedge T^0_i\right)
$$
represents a (vector-valued) bivariate function that is symmetric with respect to $(T_i, T^0_i)$.
According to Theorem \ref{lem:SENS}, we have
$$
\left(T_i, T^0_i \mid \mathbf{T}_{-i}, \mathbf{T}^0_{-i}\right) \stackrel{d}{=}\left(T^0_i, T_i \mid \mathbf{T}_{-i}, \mathbf{T}^0_{-i}\right)
$$
for $i \in \mathcal{H}_0$. Applying \eqref{equ:lem2.1}, we have
\begin{equation}\label{equ:lem2.3}
    \left(T_i, T^0_i \mid \mathbf{W}_i, \mathbf{T}_{-i}, \mathbf{T}^0_{-i}\right) \stackrel{d}{=}\left(T^0_i, T_i \mid \mathbf{W}_i, \mathbf{T}_{-i}, \mathbf{T}^0_{-i}\right).
\end{equation}
As $g\left(t;\left\{T_i, T^0_i\right\},\left(\mathbf{T}_{-i}, \mathbf{T}^0_{-i}\right)\right)$ is nonrandom with respect to 
$
\sigma\left(\left\{T_i, T^0_i\right\},\left(\mathbf{T}_{-i}, \mathbf{T}^0_{-i}\right)\right) \subset \sigma\left(\mathbf{W}_i, \mathbf{T}_{-i}, \mathbf{T}^0_{-i}\right),
$ 
it follows from \eqref{equ:lem2.3} that
$$
\left(U_i, U^0_i \mid \mathbf{W}_i, \mathbf{T}_{-i}, \mathbf{T}^0_{-i}\right) \stackrel{d}{=}\left(U^0_i, U_i \mid \mathbf{W}_i, \mathbf{T}_{-i}, \mathbf{T}^0_{-i}\right), \quad \text { for } i \in \mathcal{H}_0 .
$$
Condition \eqref{pwex:scores} follows by integrating out $\left(\mathbf{T}_{-i},\mathbf{T}^0_{-i}\right)$ and $\left(T_i \vee T^0_i, T_i \wedge T^0_i\right)$:
$$
\left(U_i, U^0_i|\mathbf{U}_{-i}, \mathbf{U}^0_{-i}\right) \stackrel{d}{=}\left(U^0_i, U_i|\mathbf{U}_{-i}, \mathbf{U}^0_{-i}\right), \quad \forall i \in \mathcal{H}_0,
$$
proving the desired result. 

\subsection{Proof of Proposition \ref{pro1}}

We first demonstrate that the proposed estimator \( \hat{f}_0 \) is symmetric. We consider two cases, which correspond to the two options in Algorithm \ref{alg:SENS}, respectively.

\textbf{Case I}: Suppose we set the option to ``JC'' then the estimator is for $f_0$ is defined in Equation \eqref{equ:JC}. As shown in Section \ref{subsec:JC} of the Supplement, we have
\[
\varphi_{2m}(t;(\mathbf{T}, \mathbf{T}^0)_{\operatorname{swap}(\mathcal{J})}) = \varphi_{2m}(t;(\mathbf{T}, \mathbf{T}^0)),\quad \forall \mathcal{J} \subset [m].
\]
According to Equation \eqref{equ:t,sig,mu}, we also have
\[
\hat{\mu}_0((\mathbf{T}, \mathbf{T}^0)_{\operatorname{swap}(\mathcal{J})}) = \hat{\mu}_0(\mathbf{T}, \mathbf{T}^0),\quad
\hat{\sigma}_0((\mathbf{T}, \mathbf{T}^0)_{\operatorname{swap}(\mathcal{J})}) = \hat{\sigma}_0(\mathbf{T}, \mathbf{T}^0),\quad \forall \mathcal{J} \subset [m].
\]
Since \(\hat{f}_0(t) = \phi_{\hat{\sigma}_0}(t - \hat{\mu}_0)\), it follows that
\[
\hat{f}_0(t; (\mathbf{T}, \mathbf{T}^0)_{\operatorname{swap}(\mathcal{J})}) = \hat{f}_0(t; (\mathbf{T}, \mathbf{T}^0)),\quad \forall \mathcal{J} \subset [m].
\]

\textbf{Case II}: If the option is set to ``KN'', then \( \tilde{\mathbf{T}}^0 \), defined in Equation \eqref{equ:camouflaged T_0}, is used to estimate \( \hat{f}_0 \). The use of \( \tilde{\mathbf{T}}^0 \) ensures that the pairwise exchangeability holds. Rewriting \( \tilde{\mathbf{T}}^0 \) as \( \tilde{\mathbf{T}}^0(\mathbf{T}, \mathbf{T}^0) \), we have:
$$
\tilde{\mathbf{T}}^0((\mathbf{T}, \mathbf{T}^0)_{\operatorname{swap}(\mathcal{J})}) = \tilde{\mathbf{T}}^0(\mathbf{T}, \mathbf{T}^0), \quad \forall \mathcal{J} \subset [m].
$$
It follows that the estimator \( \hat{f}_0 \) in Equation \eqref{f0-symm} satisfies:
\[
\hat{f}_0(t; (\mathbf{T}, \mathbf{T}^0)_{\operatorname{swap}(\mathcal{J})}) = \hat{f}_0(t; (\mathbf{T}, \mathbf{T}^0)), \quad \forall \mathcal{J} \subset [m].
\]

Next, considering the construction of \(\hat{f}_{mix}(t)\) in Equation \eqref{equ:hat_f_mix}, it is clear that:
\[
\hat{f}_{mix}(t; (\mathbf{T}, \mathbf{T}^0)_{\operatorname{swap}(\mathcal{J})}) = \hat{f}_{mix}(t; (\mathbf{T}, \mathbf{T}^0)),\quad \forall \mathcal{J} \subset [m].
\]

Since \(g(t)\) is constructed solely based on \(\hat{f}_0\) and \(\hat{f}_{mix}\), it follows that:
\[
g(t; (\mathbf{T}, \mathbf{T}^0)_{\operatorname{swap}(\mathcal{J})}) = g(t; (\mathbf{T}, \mathbf{T}^0)),\quad \forall \mathcal{J} \subset [m].
\]
The desired result follows by applying Proposition \ref{prop:score-class}.

\subsection{Proof of Proposition \ref{pro:e-SENS}}

Let $R=|\mathcal{R}|$. By the definition of $\tau$, we have
$
\frac{1+\sum_{j=1}^m \mathbb{I}\left\{G_j \leq -\tau\right\}}{R} \leq \alpha .
$
It follows that for $i \in \mathcal{R}$,
$$
e_i=\frac{m \mathbb{I}\left\{G_i \geq \tau\right\}}{1+\sum_{j=1}^m \mathbb{I}\left\{G_j \leq -\tau\right\}} \geq \frac{m}{\alpha R}.
$$
Therefore, $\hat{k}=\max \left\{i: e_{(i)} \geq \frac{m}{\alpha i}\right\} \geq R$, which implies that $i \in \mathcal{R}_{e b h}$.

Conversely, if $i \notin \mathcal{R}$ and $e_i=0$, then $i$ cannot be selected by the $e$-BH procedure and we must have $i\notin \mathcal{R}_{\text {ebh }}$.

Combining the two directions, we conclude that $\mathcal{R}=\mathcal{R}_{\text {ebh }}$.

\subsection{Proof of Theorem \ref{the:robust knockoff}}
We begin by stating two lemmas, both of which are taken from \cite{barber2020robust}. We will provide a proof for Lemma \ref{lem:knockoff1} in Section \ref{subsec:lem-proof} because our problem setup slightly differs from that of the model-X knockoffs, necessitating modifications to the notation and concepts. Conversely, the proof for Lemma \ref{lem:knockoff2} can be established by following the same arguments as given in \cite{barber2020robust}, and is therefore omitted.

\begin{lemma}\label{lem:knockoff1}
Consider \( \mathbf{G} = (G_i)_{i=1}^m \) defined in \eqref{eq:anti-sym}. Let $\mathbf{G}_{-i} = (G_1, \ldots, G_{i-1}, G_{i+1}, \ldots, G_m)$. 
Then the observed KL divergences, \( \widehat{\mathrm{KL}}_i \) [cf. Equation \eqref{equ:KL}], satisfy
\[
\mathbb{P}\left\{ G_i > 0, \widehat{\mathrm{KL}}_i \leq \epsilon \big| | G_i |, \mathbf{G}_{-i} \right\} \leq e^\epsilon \cdot \mathbb{P}\left\{ G_i < 0 \big| | G_i |, \mathbf{G}_{-i} \right\}, \quad \forall \epsilon \geq 0, \, i \in \mathcal{H}_0.
\]
\end{lemma}

\begin{lemma}\label{lem:knockoff2}
Define
$
\tau_i=\tau\left(\left(G_1, \ldots, G_{i-1},\left|G_i\right|, G_{i+1}, \ldots, G_m\right)\right)>0,
$
i.e. the threshold that we would obtain if $G_i$ were replaced with $\left|G_i\right|$, $i\in[m]$. For any $i, j \in [m]$, If $G_i \leq-\min \left\{\tau_i, \tau_j\right\}$ and $G_j \leq-\min \left\{\tau_i, \tau_j\right\}$, then $\tau_i=\tau_j$.
\end{lemma}

\noindent\textbf{Proof of Part (a) of the theorem.} For any $\epsilon \geq 0$ and $\lambda>0$, define
$
W_\epsilon(\lambda):=\frac{\sum_{i \in \mathcal{H}_0} \mathbb{I}\left\{G_i \geq \lambda, \widehat{\mathrm{K L}}_i \leq \epsilon\right\}}{1+\sum_{i \in \mathcal{H}_0} \mathbb{I}\left\{G_i \leq-\lambda\right\}} .
$
Then we have
$$
\begin{aligned}
\mathbb{E}\left[W_\epsilon(\tau)\right] & =\mathbb{E}\left[\frac{\sum_{i \in \mathcal{H}_0} \mathbb{I}\left\{G_i \geq \tau, \widehat{\mathrm{K L}}_i \leq \epsilon\right\}}{1+\sum_{i \in \mathcal{H}_0} \mathbb{I}\left\{G_i \leq-\tau\right\}}\right] \\
& {=}\sum_{i \in \mathcal{H}_0} \mathbb{E}\left[\frac{\mathbb{I}\left\{G_i \geq \tau_i, \widehat{\mathrm{K L}}_i \leq \epsilon\right\}}{1+\sum_{j \in \mathcal{H}_0, j \neq i} \mathbb{I}\left\{G_i \leq-\tau_i\right\}}\right]\\
& =\sum_{i \in \mathcal{H}_0} \mathbb{E}\left[\frac{\mathbb{I}\left\{G_i>0, \widehat{\mathrm{K L}}_i \leq \epsilon\right\} \cdot \mathbb{I}\left\{\left|G_i\right| \geq \tau_i\right\}}{1+\sum_{j \in \mathcal{H}_0, j \neq i} \mathbb{I}\left\{G_j \leq-\tau_i\right\}}\right] \\
& {=} \sum_{i \in \mathcal{H}_0} \mathbb{E}\left[\frac{\mathbb{P}\left\{G_i>0, \widehat{\mathrm{K L}}_i \leq \epsilon \big|| G_i \mid, \mathbf{G}_{-i}\right\} \cdot \mathbb{I}\left\{\left|G_i\right| \geq \tau_i\right\}}{1+\sum_{j \in \mathcal{H}_0, j \neq i} \mathbb{I}\left\{G_j \leq-\tau_i\right\}}\right]\\
& \leq e^\epsilon \cdot \sum_{i \in \mathcal{H}_0} \mathbb{E}\left[\frac{\mathbb{P}\left\{G_i<0\big|| G_i \mid, \mathbf{G}_{-i}\right\} \cdot \mathbb{I}\left\{\left|G_i\right| \geq \tau_i\right\}}{1+\sum_{j \in \mathcal{H}_0, j \neq i} \mathbb{I}\left\{G_j \leq-\tau_i\right\}}\right]. 
\end{aligned}
$$
Following the arguments in \cite{barber2020robust}, it can be shown that $\mathbb{E}\left[W_\epsilon(\tau)\right]\leq e^\epsilon$. Then for the generalized $e$-values defined by \eqref{equ:e-value}, we have 
$$
\begin{aligned}
\mathbb{E}\left[\sum_{i\in\mathcal{H}_0}e_i\right]&=\mathbb{E}\left[\frac{m\sum_{i\in\mathcal{H}_0}\mathbb{I}\{G_i\geq\tau\}}{1+\sum_{i=1}^m\mathbb{I}\{G_i\leq-\tau\}}\right]\\  
&=\mathbb{E}\left[\frac{m\sum_{i\in\mathcal{H}_0}\mathbb{I}\{G_i\geq\tau,\widehat{\mathrm{KL}}_i\leq \epsilon\}}{1+\sum_{i=1}^m\mathbb{I}\{G_i\leq-\tau\}}\right]+\mathbb{E}\left[\frac{m\sum_{i\in\mathcal{H}_0}\mathbb{I}\{G_i\geq\tau,\widehat{\mathrm{KL}}_i> \epsilon\}}{1+\sum_{i=1}^m\mathbb{I}\{G_i\leq-\tau\}}\right]\\
&\leq \mathbb{E}\left[\frac{m\sum_{i\in\mathcal{H}_0}\mathbb{I}\{G_i\geq\tau,\widehat{\mathrm{KL}}_i\leq \epsilon\}}{1+\sum_{i\mathcal{H}_0}\mathbb{I}\{G_i\leq-\tau\}}\right]+\mathbb{E}\left[\frac{m\sum_{i\in\mathcal{H}_0}\mathbb{I}\{\widehat{\mathrm{KL}}_i>\epsilon\}}{1+\sum_{i=1}^m\mathbb{I}\{G_i\leq-\tau\}}\right]\\
&\leq m \mathbb{E}\left[W_\epsilon(\tau)\right]+m\mathbb{E}\left[\sum_{i\in\mathcal{H}_0}\mathbb{I}\{\widehat{\mathrm{KL}}_i>\epsilon\}\right]\\
&\leq m \left[e^\epsilon+ \sum_{i\in\mathcal{H}_0}\mathbb{P}\left(\widehat{\mathrm{KL}}_i>\epsilon\right)\right].
\end{aligned}
$$

\noindent\textbf{Proof of Part (b) of the theorem.} 
Theorem 1 in \cite{ren2024derandomised} implies that Algorithm \ref{alg:SENS} is equivalent to the $e$-BH procedure \citep{wang2022false}. The FDR of the $e$-BH procedure based on generalized $e$-values $\{e_i\}_{i=1}^m$, which outputs the rejection set $\mathcal{R}$, satisfies the following inequality:
$$
\begin{aligned}
    \text{FDR}&=\mathbb{E}\left[\frac{|\mathcal{R} \cap \mathcal{H}_0|}{|\mathcal{R}| \vee 1}\right]=\mathbb{E}\left[\sum_{i \in \mathcal{H}_0} \frac{\mathbb{I}(i \in \mathcal{R})}{|\mathcal{R}| \vee 1}\right]\leq \mathbb{E}\left[\sum_{i \in \mathcal{H}_0} \frac{\mathbb{I}(i \in \mathcal{R}) \alpha e_i}{m}\right]\\ 
    &\leq \mathbb{E}\left[\sum_{i \in \mathcal{H}_0} \frac{\alpha e_i}{m}\right] \leq \alpha \left[e^\epsilon+ \sum_{i\in\mathcal{H}_0}\mathbb{P}\left(\widehat{\mathrm{KL}}_i>\epsilon\right)\right],
\end{aligned}
$$
proving the desired result. The first inequality holds because the $e$-BH procedure rejects hypotheses in the set \(\mathcal{R} = \{i \in [m] : e_i \geq e_{(\hat{k})}\}\), where 
$\hat{k} = \max \left\{i : \frac{i e_{(i)}}{m} \geq \frac{1}{\alpha}\right\}.$

\subsection{Proof of Lemma \ref{lem:knockoff1}}\label{subsec:lem-proof}

Our proof builds upon the approach in \cite{barber2020robust}, which addresses a regression setup involving response variables. While the underlying concepts and arguments are similar, we adapt their proof for the SSMT framework and present it here for completeness.

Without loss of generality, we can label the unordered feature pair \(\left\{U_i, U^0_i\right\}\) as \(U_i^{(0)}\) and \(U_i^{(1)}\), so that:
\begin{equation}\label{equ:lemma6.1}
\left\{\begin{array}{l}
\text { If } U_i=U_i^{(0)} \text { and } U^0_i=U_i^{(1)} \text {, then } G_i \geq 0 ; \\
\text { If } U_i=U_i^{(1)} \text { and } U^0_i=U_i^{(0)} \text {, then } G_i \leq 0 .
\end{array}\right.
\end{equation}

We can therefore write
\begin{eqnarray*}
\mathbb{P}\left\{G_i\right.  \left.>0, \widehat{\mathrm{KL}}_i \leq \epsilon\big|| G_i \mid, \mathbf{G}_{-i}\right\} & = & \mathbb{E}\left[\mathbb{P}\left\{G_i>0, \widehat{\mathrm{KL}}_i \leq \epsilon \mid U_i^{(0)}, U_i^{(1)}, \mathbf{U}_{-i}, \mathbf{U}^0_{-i}\right\}\big|| G_i \mid, \mathbf{G}_{-i}\right];\\
\mathbb{P}\left\{G_i<0\big|| G_i \mid, \mathbf{G}_{-i}\right\} & = & \mathbb{E}\left[\mathbb{P}\left\{G_i<0 \mid U_i^{(0)}, U_i^{(1)}, \mathbf{U}_{-i}, \mathbf{U}^0_{-i}\right\}\big| | G_i \mid, \mathbf{G}_{-i}\right].
\end{eqnarray*}
Therefore, it will be sufficient to prove that
$$
\mathbb{P}\left\{G_i>0, \widehat{\mathrm{KL}}_i\right. \left.\leq \epsilon \mid U_i^{(0)}, U_i^{(1)}, \mathbf{U}_{-i}, \mathbf{U}^0_{-i}\right\} \leq e^\epsilon \cdot \mathbb{P}\left\{G_i<0 \mid U_i^{(0)}, U_i^{(1)}, \mathbf{U}_{-i}, \widetilde{\mathbf{U}}_{-i}\right\}. 
$$

The bound holds trivially if $\left|G_i\right|=0$. Hence from this point on we assume that $\left|G_i\right|>0$. By the definition of $U_i^{(0)}$ and $U_i^{(1)}$ [cf. \eqref{equ:lemma6.1}], we have
$$
\begin{aligned}
& \frac{\mathbb{P}\left\{G_i>0 \mid U_i^{(0)}, U_i^{(1)}, \mathbf{U}_{-i}, \mathbf{U}^0_{-i}\right\}}{\mathbb{P}\left\{G_i<0 \mid U_i^{(0)}, U_i^{(1)}, \mathbf{U}_{-i}, \mathbf{U}^0_{-i}\right\}} \\
&= \frac{\mathbb{P}\left\{\left(U_i, U^0_i\right)=\left(U_i^{(0)}, U_i^{(1)}\right) \mid U_i^{(0)}, U_i^{(1)}, \mathbf{U}_{-i}, \mathbf{U}^0_{-i}\right\}}{\mathbb{P}\left\{\left(U_i, U^0_i\right)=\left(U_i^{(1)}, U_i^{(0)}\right) \mid U_i^{(0)}, U_i^{(1)}, \mathbf{U}_{-i}, \mathbf{U}^0_{-i}\right\}}\\
&= \frac{p^{U,U^0}_i(U_i^{(0)}, U_i^{(1)})}{p^{U^0,U}_i(U_i^{(0)}, U_i^{(1)})}=: e^{\rho_i}.
\end{aligned}
$$
According to the definition of $p^{U,U^0}_i(u,v)$ and $p^{U^0,U}_i(u,v)$, we have $p^{U,U^0}_i(u,v)=p^{U^0,U}_i(v,u)$. Consequently, combining Equation \eqref{equ:lemma6.1}, we have \(\widehat{\mathrm{KL}}_i = \rho_i\) if \(G_i > 0\), and \(\widehat{\mathrm{KL}}_i = -\rho_i\) if \(G_i < 0\). Therefore,
$$
\begin{aligned}
&\mathbb{P}\left\{G_i>0, \widehat{\mathrm{K L}}_i \leq \epsilon \mid U_i^{(0)}, U_i^{(1)}, \mathbf{U}_{-i}, \mathbf{U}^0_{-i}\right\}\\
&=\mathbb{I}\left\{\rho_i \leq \epsilon\right\} \cdot \mathbb{P}\left\{G_i>0 \mid U_i^{(0)}, U_i^{(1)}, \mathbf{U}_{-i}, \mathbf{U}^0_{-i}\right\} \\
&=\mathbb{I}\left\{\rho_i \leq \epsilon\right\} \cdot e^{\rho_i} \cdot \mathbb{P}\left\{G_i<0 \mid U_i^{(0)}, U_i^{(1)}, \mathbf{U}_{-i}, \mathbf{U}^0_{-i}\right\}.
\end{aligned}
$$
The next-to-last step holds since $\rho_i$ is a function of $U_i^{(0)}, U_i^{(1)}, \mathbf{U}_{-i}, \mathbf{U}^0_{-i}$. The desired result follows by noting that $\mathbb{I}\left\{\rho_i \leq \epsilon\right\} \cdot e^{\rho_i} \leq e^\epsilon$.

\subsection{Proof of Proposition \ref{pro3}}\label{subsec:proof of pro3}

Under model \eqref{equ:theoretical-working-model}, we can write $\operatorname{lfdr}(t)=(1-\pi)r_m(t)/(2-r_m(t))$. Since the transformation $x \mapsto (1-\pi)x /(2-x)$ is monotone, we have
$$
\mathbb{I}\left\{r_m\left(T_i\right) \leq \lambda^R\right\} \equiv \mathbb{I}\left\{\operatorname{lfdr}\left(T_i\right) \leq \frac{(1-\pi)\lambda^R}{2-\lambda^R}\right\}, \quad \forall i \in[m].
$$

Denote $\text{lfdr}_i=\operatorname{lfdr}\left(T_i\right)$ and $\lambda^\text{OR}=(1-\pi)\lambda^R /\left(2-\lambda^R\right)$.
We have
$$
\begin{aligned}
\mathbb{E}\left[\sum_{i \in \mathcal{H}_0} \mathbb{I}\left\{\text {lfdr}_i\leq \lambda\right\}\right] & =\sum_{i=1}^m \mathbb{E}\left[\mathbb{I}\left\{\text{lfdr}_i\leq \lambda\right\} \mathbb{I}\left\{H_{0,i} \text { is true}\right\}\right] \\
& =\sum_{i=1}^m \mathbb{E}\left\{\mathbb{E}\left[\mathbb{I}\left\{\text{lfdr}_i\leq \lambda\right\} \mathbb{I}\left\{H_{0,i}\text { is true}\right\} \mid \mathbf{T}\right]\right\} \\
& =\mathbb{E}\left\{\sum_{i=1}^m \mathbb{I}\left\{\text{lfdr}_i\leq \lambda\right\} \text{lfdr}_i\right\} .
\end{aligned}
$$

Since the decision $\boldsymbol{\delta}^\text{OR} =\{\mathbb(r_m\left(T_i\right) \leq \lambda^R): i \in [m]\}=\{\mathbb{I}(\text{lfdr}_i\leq \lambda^\text{OR}) : i \in [m]\}$ controls $\mathrm{mFDR}$ at $\alpha$ exactly, we have
\begin{equation}
    \mathbb{E}\left\{\sum_{i=1}^m\left(\text{lfdr}_i-\alpha\right) \mathbb{I}\left\{\text{lfdr}_i\leq \lambda^\text{OR}\right\}\right\}=0.
\label{equ:pro3.1}
\end{equation}
It follows that $\alpha\leq \lambda^\text{OR}$, otherwise the summation in \eqref{equ:pro3.1} would  be negative. Define
$$
W(\lambda)=\frac{\sum_{i \in \mathcal{H}_0} \mathbb{I}\left\{\text{lfdr}_i\leq \lambda\right\}}{\sum_{i=1}^m \mathbb{I}\left\{\text{lfdr}_i\leq \lambda\right\}}
$$
We claim that $W(\lambda)$ is monotone in $\lambda$. Let $W\left(\lambda_j\right)=\alpha_j$ for $j=1,2$. By \eqref{equ:pro3.1}, we have
\begin{equation}
    \mathbb{E}\left[\sum_{i=1}^m\left(\text{lfdr}_i-\alpha_j\right) \mathbb{I}\left\{\text{lfdr}_i\leq \lambda_j\right\}\right]=0.
\label{equ:pro3.2}
\end{equation}

To prove that \(\alpha_1 \leq \alpha_2\) if \(\lambda_1 \leq \lambda_2\) by contradiction, we proceed as follows. If $\alpha_1>\alpha_2$ for $\lambda_1\leq \lambda_2$, then
$$
\begin{aligned}
& \left(\text{lfdr}_i-\alpha_2\right) \mathbb{I}\left(\text{lfdr}_i\leq \lambda_2\right) \\
= & \left(\text{lfdr}_i-\alpha_2\right) \mathbb{I}\left(\text{lfdr}_i\leq \lambda_1\right)+\left(\text{lfdr}_i-\alpha_2\right) \mathbb{I}\left(\lambda_1 \leq  \text{lfdr}_i\leq \lambda_2\right) \\
= & \left(\text{lfdr}_i-\alpha_1\right) \mathbb{I}\left(\text{lfdr}_i\leq \lambda_1\right)+\left(\alpha_1-\alpha_2\right) \mathbb{I}\left(\text{lfdr}_i\leq \lambda_2\right)+\left(\text{lfdr}_i-\alpha_1\right) \mathbb{I}\left(\lambda_1 \leq  \text{lfdr}_i\leq \lambda_2\right),
\end{aligned}
$$
where $\mathbb{E}\left[\left(\alpha_1-\alpha_2\right) \mathbb{I}\left(\text{lfdr}_i\leq \lambda_2\right)+\left(\text{lfdr}_i-\alpha_1\right) \mathbb{I}\left(\lambda_1 \leq  \text{lfdr}_i\leq \lambda_2\right)\right]>0$. Then 
$$
\mathbb{E}\left[\sum_{i=1}^m\left(\text{lfdr}_i-\alpha_2\right) \mathbb{I}\left(\text{lfdr}_i\leq \lambda_2\right)\right]>0,
$$ 
contradicting \eqref{equ:pro3.2}.

Consider a decicion rule $\boldsymbol{\delta}_{v} = \{\mathbb{I}(v_i \leq  \lambda^{\prime}) : i \in [m]\}$ at mFDR level $\alpha$. Using a similar argument of \eqref{equ:pro3.1}, we have $\mathbb{E}\left[\sum_{i=1}^m\left(\text{lfdr}_i-\alpha\right) \mathbb{I}\left\{v_i\leq \lambda^{\prime}\right\}\right] \leq  0$, and
\begin{equation}
    \mathbb{E}\left[\sum_{i=1}^m\left(\text{lfdr}_i-\alpha\right)\left(\mathbb{I}\left\{\text{lfdr}_i\leq \lambda^\text{OR}\right\}-\mathbb{I}\left\{v_i\leq \lambda^{\prime}\right\}\right)\right] \geq 0,
    \label{equ:pro3.3}
\end{equation}
where $\mathbb{I}\left\{\text{lfdr}_i\leq \lambda^\text{OR}\right\}=\mathbb{I}\left\{\frac{\text{lfdr}_i-\alpha}{1-\text{lfdr}_i}\leq \kappa^\text{OR}\right\}$ and $\kappa^\text{OR}=\frac{\lambda^\text{OR}-\alpha}{1-\lambda^\text{OR}}$. 
It follows that
\begin{eqnarray*}
\text{lfdr}_i-\alpha-\kappa^\text{OR}\left(1-\text{lfdr}_i\right)\leq 0, \quad \text { if } \mathbb{I}\left\{\text{lfdr}_i\leq \lambda^\text{OR}\right\}>\mathbb{I}\left\{v_i\leq \lambda^{\prime}\right\},\\
\text{lfdr}_i-\alpha-\kappa^\text{OR}\left(1-\text{lfdr}_i\right) \geq 0, \quad \text { if } \quad \mathbb{I}\left\{\text{lfdr}_i\leq \lambda^\text{OR}\right\}\leq \mathbb{I}\left\{v_i\leq \lambda^{\prime}\right\}.
\end{eqnarray*}

We conclude that the following inequality holds for all $i \in[m]$,
$$
\left[\mathbb{I}\left\{\text{lfdr}_i\leq \lambda^\text{OR}\right\}-\mathbb{I}\left\{v_i\leq \lambda^{\prime}\right\}\right]\left[\text{lfdr}_i-\alpha-\kappa^\text{OR}\left(1-\text{lfdr}_i\right)\right] \leq  0 .
$$
Summing over $i$ and taking expectation, we have 
\begin{equation}
    \mathbb{E}\left\{\sum_{i=1}^m\left(\mathbb{I}\left\{\text{lfdr}_i\leq \lambda^\text{OR}\right\}-\mathbb{I}\left\{v_i\leq \lambda^{\prime}\right\}\right)\left[\text{lfdr}_i-\alpha-\kappa^\text{OR}\left(1-\text{lfdr}_i\right)\right]\right\} \leq  0.
    \label{equ:pro3.4}
\end{equation}
Combining \eqref{equ:pro3.3} and \eqref{equ:pro3.4}, we have
$$
\begin{aligned}
& \kappa^\text{OR} \mathbb{E}\left\{\sum_{i=1}^m\left(\mathbb{I}\left\{\text{lfdr}_i\leq \lambda^\text{OR}\right\}-\mathbb{I}\left\{v_i\leq \lambda^{\prime}\right\}\right)\left(1-\text{lfdr}_i\right)\right\} \\
& \geq \mathbb{E}\left\{\sum_{i=1}^m\left(\mathbb{I}\left\{\text{lfdr}_i\leq \lambda^\text{OR}\right\}-\mathbb{I}\left\{v_i\leq \lambda^{\prime}\right\}\right)\left(\text{lfdr}_i-\alpha\right)\right\} \geq 0 .
\end{aligned}
$$
Finally, note that $\kappa^\text{OR}>0$ and the expected number of true positives for $(\delta_i)_{i=1}^m$ is given by $\mathbb{E}\left\{\sum_{i=1}^m \delta_i\left(1-\text{lfdr}_i\right)\right\}$. The proof is complete by noting that
$$
\mathbb{E}\left\{\sum_{i=1}^m \mathbb{I}\left\{\text{lfdr}_i\leq \lambda^\text{OR}\right\}\left(1-\text{lfdr}_i\right)\right\} \geq \mathbb{E}\left\{\sum_{i=1}^m \mathbb{I}\left\{v_i\leq \lambda^{\prime}\right\}\left(1-\text{lfdr}_i\right)\right\}.
$$

\subsection{Proof of Theorem \ref{the:Asymptotic optimality}}

Let $\theta_i=\mathbb{I}\{H_{0,i} \text{ is false}\}$ denote a Bernoulli variable with success probability $\pi$. To facilitate the subsequent analysis, we rewrite the working model presented in equation \eqref{equ:theoretical-working-model}, which coincides with the true data-generating model, as follows:
\begin{equation}
    \begin{aligned}
        &T_i\mid \theta_i\stackrel{ind.}{\sim}(1-\theta_i)f_0+\theta_i f_{1m},\quad T^0_i\stackrel{i.i.d.}{\sim}f_0,\\
         &\theta_i\stackrel{i.i.d.}{\sim}\text{Bernoulli}(\pi),\quad i\in[m],\\
         &(T_i, T_i^0)\text{ are mutually independent for } i \in [m].
    \end{aligned}
    \label{equ:rewritten theoretical working model}
    \end{equation}
Consider the lfdr given by 
\[
\operatorname{lfdr}(t) = \frac{(1 - \pi) f_0(t)}{f_m(t)} = \frac{(1 - \pi) r_m(t)}{2 - r_m(t)}.
\]
Since \(\pi\) is a constant for all \(i \in [m]\) and only the ranking is relevant, we instead use 
$
L_i = \frac{\operatorname{lfdr}(T_i)}{1 - \pi} 
$
in our theoretical analysis. We denote the corresponding calibration statistic as 
$
L^0_i = \frac{\operatorname{lfdr}(T^0_i)}{1 - \pi}.
$
The estimated values for \(L_i\) and \(L^0_i\) are represented as 
$
\hat{L}_i = \frac{U_i}{2 - U_i}$ and $\hat{L}^0_i = \frac{U^0_i}{2 - U^0_i}, 
$
respectively. 

The SENS algorithm employs the decision $\mathbb{I}(U_i \leq U_i^0 \wedge \tau^\prime)$, where $\tau^\prime$ is defined in Equation \eqref{equ:tau^prime}. As shown in Lemma \ref{lem:G=U}, the decision rule is equivalent to the rule $\mathbb{I}(\gamma(U_i, U^0_i) \geq \tau)$, with $\tau$ defined in Equation \eqref{equ:tau}.

Let $\hat{\mathbf{L}}=\{\hat L_i:i\in[m]\}$ and $\hat{\mathbf{L}}^0=\{\hat L^0_i:i\in[m]\}$. 
Denote $\tau^{\prime\prime}=(1-\pi)\tau^\prime/(2-\tau^\prime)$. Since the transformation $x \mapsto (1-\pi)x /(2-x)$ is monotone, we have 
\begin{equation}
\tau^{\prime\prime}=\sup \left\{\lambda \in \hat{\mathbf{L}} \cup \hat{\mathbf{L}}^0: \frac{1+\sum_{j=1}^m \mathbb{I}\left\{\hat L^0_i \leq \hat L_i\wedge \lambda\right\}}{\left[\sum_{j=1}^m \mathbb{I}\left\{\hat L_i \leq \hat L^0_i \wedge \lambda\right\}\right] \vee 1} \leq \alpha\right\}
\label{equ:tau^prime prime}
\end{equation}
and $\mathbb{I}(U_i\leq U_i^0 \wedge \tau^\prime)=\mathbb{I}(\hat{L}_i\leq \hat{L}_i^0 \wedge \tau^{\prime\prime})$. The proof is similar to that of Lemma \ref{lem:G=U} in Section \ref{sec:G=U} and hence omitted. Next, denote 
\begin{equation}\label{delta-dd}
\boldsymbol{\delta}^\text{DD}=\{\mathbb{I}(\hat{L}_i\leq \hat{L}_i^0 \wedge \tau^{\prime\prime}):i\in[m]\},
\end{equation}
where $\tau^{\prime\prime}$ is calculated via \eqref{equ:tau^prime prime}. Then the SENS Algorithm is equivalent to $\boldsymbol{\delta}^\text{DD}$. Our goal is to show that $\text{ETP}_{\boldsymbol{\delta}^\text{DD}}/\text{ETP}_{\boldsymbol{\delta}^\text{OR}}=1+o(1)$. 

\medskip

\textbf{Summary of Notations.}

\begin{itemize}
 \item Generic random variables:
 \begin{itemize}
    \item $L$: a generic member from $\{L_i:i\in[m]\}$.
    \item $L^0$: a generic member from $\{L^0_i:i\in[m]\}$. 
    \item $\theta$: a generic member from $\{\theta_i:i\in[m]\}$. 
 \end{itemize}
 
  \item Different $Q(\lambda)$'s:
  \begin{itemize}
    \item $Q^\text{OR}(\lambda)=\mathbb{E}[(1-\theta-\alpha)\mathbb{I}(L\leq \lambda)]$;
    \item $Q^*(\lambda)=\mathbb{E}[\theta\mathbb{I}(L^0\leq L\wedge \lambda)+(1-\theta-\alpha)\mathbb{I}(L\leq L^0\wedge \lambda)]$;
    \item $Q(\lambda)=m^{-1}+m^{-1}\sum_{i}\mathbb{I}(L^0_i\leq L_i\wedge \lambda)-m^{-1}\sum_{i}\alpha\mathbb{I}(L_i\leq L^0_i\wedge \lambda)$;
    \item $\hat{Q}(\lambda)=m^{-1}+m^{-1}\sum_{i}\mathbb{I}(\hat{L}^0_i\leq \hat{L}_i\wedge \lambda)-m^{-1}\sum_{i}\alpha\mathbb{I}(\hat{L}_i\leq \hat{L}^0_i\wedge \lambda)$.
   
    \item Denote $(L^*_1,\ldots,L^*_{2m})=(\hat{L}_1,\ldots,\hat{L}_m,\hat{L}^0_1,\ldots,\hat{L}^0_m)$. Let $L^*_{(1)}\leq \cdots \leq L^*_{(2m)}$ denote the ordered values of $(L^*_1,\ldots,L^*_{2m})$. For $L^*_{(k)}<\lambda<L^*_{(k+1)}$, define 
$$
\hat{Q}_C(\lambda)=\frac{\lambda-L^*_{(k)}}{L^*_{(k+1)}-L^*_{(k)}}\hat{Q}_{k+1}+\frac{L^*_{(k+1)}-\lambda}{L^*_{(k+1)}-L^*_{(k)}} \hat{Q}_{k},
$$
where $\hat{Q}_k=\hat{Q}(L^*_{(k)})$. Note that $\hat{Q}_C(\lambda)$ is continuous.
   \end{itemize}

    \item Different decision rules:
   \begin{itemize}    
     \item $\boldsymbol{\delta}^\text{OR}=\{\mathbb{I}(L_i\leq \tau^\text{OR}):i\in [m]\}$, where $\tau^\text{OR}=\text{sup}\{\lambda\in(0,1):Q^\text{OR}(\lambda)\leq 0\}$. It is another form of $\boldsymbol{\delta}^\text{OR}$ according to the proof of Proposition \ref{pro3} in Section \ref{subsec:proof of pro3}.
     
   \item    $\boldsymbol{\delta}^*=\{\mathbb{I}(L_i\leq L^0_i\wedge \tau^*):i\in [m]\}$, where $\tau^*=\text{sup}\{\lambda\in(0,1):Q^*(\lambda)\leq 0\}$.
   
   \item $\boldsymbol{\delta}^\text{DD}=\{\mathbb{I}(\hat{L}_i\leq \hat{L}^0_i\wedge \tau^\text{DD}):i\in [m]\}$, where $\tau^\text{DD}=\text{sup}\{\lambda\in(0,1):\hat{Q}_C(\lambda)\leq 0\}$. It can be shown that this decision rule is identical to the decision rule defined in \eqref{delta-dd} and therefore we have utilized the same notation  $\boldsymbol{\delta}^\text{DD}$.   
   \item   $\boldsymbol{\delta}^\text{M}=\{\mathbb{I}(\hat{L}_i\leq \hat{L}^0_i\wedge \tau^*):i\in [m]\}$.
   
   \end{itemize} 
    
\end{itemize}

We first state two lemmas, which are proved in Sections \ref{sec:compatible} and \ref{sec:proof-eta}, respectively.

\begin{lemma}\label{lem:compatible}
    Consider model \eqref{equ:rewritten theoretical working model} and suppose Assumption \ref{ass:optimality} holds. Then we have
  \begin{equation*}  
 \mbox{$\hat{L}_i-L_i\xrightarrow{p}0$ and $\hat{L}^0_i-L^0_i\xrightarrow{p}0$.}
   \end{equation*}
\end{lemma}
\begin{lemma}\label{lem:eta}
    Let $\eta_i=\mathbb{I}(L^0_i\leq L_i)\mathbb{I}(L^0_i\leq \lambda)-\alpha\mathbb{I}(L_i\leq L^0_i)\mathbb{I}(L_i\leq \lambda)$ and $\hat{\eta}_i=\mathbb{I}(\hat{L}^0_i\leq \hat{L}_i)\mathbb{I}(\hat{L}^0_i\leq \lambda)-\alpha\mathbb{I}(\hat{L}_i\leq \hat{L}^0_i)\mathbb{I}(\hat{L}_i\leq \lambda)$. Then $\mathbb{E}(\hat{\eta}_i-\eta_i)^2=o(1)$.
\end{lemma}

\textbf{Proof of the theorem.} 

We first prove the following two results in turn: (i) $\hat{Q}_C(\lambda)-Q^*(\lambda)\stackrel{p}{\rightarrow}0$ and (ii) $\mathbb{P}(\tau^\text{DD}<\tau^*)=o(1)$.

\noindent\emph{Proof of Claim (i).} We first decompose $\hat{Q}_C(\lambda)-Q^*(\lambda)$ as
\begin{equation}\label{hatQ-Qstar}
\hat{Q}_C(\lambda)-Q^*(\lambda)=\{\hat{Q}_C(\lambda)-\hat Q(\lambda)\}+\{\hat{Q}(\lambda)-Q(\lambda)\}+\{{Q}(\lambda)-Q^*(\lambda)\}. 
\end{equation}

By the definitions of $\hat{Q}_C(\lambda)$ and $\hat{Q}(\lambda)$, we have 
\begin{equation}\label{Q1}
|\hat{Q}_C(\lambda)-\hat{Q}(\lambda)| \leq (2m)^{-1}.
\end{equation} 

Next, using the pairwise exchangeability of $L_i$ and $L^0_i$ for $i\in\mathcal H_0$, we have
\begin{eqnarray*}
&& \mathbb{E}[\theta_i\mathbb{I}(L^0_i\leq L_i\wedge \lambda)+(1-\theta_i)\mathbb{I}(L^0_i\leq L_i\wedge \lambda)-\alpha\mathbb{I}(L_i\leq L^0_i\wedge \lambda)] \\ & = & \mathbb{E}[\theta\mathbb{I}(L^0\leq L\wedge \lambda)+(1-\theta-\alpha)\mathbb{I}(L\leq L^0\wedge \lambda)]
~=~ Q^*(\lambda). 
\end{eqnarray*}
According to Khinchin's Law of Large Numbers, we have 
\begin{equation}\label{Q2}
Q(\lambda)-Q^*(\lambda)\stackrel{p}{\rightarrow}0.
\end{equation}
By Lemma \ref{lem:eta} and Cauchy-Schwartz inequality, $\mathbb{E}\{(\hat{\eta}_i-\eta_i)(\hat{\eta}_j-\eta_j)\}=o(1)$. Let $S_m=\sum_i(\hat{\eta}_i-\eta_i)$. It follows that
\begin{equation*}
\operatorname{Var}\left(m^{-1} S_m\right) \leq m^{-2} \sum_{i=1}^m \mathbb{E}\left\{\left(\hat{\eta}_i-\eta_i\right)^2\right\}+O\left(m^{-2} \sum_{i, j: i \neq j} \mathbb{E}\left\{\left(\hat{\eta}_i-\eta_i\right)\left(\hat{\eta}_j-\eta_j\right)\right\}\right)=o(1).
\end{equation*}
According to Lemma \ref{lem:compatible}, we have $\mathbb{E}\left(m^{-1} S_m\right) \rightarrow 0$. Next, applying Chebyshev's inequality, we obtain $m^{-1} S_m=$ $\hat{Q}(\lambda)-Q(\lambda) \xrightarrow{p} 0$. We claim that 
\begin{equation}\label{Q3}
\hat{Q}_C(\lambda)-\hat{Q}(\lambda) \xrightarrow{p} 0.
\end{equation}
Based on the decomposition  \eqref{hatQ-Qstar}, it follows from \eqref{Q1}, \eqref{Q2}, and \eqref{Q3} that 
\[
\hat{Q}_C(\lambda) - Q^*(\lambda) \stackrel{p}{\rightarrow} 0,
\]
thus Claim (i) is proven.

\bigskip

\noindent\emph{Proof of Claim (ii).}  According to the definitions of $\tau^*$ and $\tau^\text{DD}$ we have 
\begin{equation}\label{event}
    \mathbb{P}(\tau^\text{DD}<\tau^*)\leq \mathbb{P}(\hat{Q}_C(\tau^*)>0).
\end{equation}

The previous inequality holds because \(\tau^{DD}\) is defined as the largest value for which \(\hat{Q}_C(t) \leq 0\). If the event \(\tau^*>\tau^{DD}\) occurs, then the event \(\hat{Q}_C(\tau^*) > 0\) must also occur, thereby proving the inequality in \eqref{event}. We emphasize that \(\tau^*\) is a non-random constant. Consequently, based on Claim (i), we have 
\[
\hat{Q}_C(\tau^*) - Q^*(\tau^*) \stackrel{p}{\rightarrow} 0.
\]
Furthermore, given that \(Q^*(\tau^*) \leq 0\), it follows that \(\mathbb{P}(\hat{Q}_C(\tau^*) > 0) = o(1)\), which proves Claim (ii).

Our target power is $\operatorname{ETP}:=\mathbb{E} [\sum_i\theta_i \delta_i]$. Recall that our goal is to show that 
$$
\text{ETP}_{\boldsymbol{\delta}^\text{DD}}/\text{ETP}_{\boldsymbol{\delta}^\text{OR}}=1+o(1).
$$ 
Consider the following decomposition: 
\begin{equation*}
\begin{aligned}
& m^{-1} \operatorname{Regret}\left(\boldsymbol{\delta}^\text{OR}, \boldsymbol{\delta}^\text{DD}\right)=m^{-1}\left\{\sum_i\theta_i \delta_i^\text{OR}-\sum_i\theta_i \delta_i^\text{DD}\right\}\\
&=m^{-1}\left\{\sum_i\theta_i \delta_i^\text{OR}-\sum_i\theta_i \delta_i^*\right\}+m^{-1}\left\{\sum_i\theta_i \delta_i^*-\sum_i\theta_i \delta_i^\text{M}\right\}+m^{-1}\left\{\sum_i\theta_i \delta_i^\text{M}-\sum_i\theta_i \delta_i^\text{DD}\right\}\\
&\coloneqq m^{-1}\operatorname{Regret}\left(\boldsymbol{\delta}^\text{OR}, \boldsymbol{\delta}^*\right)+m^{-1}\operatorname{Regret}\left(\boldsymbol{\delta}^*, \boldsymbol{\delta}^\text{M}\right)+m^{-1}\operatorname{Regret}\left(\boldsymbol{\delta}^\text{M}, \boldsymbol{\delta}^\text{DD}\right).
\end{aligned}
\end{equation*}
Next, we shall show in turn that the expectations of all three sums on the RHS are vanishingly small.

Using Assumption \ref{ass:optimality}, we can show that
\begin{equation*}
\begin{aligned}
    &\mathbb{P}(L_i > L^0_i \mid \theta_i = 1)=\mathbb{P}\left(r_m(T_i)>r_m(T^0_i)\big|H_{0,i}\text{ is false}\right)\\
    &\stackrel{*}{=}\mathbb{P}\left(\frac{r_m(T_i)}{2 - (2 - \pi) r_m(T_i)}>\frac{r_m(T^0_i)}{2 - (2 - \pi) r_m(T^0_i)}\bigg|H_{0,i}\text{ is false}\right)\\
    & \stackrel{**}{=} \mathbb{P}\left(\frac{f_0(T_i)}{f_{1m}(T_i)} > \frac{f_0(T^0_i)}{f_{1m}(T^0_i)} \bigg| H_{0,i}\text{ is false}\right)\\
    &=\mathbb{P}\left(\frac{\phi_{\sigma_0}(T_i-\mu_0)}{\phi_{\sigma_0}(T_i-\mu_m)} > \frac{\phi_{\sigma_0}(T^0_i-\mu_0)}{\phi_{\sigma_0}(T^0_i-\mu_m)} \bigg| H_{0,i}\text{ is false}\right)\\
    &=\mathbb{P}\left((\mu_m-\mu_0) T_i < (\mu_m-\mu_0) T^0_i \mid H_{0,i}\text{ is false}\right)\\
    &=o(1).
\end{aligned}
\end{equation*}
Here, the equality $\stackrel{*}{=}$ follows from the facts that (i) the transformation 
$$x \mapsto \frac{x}{2 - (2 - \pi) x}$$ is monotone for $x<\frac 2{2-\pi}$, and that (ii) 
$$r_m(t) = \frac{2f_0(t)}{f_0(t) + f_m(t)}<\frac 2{2-\pi}.$$ The equality $\stackrel{**}{=}$ follows from the  facts that $\big(T_i~|~\mbox{$H_{0,i}$ is false}\big)\sim\mathcal{N}(\mu_m, \sigma_0^2)$, $T^0_i\sim \mathcal{N}(0, \sigma_0^2)$ for all $i\in[m]$, and that
$$
\frac{f_0(t)}{f_{1m}(t)} = \frac{r_m(t)}{2 - (2 - \pi) r_m(t)}.
$$

We begin by explaining the relationship between $\boldsymbol{\delta}^*$ and $\boldsymbol{\delta}^\text{OR}$. Recall from Proposition \ref{pro1} that $\boldsymbol{\delta}^\text{OR}$ achieves the highest $\text{ETP}$ among all decision rules at level $\alpha$. To proceed, we decompose $Q^*(\lambda)$ as follows:
\begin{equation*}
\begin{aligned}
    Q^*(\lambda) &= \mathbb{E}[\theta \mathbb{I}(L^0 \leq L \wedge \lambda) + (1 - \theta - \alpha) \mathbb{I}(L \leq L^0 \wedge \lambda)] \\
    &= \mathbb{E}[\theta \mathbb{I}(L^0 \leq L \wedge \lambda)] + \mathbb{E}[(1 - \theta - \alpha) \mathbb{I}(L \leq L^0 \wedge \lambda)] \\
    &= \mathbb{P}(\theta = 1) \mathbb{P}(L^0 \leq L \mid \theta = 1) \mathbb{E}[\mathbb{I}(L^0 \leq \lambda) \mid L^0 \leq L, \theta = 1] + \mathbb{E}[(1 - \theta - \alpha) \mathbb{I}(L \leq L^0 \wedge \lambda)] \\
    &= |o(1)| + \mathbb{E}[(1 - \theta - \alpha) \mathbb{I}(L \leq L^0 \wedge \lambda)] \\
    &= |o(1)|\mathbb{E}[\mathbb{I}(L \leq L^0 \wedge \lambda)] + \mathbb{E}[(1 - \theta - \alpha) \mathbb{I}(L \leq L^0 \wedge \lambda)] \\
    &= \mathbb{E}[(1 - \theta - \alpha + |o(1)|)\mathbb{I}(L \leq L^0 \wedge \lambda)].
\end{aligned}
\end{equation*}

From this, we observe that $\boldsymbol{\delta}^*$ is a decision rule that controls $\text{mFDR}$ below $\alpha - |o(1)| < \alpha$. Coupling this fact with the optimality of the oracle rule $\boldsymbol{\delta}^\text{OR}$, we conclude that  
\[
\mathbb{E}\left(m^{-1} \operatorname{Regret}\left(\boldsymbol{\delta}^\text{OR}, \boldsymbol{\delta}^*\right)\right) \geq 0.
\]

Furthermore, we have
\begin{equation}\label{*OR}
\begin{aligned}
&0\leq\mathbb{E}\left(m^{-1}\operatorname{Regret}\left(\boldsymbol{\delta}^\text{OR}, \boldsymbol{\delta}^*\right)\right)=m^{-1}\mathbb{E}\left\{\sum_i \theta_i \delta_i^\text{OR}-\sum_i \theta_i \delta_i^*\right\}\\
&=m^{-1}\mathbb{E}\left\{\sum_i \theta_i \mathbb{I}(L_i\leq \tau^\text{OR})-\sum_i \theta_i \mathbb{I}(L_i\leq L^0_i)\mathbb{I}(L_i\leq \tau^*)\right\}\\
&=\mathbb{E}\left\{\theta \mathbb{I}(L\leq \tau^\text{OR})- \theta \mathbb{I}(L\leq L^0)\mathbb{I}(L\leq \tau^*)\right\}\\
&=\mathbb{P}(\theta=1)\mathbb{E}\left\{\mathbb{I}(L\leq \tau^\text{OR})-\mathbb{I}(L\leq L^0)\mathbb{I}(L\leq \tau^*) \Big| \theta=1 \right\}\\
&=\mathbb{P}(\theta=1)\left[\mathbb{E}\left\{\mathbb{I}(L\leq \tau^\text{OR})-\mathbb{I}(L\leq L^0)\mathbb{I}(L\leq \tau^*) \Big| \theta=1,L\leq L^0 \right\}\mathbb{P}(L\leq L^0|\theta=1)+  \right.\\
&\left.\mathbb{E}\left\{\mathbb{I}(L\leq \tau^\text{OR})-\mathbb{I}(L\leq L^0)\mathbb{I}(L\leq \tau^*) \Big| \theta=1,L>L^0 \right\}\mathbb{P}(L>L^0|\theta=1)
\right]\\
&=\mathbb{P}(\theta=1)\mathbb{E}\left\{\mathbb{I}(L\leq \tau^\text{OR})-\mathbb{I}(L\leq \tau^*) \Big| \theta=1,L\leq L^0 \right\}\mathbb{P}(L\leq L^0|\theta=1)+o(1).
\end{aligned}    
\end{equation}

Next we prove that there exists $M>0$ such that $\tau^\text{OR}\leq \tau^*$ for $m>M$. We argue by contradiction. Suppose $\tau^\text{OR}> \tau^*$, then it follows that 
$$
Q^*(\tau^\text{OR})>0\geq Q^\text{OR}(\tau^\text{OR}).
$$
However, we have the following at the same time:
\begin{equation*}
\begin{aligned}
    &Q^*(\tau^\text{OR})-Q^\text{OR}(\tau^\text{OR})\\
    &=\mathbb{E}\left[\theta\mathbb{I}(L^0\leq L\wedge \tau^\text{OR})\right]+\mathbb{E}\left[(1-\theta-\alpha)\mathbb{I}(L\leq L^0\wedge \tau^\text{OR})-(1-\theta-\alpha)\mathbb{I}(L\leq \tau^\text{OR})\right]\\
    &=\mathbb{P}(\theta=1)\mathbb{E}\left[\theta\mathbb{I}(L^0\leq L)\mathbb{I}(L^0\leq \tau^\text{OR})\Big| \theta=1,L^0\leq L\right]\mathbb{P}(L^0\leq L\mid\theta=1)\\
    &-\mathbb{E}\left[(1-\theta-\alpha)\mathbb{I}(L> L^0)\mathbb{I}(L\leq \tau^\text{OR})\right]\\
    &=o(1)-\mathbb{P}(\theta=1)\mathbb{E}\left[(1-\theta-\alpha)\mathbb{I}(L> L^0)\mathbb{I}(L\leq \tau^\text{OR})\Big| \theta=1,L>L^0\right]\mathbb{P}(L>L^0\mid \theta=1)\\
    &-\mathbb{P}(\theta=0)\mathbb{E}\left[(1-\theta-\alpha)\mathbb{I}(L>L^0)\mathbb{I}(L\leq \tau^\text{OR}))\Big| \theta=0\right]\\
    &=o(1)-\mathbb{P}(\theta=0)\mathbb{E}\left[(1-\theta-\alpha)\mathbb{I}(L>L^0)\mathbb{I}(L\leq \tau^\text{OR}))\Big| \theta=0\right].
\end{aligned}
\end{equation*}

Therefore, there exists $M>0$ such that $ Q^*(\tau^\text{OR})-Q^\text{OR}(\tau^\text{OR})\leq 0$ for $m>M$, which contradicts our previous assumption. Therefore we must have $\tau^\text{OR}\leq \tau^*$. Back to \eqref{*OR}, we have $m^{-1}\operatorname{Regret}\left(\boldsymbol{\delta}^\text{OR}, \boldsymbol{\delta}^*\right)\leq o(1)$ for enough large m.

Since $\tau^*$ is a non-random constant, according to Lemma \ref{lem:compatible}, we have 
$
\mathbb{E}\left(m^{-1}\operatorname{Regret}\left(\boldsymbol{\delta}^*, \boldsymbol{\delta}^\text{M}\right)\right)=\mathbb{E}\left\{m^{-1}\sum_i\theta_i [\mathbb{I}(L_i\leq L^0_i\wedge \tau^*)-
\mathbb{I}(\hat{L}_i\leq \hat{L}^0_i\wedge \tau^*)]\right\}=o(1). 
$
Moreover, 
\begin{equation*}
\begin{aligned}
&\mathbb{E}\left(m^{-1}\operatorname{Regret}\left(\boldsymbol{\delta}^\text{M}, \boldsymbol{\delta}^\text{DD}\right)\right)=\mathbb{E}\left(m^{-1}\left\{\sum_i\theta_i \delta_i^\text{M}-\sum_i\theta_i \delta_i^\text{DD}\right\}\right)\\
&=\mathbb{E}\left(m^{-1}\left\{\sum_i\theta_i \mathbb{I}(\hat{L}_i\leq\hat{L}^0_i\wedge \tau^*)-\sum_i\theta_i\mathbb{I}(\hat{L}_i\leq\hat{L}^0_i\wedge \tau^\text{DD})\right\}\right)\\
&=\mathbb{E}\left(m^{-1}\left\{\sum_i\theta_i \mathbb{I}(\hat{L}_i\leq\hat{L}^0_i\wedge \tau^*)-\sum_i\theta_i\mathbb{I}(\hat{L}_i\leq\hat{L}^0_i\wedge \tau^\text{DD})\right\}\Big|\tau^*\leq \tau^\text{DD}\right)\mathbb{P}\left(\tau^*\leq \tau^\text{DD}\right)\\
&+\mathbb{E}\left(m^{-1}\left\{\sum_i\theta_i \mathbb{I}(\hat{L}_i\leq\hat{L}^0_i\wedge \tau^*-\sum_i\theta_i\mathbb{I}(\hat{L}_i\leq\hat{L}^0_i\wedge \tau^\text{DD})\right\}\Big|\tau^*>\tau^\text{DD}\right)\mathbb{P}\left(\tau^*> \tau^\text{DD}\right)\\
&\leq 0+\mathbb{P}\left(\tau^*> \tau^\text{DD}\right).
\end{aligned}
\end{equation*}

Combining the above results and invoking the optimality of $\boldsymbol{\delta}^\text{OR}$ again, we have 
$$
0\leq\mathbb{E}\left( m^{-1} \operatorname{Regret}\left(\boldsymbol{\delta}^\text{OR}, \boldsymbol{\delta}^\text{DD}\right)\right)\leq o(1).
$$
Therefore we have shown that 
$$
\mathbb{E}\left(m^{-1}\operatorname{Regret}\left(\boldsymbol{\delta}^\text{OR}, \boldsymbol{\delta}^\text{DD}\right)\right)= o(1). 
$$

Following from Theorem 1 in
\cite{tony2019covariate}, $\tau^\text{OR}\geq \alpha$. It follows that 
$$
\mathbb{E}\left\{m^{-1}\sum_{i}\theta_i\mathbb{I}(L_i\leq \tau^\text{OR})\right\}=E\left\{\theta\mathbb{I}(L\leq \tau^\text{OR})\right\}=\mathbb{P}(\theta=1)\mathbb{P}_{H_1}(L\leq \tau^\text{OR})
$$ is bounded below by a nonzero constant for any non-vanishing $\alpha$. Thus we have established that $\text{ETP}_{\boldsymbol{\delta}^\text{DD}}/\text{ETP}_{\boldsymbol{\delta}^\text{OR}}=1+o(1)$.

\subsection{Proof of Lemma \ref{lem:G=U}}\label{sec:G=U}
Since $\mathrm{exp}(-\cdot)$ is a monotonically strictly decreasing function, we have
$$
G_i=\mathrm{sign}(U_i^0-U_i)\left[\mathrm{exp}(-U_i)\vee \mathrm{exp}(-U^0_i)\right]=\mathbb{I}(U_i\leq U^0_i)\mathrm{exp}(-U_i)-\mathbb{I}(U^0_i\leq U_i)\mathrm{exp}(-U^0_i)
$$
Subsequently, we have \(\mathbb{I}\left(G_i \geq \lambda\right) = \mathbb{I}\left(\exp(-U_i)\geq \lambda \right)\mathbb{I}(U_i\leq U^0_i) = \mathbb{I}\left\{U_i \leq -\ln(\lambda)\right\} \mathbb{I}\left(U_i\leq U^0_i\right)\) and \(\mathbb{I}\left(G_i \leq -\lambda\right) = \mathbb{I}\left(\exp(-U^0_i)\geq \lambda \right)\mathbb{I}(U^0_i\leq U_i) = \mathbb{I}\left(U^0_i \leq -\ln(\lambda)\right) \mathbb{I}\left(U^0_i\leq U_i\right)\). Then we have
\[
\frac{1+\sum_{i \in [m]} \mathbb{I}\left(G_i \leq -\lambda\right)}{\left(\sum_{i \in [m]} \mathbb{I}\left(G_i \geq \lambda\right)\right) \vee 1} = \frac{1+\sum_{i \in [m]} \mathbb{I}\left(U^0_i \leq -\ln(\lambda)\right) \mathbb{I}\left(U^0_i \leq U_i\right)}{\left(\sum_{i \in [m]} \mathbb{I}\left\{U_i \leq -\ln(\lambda) \mathbb{I}\left(U_i \leq U^0_i\right)\right) \right) \vee 1}.
\]

Denote $\mathcal{U}=\{U_i\}_{i=1}^m$, $\mathcal{U}^0=\{U_i^0\}_{i=1}^m$, $\mathcal{E} = \left\{\exp(-U_i)\right\}_{i=1}^m$ and $\mathcal{E}^0 = \left\{\exp(-U^0_i)\right\}_{i=1}^m$. Since
$$
\begin{aligned}
\tau^\prime=&\sup \left\{\lambda\in \mathcal{U} \cup \mathcal{U}^0: \frac{1+\sum_{j=1}^m \mathbb{I}\left(U^0_i \leq  U_i\right)\mathbb{I}\left( U^0_i\leq\lambda\right)}{\left[\sum_{j=1}^m \mathbb{I}\left(U_i \leq  U^0_i\right)\mathbb{I}\left( U_i\leq\lambda\right)\right] \vee 1} \leq \alpha\right\}\\
&\stackrel{(1)}{=}-\ln\left(\inf \left\{\lambda^\prime\in \mathcal{E} \cup \mathcal{E}^0: \frac{1+\sum_{j=1}^m \mathbb{I}\left(U^0_i \leq  U_i\right)\mathbb{I}\left( U^0_i\leq -\ln(\lambda^\prime)\right)}{\left[\sum_{j=1}^m \mathbb{I}\left(U_i \leq  U^0_i\right)\mathbb{I}\left( U_i\leq-\ln(\lambda^\prime)\right)\right] \vee 1} \leq \alpha\right\}\right)\\
&=-\ln\left(\inf \left\{\lambda^\prime\in \mathcal{E} \cup \mathcal{E}^0: \frac{1+\sum_{j=1}^m \mathbb{I}\left(G_i\leq -\lambda^\prime\right)}{\left[\sum_{j=1}^m \mathbb{I}\left(G_i\geq \lambda^\prime\right)\right] \vee 1} \leq \alpha\right\}\right)\\
&\stackrel{(2)}{=}-\ln\left(\inf \left\{\lambda^\prime\in \{|G_i|\}_{i\in[m]}: \frac{1+\sum_{j=1}^m \mathbb{I}\left(G_i\leq -\lambda^\prime\right)}{\left[\sum_{j=1}^m \mathbb{I}\left(G_i\geq \lambda^\prime\right)\right] \vee 1} \leq \alpha\right\}\right)\\
&=-\ln\left(\tau\right),
\end{aligned}
$$
where \(\stackrel{(1)}{=}\) and \(\stackrel{(2)}{=}\) hold because \(\exp(-\cdot)\) is strictly decreasing and always positive.

Therefore, we have \(\mathbb{I}(G_i \geq \tau) = \mathbb{I}(U_i \leq -\ln(\tau)) \mathbb{I}(U_i \leq U^0_i) = \mathbb{I}(U_i \leq \tau^\prime) \mathbb{I}(U_i \leq U^0_i)\), completing the proof.

\subsection{Proof of Lemma \ref{lem:compatible}}\label{sec:compatible}
We begin by rewriting the sequence $(T_1,T^0_1,\cdots,T_m,T^0_m)$ as $\mathbf{D}=(D_1,D_2,\cdots,D_{2m-1},D_{2m})$. Under Model \eqref{equ:rewritten theoretical working model}, dependencies can only exist between $D_{2i-1}$ and $D_{2i}$ for $i\in[m]$. Following \cite{jin2007estimating}, this dependent structure of $\mathbf{D}$ corresponds to the strongly $\alpha$-mixing case. The strong mixing coefficients are defined as:
\[
\alpha(k) = \sup_{1 \leq t \leq 2m} \alpha\left(\sigma(D_s, s \leq t), \sigma(D_s, s \geq t+k)\right)
\]
where $\sigma(\cdot)$ denotes the $\sigma$-algebra generated by specified random variables, and for any two $\sigma$-algebras $\Sigma_1$ and $\Sigma_2$,
\[
\alpha(\Sigma_1, \Sigma_2) = \sup_{\substack{E_1 \in \Sigma_1 \\ E_2 \in \Sigma_2}} \left|P(E_1 \cap E_2) - P(E_1)P(E_2)\right|.
\]

For the data $\mathbf{D}$, we observe that $\alpha(1) \leq 2$ while $\alpha(k) = 0$ for $k \geq 2$. Consequently, these coefficients satisfy $\alpha(k) \leq B k^{-d}$ for some positive constants $B$ and $d$. Through Theorems 2.3 and 2.4 in \cite{jin2007estimating}, the estimators $\hat{\mu}_0$ and $\hat{\sigma}_0$ for parameters $\mu_0$ and $\sigma_0$ satisfy
\begin{equation*}
\mbox{$\hat{\mu}_0 \xrightarrow{p} \mu_0$ and $\hat{\sigma}_0 \xrightarrow{p} \sigma_0$. }
\end{equation*}
It follows that $\mathbb{E}\|\hat{f}_0(t) - f_0(t)\|^2 =\mathbb{E}\|\phi_{\hat{\sigma}_0}(t-\hat{\mu}_0) - \phi_{\sigma_0}(t-\mu_0)\|^2\rightarrow 0$. 

Additionally, since the dependency structure satisfies rapid $\alpha$-mixing decay, based on the work of \cite{wand1994kernel}, \cite{bosq2012nonparametric}, the kernel estimator for the density function in Equation \eqref{equ:hat_f_mix} satisfies 
$$
\mathbb{E}\|\hat{f}_{mix}(t) - (\phi_{\sigma_0}(t-\mu_0) + f_m(t))/2\|^2 \rightarrow 0.
$$

Following from Lemma A.1 in Sun and Cai (2007), we have $\mathbb{E}\|g(t)- r_m(t)\|^2 \rightarrow 0$, where $g(t)=\hat{f}_0(t)/\hat{f}_{mix}(t)=\phi_{\sigma_0}(t-\mu_0)/\hat{f}_{mix}(t)$ and 
$$
r_m(t) =\phi_{\sigma_0}(t-\mu_0) / [(\phi_{\sigma_0}(t-\mu_0) + f_m(t))/2]=f_0(t)/f_\text{mix}(t). 
$$
Furthermore, according to Lemma A.2 in \cite{sun2007oracle}, we can show that 
\begin{equation*}
\mbox{$g(T_i) - r_m(T_i)=U_i-r_m(T_i)\xrightarrow{p} 0$ and $g(T^0_i) - r_m(T^0_i)=U^0_i-r_m(T^0_i)\xrightarrow{p} 0$.}
\end{equation*}

Recall the definitions of $L_i$, $L^0_i$, $\hat{L}_i$ and $\hat{L}^0_i$, we have 
$$
\mbox{$\hat{L}_i - L_i \xrightarrow{p} 0$ and $\hat{L}_i^0 - L_i^0 \xrightarrow{p} 0$,}
$$
proving the desired result. 

\subsection{Proof of Lemma \ref{lem:eta}}\label{sec:proof-eta}
We first write  $(\hat{\eta}_i-\eta_i)^2$ as the following expression:
\begin{equation*}
\begin{aligned}
& \left\{\left[\mathbb{I}(\hat{L}^0_i\leq \hat{L}_i)\mathbb{I}(\hat{L}^0_i\leq \lambda)-\mathbb{I}(L^0_i\leq L_i)\mathbb{I}(L^0_i\leq \lambda)\right]-\alpha\left[\mathbb{I}(\hat{L}_i\leq \hat{L}^0_i)\mathbb{I}(\hat{L}_i\leq \lambda)-\mathbb{I}(L_i\leq L^0_i)\mathbb{I}(L_i\leq \lambda)\right]\right\}^2\\
    &=\left[\mathbb{I}(\hat{L}^0_i\leq \hat{L}_i)\mathbb{I}(\hat{L}^0_i\leq \lambda)-\mathbb{I}(L^0_i\leq L_i)\mathbb{I}(L^0_i\leq \lambda)\right]^2-2\alpha\left[\mathbb{I}(\hat{L}^0_i\leq \hat{L}_i)\mathbb{I}(\hat{L}^0_i\leq \lambda)-\mathbb{I}(L^0_i\leq L_i)\mathbb{I}(L^0_i\leq \lambda)\right]\\
    &+\alpha^2\left[\mathbb{I}(\hat{L}_i\leq \hat{L}^0_i)\mathbb{I}(\hat{L}_i\leq \lambda)-\mathbb{I}(L_i\leq L^0_i)\mathbb{I}(L_i\leq \lambda)\right]^2.
\end{aligned}
\end{equation*}

The first part of the RHS of the previous equation can be decomposed as:
\begin{equation*}
\begin{aligned}
    &\left[\mathbb{I}(\hat{L}^0_i\leq \hat{L}_i)\mathbb{I}(\hat{L}^0_i\leq \lambda)-\mathbb{I}(L^0_i\leq L_i)\mathbb{I}(L^0_i\leq \lambda)\right]^2\\
    &=\left[\mathbb{I}(\hat{L}^0_i\leq\hat{L}_i)-\mathbb{I}(L^0_i\leq L_i)\right]^2\mathbb{I}(\hat{L}^0_i\leq \lambda,L^0_i\leq \lambda)+\mathbb{I}(\hat{L}^0_i\leq\hat{L}_i)\mathbb{I}(\hat{L}^0_i\leq \lambda,L^0_i>\lambda)\\
    &+\mathbb{I}(L^0_i\leq L_i)\mathbb{I}(\hat{L}^0_i>\lambda,L^0_i \leq \lambda)
\end{aligned}
\end{equation*}
Denote the three sums on the RHS as $I,II, III$. respectively.

Note that both $\mathbb{I}(\hat{L}^0_i\leq\hat{L}_i)$ and $\mathbb{I}(L^0_i\leq L_i)$ are bounded above by 1, according to $\hat{L}_i-L_i\xrightarrow{p}0$ and $\hat{L}^0_i-L^0_i\xrightarrow{p}0$, it's obvious that 
$$\mathbb{E}\left[\mathbb{I}(\hat{L}_i\leq \hat{L}^0_i)-\mathbb{I}(L_i\leq L^0_i)\right]^2\rightarrow0.
$$ 
As a result, we have established that $\mathbb{E}(I)=o(1)$.

Next we consider 
\begin{equation*}
\begin{aligned}
P\left(\hat{L}^0_i \leq \lambda, L^0_i>\lambda\right) & \leq P\left(\hat{L}^0_i \leq \lambda, L^0_i \in(\lambda, \lambda+\epsilon)\right)+P\left(\hat{L}^0_i \leq \lambda, L^0_i \geq \lambda+\epsilon\right) \\
& \leq P\left\{L^0_i \in(\lambda, \lambda+\epsilon)\right\}+P\left(\left|\hat{L}^0_i-L^0_i\right|>\epsilon\right)
\end{aligned}    
\end{equation*}
The first term on the right hand is vanishingly small as $\epsilon \rightarrow 0$ because $\hat{L}^0_i$ is a continuous random variable. The second term converges to 0. Hence we conclude that $\mathbb{E}(II)=o(1)$. 

In a similar fashion, we can show that $\mathbb{E}(III)=o(1)$. It follows that
$$\mathbb{E}\left[\mathbb{I}(\hat{L}^0_i\leq \hat{L}_i)\mathbb{I}(\hat{L}^0_i\leq \lambda)-\mathbb{I}(L^0_i\leq L_i)\mathbb{I}(L^0_i\leq \lambda)\right]^2=o(1).
$$

Similarly, we can show that 
$$
\mathbb{E}\left\{\alpha^2\left[\mathbb{I}(\hat{L}_i\leq \hat{L}^0_i)\mathbb{I}(\hat{L}_i\leq \lambda)-\mathbb{I}(L_i\leq L^0_i)\mathbb{I}(L_i\leq \lambda)\right]^2\right\} = o(1). 
$$

Since $\hat{L}_i-L_i\xrightarrow{p}0$ and $\hat{L}^0_i-L^0_i\xrightarrow{p}0$, we can show that 
$$
\mathbb{E}\left\{-2\alpha\left[\mathbb{I}(\hat{L}^0_i\leq \hat{L}_i)\mathbb{I}(\hat{L}^0_i\leq \lambda)-\mathbb{I}(L^0_i\leq L_i)\mathbb{I}(L^0_i\leq \lambda)\right]\right\}=o(1),
$$
thus proving the lemma.

\subsection{Proof of Proposition \ref{pro:counter example}}\label{proof:pro counter example}

Previously, we have shown that \( \bar{X}_{i1} + \bar{X}_{i2} \stackrel{d}{=} \bar{X}_{i1} - \bar{X}_{i2} \). To demonstrate the lack of pairwise exchangeability, our strategy is to show that (a) \( \bar{X}_{i1} + \bar{X}_{i2} \) and \( S^*_i \) are mutually independent, whereas (b) \( \bar{X}_{i1} - \bar{X}_{i2} \) and \( S^*_i \) are correlated.

\noindent \textbf{Proof of Claim (a).} Let $\mathbf{X}_i=\left(X_{i1}, X_{i2}, \cdots, X_{in_i}\right)^T, \mathbf{Y}_i=\left(Y_{i1}, Y_{i2}, \cdots, Y_{in_i}\right)^T=A_i \mathbf{X}_i$, for $i \in \mathcal{H}_0$, where $A_i$ is an orthonormal matrix with the following form
$$
A_i=\left(\begin{array}{cccccc}
\frac{1}{\sqrt{n_i}} & \frac{1}{\sqrt{n_i}} & \frac{1}{\sqrt{n_i}} & \cdots & \frac{1}{\sqrt{n_i}} & \frac{1}{\sqrt{n_i}} \\
\frac{1}{\sqrt{2 \cdot 1}} & \frac{-1}{\sqrt{2 \cdot 1}} & 0 & \cdots & 0 & 0 \\
\frac{1}{\sqrt{3 \cdot 2}} & \frac{1}{\sqrt{3 \cdot 2}} & \frac{-2}{\sqrt{3 \cdot 2}} & \cdots & 0 & 0 \\
\vdots & \vdots & \vdots & & \vdots & \vdots \\
\frac{1}{\sqrt{n_i(n_i-1)}} & \frac{1}{\sqrt{n_i(n_i-1)}} & \frac{1}{\sqrt{n_i(n_i-1)}} & \cdots & \frac{1}{\sqrt{n_i(n_i-1)}} & \frac{-(n_i-1)}{\sqrt{n_i(n_i-1)}}
\end{array}\right).
$$
Then it is easy to see that $Y_{i1}=\frac{1}{\sqrt{n_i}} \sum_{j=1}^{n_i} X_{ij}=\sqrt{n_i} \bar{X}_i, Y_{i1}^2=n _i\bar{X}_i^2$. Moreover, we have
$$
\begin{aligned}
   \sum_{j=1}^{n_i} Y_{ij}^2=\mathbf{Y}_i^T\mathbf{Y}_i=\mathbf{X}_i^T A_i^T A_i \mathbf{X}_i=\mathbf{X}_i^T\mathbf{X}_i&=\sum_{j=1}^{n_i} X_{ij}^2=\sum_{j=1}^{n_i}\left(X_{ij}-\bar{X}_i\right)^2+n_i \bar{X}_i^2,\\
(n_i-1) {S^*_i}^2=\sum_{j=1}^{n_i}\left(X_{ij}-\bar{X}_i\right)^2&=\sum_{j=1}^{n_i} Y_{ij}^2-Y_{i1}^2=\sum_{j=2}^n Y_{ij}^2, 
\end{aligned}
$$

The expression \( \mathbf{Y}_i = A_i \mathbf{X}_i \) implies that \( Y_{ij} \) represents a linear combination of independent normal random variables \( X_{ij} \) for \( j = 1, 2, \ldots, n_i \). Consequently, the variables \( Y_{ij} \) for \( j = 1, 2, \ldots, n_i \) are all normally distributed. It is easy to see that 
$$
E[\mathbf{Y}_i] = (E[Y_{i1}], E[Y_{i2}], \ldots, E[Y_{in_i}])^T = A_i E[\mathbf{X}_i] = (0, 0, \ldots, 0)^T.
$$
Moreover, we can see that
\[
\text{Cov}(\mathbf{Y}_i) = \text{Cov}(A _i\mathbf{X}_i) = A_i \text{Cov}(\mathbf{X}_i) A_i^T =A_i A_i^T = I.
\]  
Thus, \( \text{Var}(Y_{ij}) = 1 \) for \( j = 1, 2, \ldots, n_i \), and \( Y_{ij}, Y_{ik} \) (for \( j \neq k \)) are uncorrelated. Therefore, \( Y_{ij} \), for \( j = 1, 2, \ldots, n_i \), are mutually independent. Let \( \bar{X}_{i1} \) and \( \bar{X}_{i2} \) be the sample means of the elements in the sets \( \mathcal{N}_{i1} \) and \( \mathcal{N}_{i2} \), respectively. Then we have
$$
\mbox{\( Y_{i1}^2 = n \bar{X}_i^2= n \left(\frac{\bar{X}_{i1}+\bar{X}_{i2}}{2}\right)^2\) and \( \sum_{j=2}^{n_i} Y_{ij}^2 = (n_i-1) {S^*_i}^2 \) are mutually independent.}
$$
It follows that $\bar{X}_{i1}+\bar{X}_{i2}$ and $S^*_i$ are mutually independent, proving Claim (a).

\noindent \textbf{Proof of Claim (b).}  Define \( \tilde{\mathbf{Y}}_i = \left( \tilde{Y}_{i1}, \tilde{Y}_{i2}, \cdots, \tilde{Y}_{in_i} \right)^T = \tilde{A}_i \mathbf{X}_i \), where \( \tilde{A}_i \) is a modified version of \( A_i \) that is obtained by changing the sign of the \( j \)-th element in the first row of \( A_i \) for \( j \in \mathcal{N}_{i2} \). This leads to the following: 
\[
\tilde{Y}_{i1} = \frac{1}{\sqrt{n_i}} \left( \sum_{j \in \mathcal{N}_{i1}} X_{ij} - \sum_{j \in \mathcal{N}_{i2}} X_{ij} \right) = \frac{\sqrt{n_i}}{2} (\bar{X}_{i1} - \bar{X}_{i2}).
\]
Thus, we have \(\tilde{Y}_{i1}^2 = \frac{n_i}{4} (\bar{X}_{i1} - \bar{X}_{i2})^2\). 
Since \( \tilde{A}_i \) and \( A_i \) differ only in the first row, for \( j \in [n_i] \setminus \{1\} \), we have \( \tilde{Y}_{ij} = Y_{ij} \), meaning that for all indices \( j \) other than 1, the values of \( \tilde{Y}_{ij} \) and \( Y_{ij} \) are identical. Therefore, the sum of squares for the remaining components is:
\[
\sum_{j=2}^{n_i} \tilde{Y}_{ij}^2=\sum_{j=2}^{n_i} Y_{ij}^2 = (n_i - 1) {S^*_i}^2.
\]
Then, we compute the covariance:
\[
\text{Cov}(\tilde{\mathbf{Y}}_i) = \text{Cov}(\tilde{A}_i \mathbf{X}_i) = \tilde{A}_i \text{Cov}(\mathbf{X}_i) \tilde{A}_i^T = \tilde{A}_i \tilde{A}_i^T.
\]
Notice that the first row of \( A_i \) has changed, so \( \tilde{A}_i \) is no longer an orthonormal matrix. As a result, \( \tilde{A}_i \tilde{A}_i^T \) is no longer a diagonal matrix. This means that \( \tilde{Y}_{i1} \) and \(\tilde{Y}_{ij} \) (for \( j \in [n_i] \setminus \{1\} \)) are correlated. Since we have proven that \( \sum_{j=2}^{n_i} \tilde{Y}^2 = (n_i - 1) {S^*_i}^2 \), and combining the correlation between \( \tilde{Y}_{i1} \) and \( \left(\tilde{Y}_{ij}: j \in [n_i] / \{1\}\right) \), along with the expression \( \tilde{Y}_{i1} = \frac{\sqrt{n_i}}{2} (\bar{X}_{i1} - \bar{X}_{i2}) \), we conclude that \( \bar{X}_{i1} - \bar{X}_{i2} \) and \( S^*_i \) are correlated. This proves Claim (b).

Combining Claims (a) and (b), which says \( \bar{X}_{i1} + \bar{X}_{i2} \) and \( S^*_i \) are mutually independent, while \( \bar{X}_{i1} - \bar{X}_{i2} \) and \( S^*_i \) are correlated. It follows that
\[
\frac{\bar{X}_{i1} + \bar{X}_{i2}}{S^*_i} \not\stackrel{d}{=} \frac{\bar{X}_{i1} - \bar{X}_{i2}}{S^*_i}.
\]
This implies that \( T^*_i \not\stackrel{d}{=} T^{0,*}_i \). Since if \((T^*_i, T^{0,*}_i \mid \mathbf{T}^*_{-i}, \mathbf{T}^{0,*}_{-i}) \stackrel{d}{=} (T^{0,*}_i, T^*_i \mid \mathbf{T}^*_{-i}, \mathbf{T}^{0,*}_{-i})\), it must follow that \( T^*_i \not\stackrel{d}{=} T^{0,*}_i \). Thus, we conclude that
\[
(T^*_i, T^{0,*}_i \mid \mathbf{T}^*_{-i}, \mathbf{T}^{0,*}_{-i}) \not\stackrel{d}{=} (T^{0,*}_i, T^*_i \mid \mathbf{T}^*_{-i}, \mathbf{T}^{0,*}_{-i}).
\]
Therefore the pairwise exchangeability fails to hold. This counter example shows that $S_i^*$ are not suitable for constructing the test and calibration samples. 

\setcounter{equation}{0}
\renewcommand{\theequation}{C.\arabic{equation}}
\setcounter{figure}{0}
\renewcommand{\thefigure}{C.\arabic{figure}}

\section{Auxiliary Numerical Results}\label{simu:aux}

\subsection{Impacts of distribution shifts: more comparisons under the SSMT setup}\label{simu:SSMT}

This section provides additional simulation results directly under the SSMT setup. We compare SENS with the conformal BH (CBH) method described in \cite{marandon2024adaptive}. The CBH method can be implemented via two strategies: (i) CBH\_TN: simulating calibration samples based on the theoretical null and (ii) CBH\_EEN: simulating calibration samples based on the estimated empirical null in \cite{jin2007estimating}. Meanwhile, SENS is directly implemented using samples generated from the following bivariate Gaussian model: 
\begin{equation}
   \begin{aligned}  
    &(T_i,T^0_i) \mid (\mu_i,\sigma_0,\rho) \stackrel{ind.}{\sim} \mathcal{N}_2\left(\left[
\begin{array}{c}
\mu_i\\
\mu_0\\
\end{array}
\right], \left[\begin{array}{cc}
\sigma_0^2 & \rho\sigma_0^2\\
\rho\sigma_0^2 & \sigma_0^2 \\
\end{array}
\right]\right),\\
    &\mu_i \stackrel{i.i.d.}{\sim}(1 - \pi) \delta_{\mu_0} + \pi\delta_{\mu_a},\quad i\in[m].
  \end{aligned}    
\end{equation}

The pairwise exchangeability holds for all correlations \(\rho\), which reflects the correlation between the test and calibration samples in our SENS Algorithm, providing a closer approximation to reality. The power of SENS is not affected by the value of \(\rho\); hence, we only present the results for \(\rho = 0.5\). Additionally, we set \(\pi = 0.1\), \(\mu_a = 5\) and \(m=2000\). In our simulation study, with 100 replications, we examine three scenarios: (a) \(\mu_0 = 0\), \(\alpha = 0.05\), with \(\sigma_0\) varying; (b) \(\sigma_0 = 1\), \(\alpha = 0.05\), with \(\mu_0\) varying; and (c) \(\mu_0 = 0.5\), \(\sigma_0 = 1.5\), with the FDR level \(\alpha\) varying. 

To implement CBH\_TN, we simulate \( m \) calibration samples \(  \mathbf{T}^{0,TN}=(T_i^{0,TN})_{i=1}^m \) from the theoretical null \( \mathcal{N}(0,1) \), without considering the training samples, and use \( \mathbf{T}=(T_i)_{i=1}^m \) as the test samples. The score function is constructed as $g(t) = \frac{\phi(t)}{\hat{f}^{TN}_{mix}(t)}$, where \( \phi(t) \) is the density function of \( \mathcal{N}(0,1) \), and
\[
\hat{f}^{TN}_{mix}(t) = \frac{1}{2m} \sum_{i=1}^m \left[ K_{h^{TN}_{mix}}(t - T_i) + K_{h^{TN}_{mix}}(t - T^{0,TN}_i) \right],
\]
where $h^{TN}_{mix}$ is a bandwidth satisfies \( h^{TN}_{mix}((\mathbf{T}, \mathbf{T}^{0,TN})_{\Pi}) = h_{mix}(\mathbf{T}, \mathbf{T}^{0,TN}) \).

To implement CBH\_EEN, we simulate \( m \) calibration samples \( \mathbf{T}^{0,EEN}=(T_i^{0,EEN})_{i=1}^m \) from the estimated empirical null \( \mathcal{N}(\hat{\mu}_0, \hat{\sigma}^2_0) \), as described in \cite{jin2007estimating}, using \( \mathbf{T} \) as the test samples. The score function is $g(t) = \frac{\phi_{\hat{\sigma}_0}(t - \hat{\mu}_0)}{\hat{f}^{EEN}_{mix}(t)}$, where \( \phi_{\hat{\sigma}_0}(t - \hat{\mu}_0) \) is the density function for \( \mathcal{N}(\hat{\mu}_0, \hat{\sigma}^2_0) \), and
\[
\hat{f}^{EEN}_{mix}(t) = \frac{1}{2m} \sum_{i=1}^m \left[ K_{h^{EEN}_{mix}}(t - T_i) + K_{h^{EEN}_{mix}}(t - T^{0,EEN}_i) \right],
\]
where $h^{EEN}_{mix}$ is a bandwidth satisfies \( h^{EEN}_{mix}((\mathbf{T}, \mathbf{T}^{0,EEN})_{\Pi}) = h_{mix}(\mathbf{T}, \mathbf{T}^{0,EEN}) \).

\begin{figure}[htbp!]
    \centering
    \includegraphics[scale=0.7]{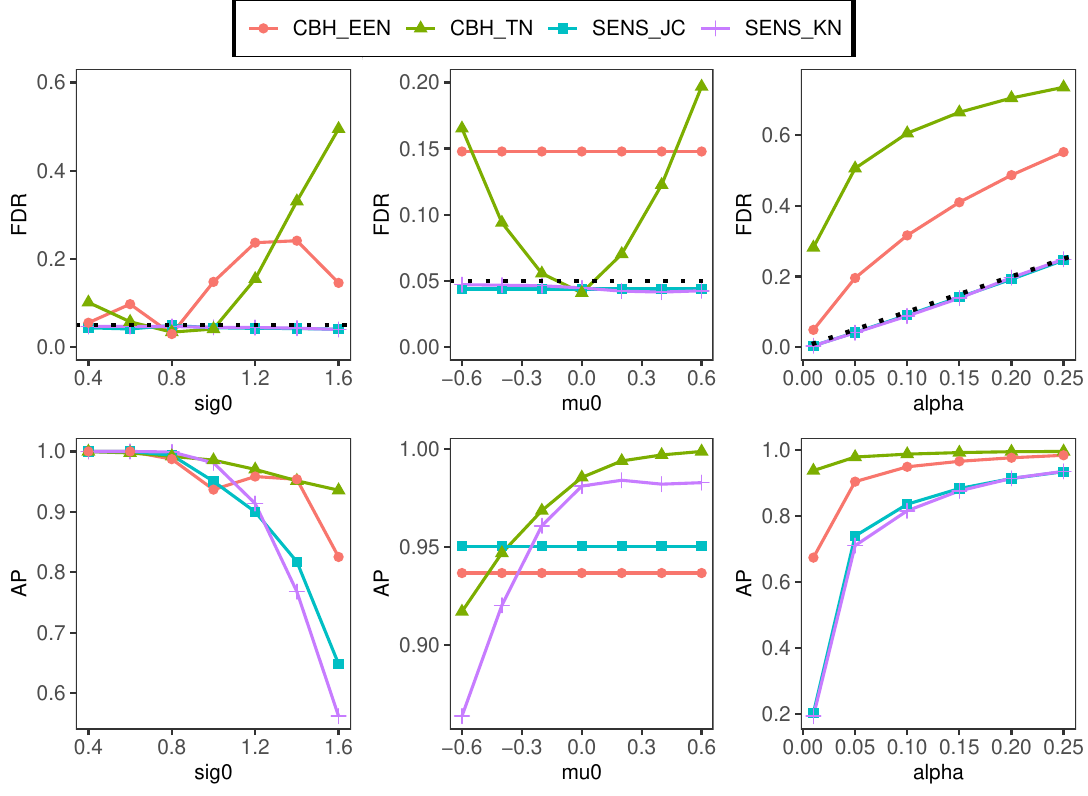}
    \caption{FDR and AP comparison for SENS with conformal BH using the theoretical null \(\mathcal{N}(0,1)\) and the estimated empirical null. The left, middle, and right columns correspond to (a), (b), and (c), respectively.}
    \label{fig:setting2}
\end{figure}

Then we calculate conformal $p$-values using equation \eqref{eq:cp} and apply the BH method to the resulting conformal $p$-values for the two CBH methods, respectively. The simulation results are presented in Figure \ref{fig:setting2}.

In the absence of distribution shifts, AdaDetect, when combined with the BH procedure, can be viewed as a conformalized empirical Bayes approach that employs lfdr statistics as its basic operational units. Thus, this comparison within the SSMT setup provides valuable insights into the advantages of SENS over conformalized empirical Bayes FDR methods. Specifically, strategies (i) and (ii) correspond to the conformalized versions of the method described in \cite{sun2007oracle}, where the theoretical null and the estimated empirical null are used, respectively, to compute the lfdr statistics.

Figure \ref{fig:setting2} illustrates the effects of distribution shifts between the theoretical null and the empirical null by examining the performance of CBH\_TN across various settings. It is evident that the FDR levels of CBH\_TN deviate more significantly from the nominal level as \(\sigma_0\) moves away from 1 or as \(\mu_0\) moves away from 0. In these circumstances, neither CBH\_EEN nor CBH\_TN can effectively control the FDR. In contrast, our proposed methods, SENS\_JC and SENS\_KN, successfully control the FDR at the nominal level. Notably, SENS\_JC outperforms SENS\_KN in terms of power when the Gaussian distribution provides a good approximation for the empirical null. Although SENS\_JC becomes less effective when the Gaussian assumption is violated, it remains valid in FDR levels.

\subsection{Simulation results for derandomized SENS}\label{simu:dsens}

We investigate the issues related to selecting hyperparameters in Derandomized SENS, which involves averaging \(e\)-values obtained from multiple sample splittings.

\subsubsection{Choice of the hyperparameters}

Derandomized SENS requires specifying hyperparameters \(N\) and \((\alpha_k)_{k=1}^N\). This remains an open question that warrants further research. We provide some preliminary investigations to explore how varying these hyperparameters affects performance. To simplify the discussion, we set \(\alpha^*\) to be proportional to the target FDR level \(\alpha = 0.05\), and choose \(\alpha_1 = \cdots = \alpha_N \equiv \alpha^*\). Although it is possible to use varied \(\alpha_k\) levels across replications, this introduces additional complexity that is beyond the scope of this research.

The primary goal of derandomization is to reduce uncertainty in decision-making, thereby enhancing algorithm stability. To quantify this stability, we use a variability metric from \cite{bashari2023derandomized}, which evaluates the consistency of rejection decisions across different splittings. Specifically, let 
\[
R_{i, r} = \begin{cases}
1, & \text{if the } i\text{-th null hypothesis is rejected in the } r\text{-th analysis}, \\
0, & \text{otherwise}.
\end{cases}
\]

In this context, let \(\text{Rep}\) represent the number of replicated decisions for each of the unit, which we set to 200. For each dataset, SENS requires 200 splittings to assess the consistency of decisions. Meanwhile, derandomized SENS requires \(200 N\) splittings per assessment, as each decision involves aggregating results across \(N\) splittings.

The variability of the rejection decisions across different replications is then evaluated using the following metric: 
  \begin{equation}\label{equ:variance}
  \widehat{\text{Variance}} = \frac{1}{m} \sum_{i=1}^{m} \frac{1}{Rep-1} \sum_{r=1}^{Rep} \left( R_{i, r} - \bar{R}_i \right)^2,
  \end{equation}
  where \( m \) is the number of units being tested and \( \bar{R}_i = \frac{1}{Rep} \sum_{r=1}^{Rep} R_{i, r} \).
 
We generate 50 observations for each unit according to the following model:
\begin{equation}\label{model:simu-var1}
\begin{aligned}
X_{ij}& = \mu_i + \epsilon_{ij}, \quad\mu_i \stackrel{i.i.d.}{\sim}(1 - \pi) \delta_0 + \pi \mathcal{N}(-\mu,\mu^2),\quad \sigma_i\stackrel{i.i.d.}{\sim} \mathcal{U}(0.05,\sigma_{\text{max}}),\\
\epsilon_{ij}\mid \sigma_i &\stackrel{i.i.d.}{\sim} (1-\beta)\mathcal{N}(0, \sigma_i^2) + \frac{3\beta}{4}\mathcal{U}(-\sqrt{3}\sigma_i, \sqrt{3}\sigma_i)+\frac{\beta}{4}\text{Laplace}(0, \sigma_i/\sqrt{2}).
\end{aligned}
\end{equation}
We set \(\pi = 0.1\), \(\beta = 0.5\), \(\sigma_{\text{max}} = 0.45\), and \(\mu = 0.1\), and present results for the 'KN' option in Algorithm 2, focusing on FDR, AP, and variance, while varying \(N\) and \(\alpha^*\).

\begin{figure}[htbp!]
    \center
    \includegraphics[width=6in]{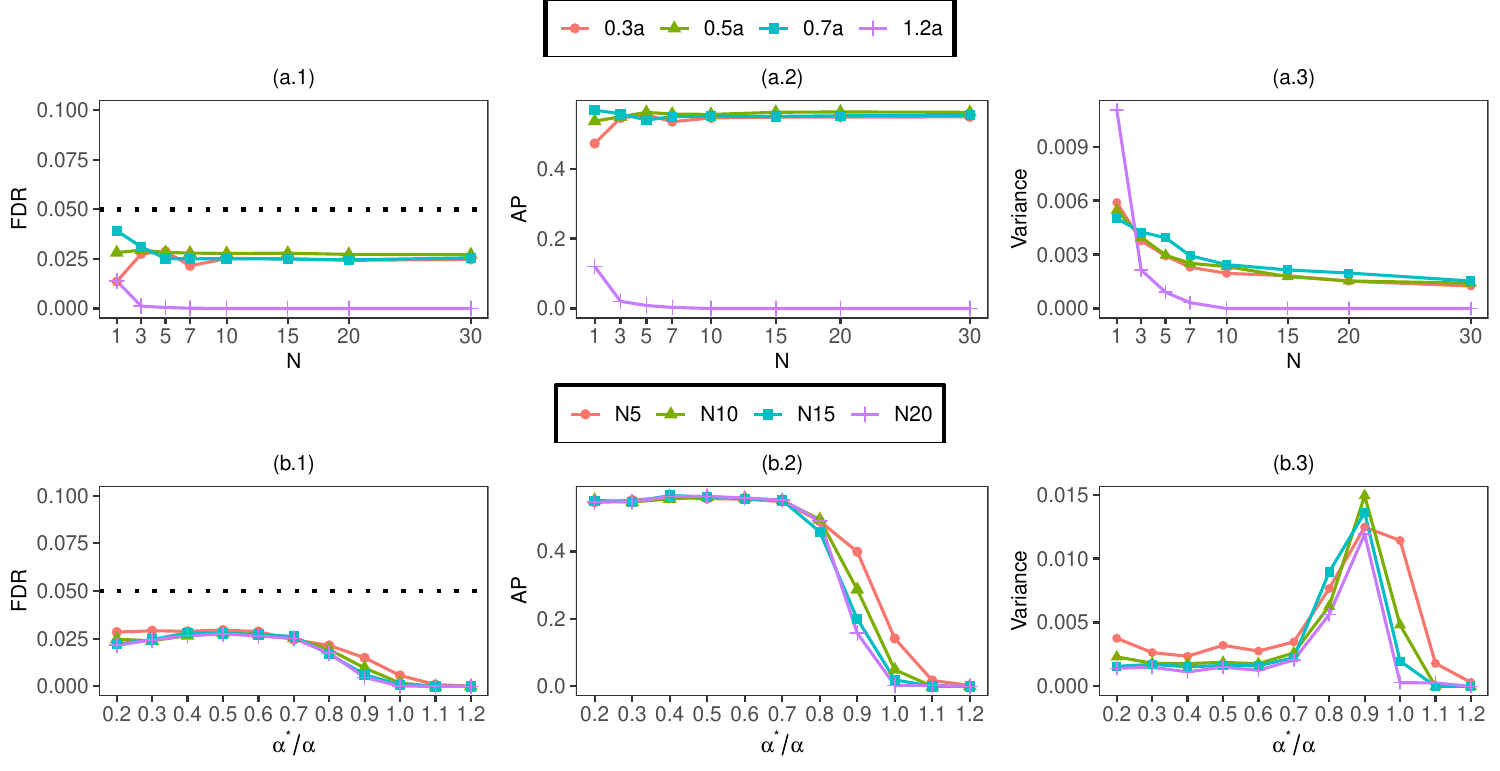}
    \caption{Comparison of FDR, AP, and Variance for Derandomized SENS (``KN'' option). The top row shows results for varying $N$, while the bottom row shows the effects of varying $\alpha^*$ ($\alpha$ is constant).} 
    \label{fig:dSENS_pair}
\end{figure}

The results are shown in Figure \ref{fig:dSENS_pair}, revealing several notable patterns.

First, Panel (a.2) indicates that the number of replications \(N\) has little impact on the power performance of Derandomized SENS. In Panel (a.3), it is evident that as \(N\) increases, the variance of the Derandomized SENS procedure decreases. However, beyond \(N = 10\), there is little gain in variability reduction. Additionally, because larger \(N\) entails higher computational costs, we recommend using a moderate value (e.g., \(N = 10\)).

Second, Panel (b.2) shows that the ratio \(\alpha^*/\alpha\) has little impact on power when it is below 0.7. However, there is a noticeable decline in power as the ratio surpasses 0.7, with power nearly dropping to zero once the ratio exceeds 1. Panel (b.3) illustrates that when the ratio is below 0.7, the variance remains relatively stable and low. As the ratio goes beyond 0.7, the variance increases, peaking near \(\alpha^*/\alpha = 1\). Beyond this point, as the ratio exceeds 1, the variance decreases towards zero due to the nearly zero AP. Thus, we recommend using a value for \(\alpha^*/\alpha\) below 0.7, such as \(\alpha^* = 0.5\alpha\).

Intuitively, top-ranked non-null hypotheses with larger effect sizes are more consistently reproducible across different experiments. Therefore, selecting a smaller \(\alpha^*\) is advisable as it stabilizes the algorithm's outputs. Additionally, a smaller \(\alpha^*\) enhances power because a more stable rejection set aids in constructing e-values that are more distinguishable under null and alternative hypotheses, as insightfully noted by \cite{ren2024derandomised}.

However, we emphasize that our current understanding of selecting the hyperparameters \(\alpha^*\) and \(N\) is still in its early stages and not yet definitive. We believe that further investigation into their optimal selection is both necessary and promising.

\subsubsection{Comparison of SENS and Derandomized SENS: Simulation Results}

We compare SENS and Derandomized SENS using metrics such as FDR, AP, and Variance, with \(N = 10\) and \(\alpha_k = 0.5\alpha\). To highlight the advantages of including a derandomization step, we introduce another metric: the average ranking (AR). Derandomized methods generally exhibit more conservative behavior compared to their random counterparts. To offer an alternative perspective and evaluate methods on equal footing, we present a plot of the AR metric, which displays the true positives against the number of rejections, thereby indicating the ranking efficiency of each method. Suppose we have selected a subset \(\mathcal{S}_k\) containing the top \(k\) candidates. The AR can then be computed as:
$\mbox{AR}_k=\mathbb{E}\left\{\frac{1}{k}\sum_{j\in\mathcal{S}_k}\mathbb{I}(H_{j,0}\ \text{is false})\right\}$. 

\begin{figure}[htbp!]
    \center
    \includegraphics[scale=0.8]{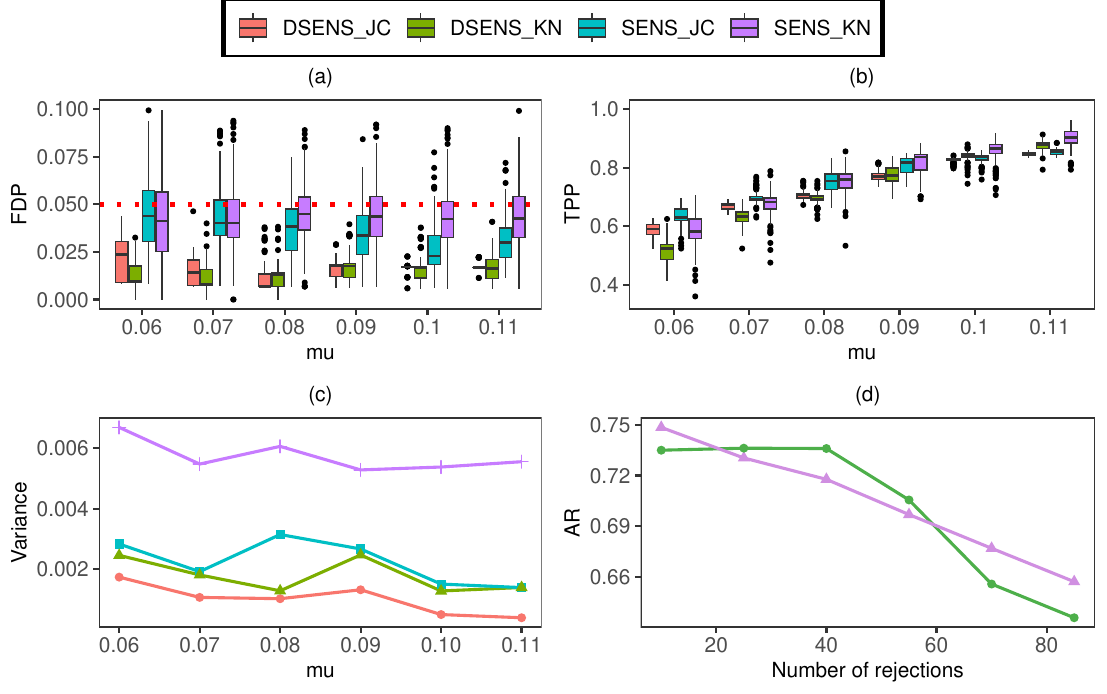}
    \caption{FDR, AP, Variance and AR comparison: SENS vs. Drandomized SENS with $N=10$ and $\alpha^* = 0.5\alpha$. The colors of the lines in the bottom row align with the colors of the boxes in the top row. In Panel (b), TPP (the label on y-axis) indicates the true positive proportion. }
    \label{fig:dSENS}
\end{figure}

We present simulation results comparing  SENS and Derandomized SENS using Model \eqref{model:simu-var1}. A summary of the results is shown in Figure \ref{fig:dSENS}. In Panels (a)-(c), we vary \(\mu\) while keeping other parameters constant to evaluate the effectiveness of both methods using the metrics of FDR, AP, and variance. In Panel (d), we fix \(\mu = 0.07\) and vary the number of rejections to assess effectiveness using the AR metric.

The FDP boxes in Panel (a) indicate that DSENS\_JC and DSENS\_KN are more conservative but less variable than SENS\_JC and SENS\_KN. Panel (b) shows that while derandomized SENS has lower power, it significantly reduces variability in the TPP. Panel (c) further demonstrates that derandomization enhances the consistency of rejection decisions. Panel (d) illustrates that ranking efficiency can sometimes be improved by derandomization. In summary, derandomization enhances the replicability and reliability of the proposed SENS algorithm, which is its primary motivation.

\subsection{Additional comparisons with distribution-free FDR methods}\label{sec:add-simu-res}

\subsubsection{Comparison with sfBH}\label{subsec:sfBH}
\textbf{1. Detailed description of the sfBH procedure.}
Below we first provide revised and detailed description of the sfBH procedure, which applies the vanilla BH procedure to sign-flipping \(p\)-values constructed following the proposal in \cite{arlot2010some}, under model \eqref{equ:model}.

\noindent \textbf{Step 1}. Transform the observed data corresponding to unit \(i\) into \(z\)-statistics using
$
Z^{*(0)}_i = \Phi^{-1}\left\{G_{t,n_i-1}\left(\frac{\sqrt{n_i}\,\bar{X}_i}{S_i}\right)\right\},
$
where \(\bar{X}_i\) and \(S_i\) denote the sample mean and sample standard deviation, respectively, for \(i\in[m]\).

\noindent \textbf{Step 2}. Generate random signs \(\text{Sign}^{(b)}_{ij} \in \{-1,1\}\) with equal probability, i.e.,
$
\mathbb{P}\left(\text{Sign}^{(b)}_{ij} = -1\right) = \mathbb{P}\left(\text{Sign}^{(b)}_{ij} = 1\right) = \frac{1}{2},
$
for \(b \in [B]\), \(j \in [n_i]\), and \(i \in [m]\). To avoid too many ties, we require that  \(B \leq 2^{n_i}\). Define the sign-perturbed data as
$
X^{(b)}_{ij} = \text{Sign}^{(b)}_{ij} \cdot X_{ij}.
$
The corresponding \(z\)-statistics \(\{Z^{*(b)}_i: i\in[m]\}\) are computed using the same transformation as in Step 1.

\noindent \textbf{Step 3}. Compute the sign-flipping \(p\)-values with a smoothing correction:
\begin{equation}\label{equ:valid p values with g}
p_i = \frac{1 + \sum_{b=1}^B \mathbb{I}\left\{g_i(Z_i^{*(b)}) \leq g_i(Z^{*(0)}_i)\right\}}{B + 1}, \quad i\in[m], 
\end{equation}
where \(g_i(\cdot)\) are score functions, with lower values providing stronger evidence against the null hypothesis.

\noindent \textbf{Step 4}. Apply the BH procedure to the conformal \(p\)-values constructed via \eqref{equ:valid p values with g}.

\noindent\textbf{2. The choice of the score functions.} Next we discuss two strategies for constructing score functions \(g_i(\cdot)\) in detail. Let \(\mathbf{Z}^*_i = \left(Z^{*(0)}_i, Z^{*(1)}_i, \dots, Z^{*(B)}_i\right)\)
for \(i\in[m]\) and denote \(\mathbf{Z}^* = (\mathbf{Z}^*_i)_{i=1}^m\). Consider a general class of score functions \(\{g_i(\cdot; \mathbf{Z}^*): i\in[m]\}\). A fundamental  condition for ensuring the validity of the \(p\)-values defined in \eqref{equ:valid p values with g} is:
\begin{equation}\label{equ:property of g_i}
g_i(\cdot; \mathbf{Z}^*_{\Pi_i}) = g_i(\cdot; \mathbf{Z}^*),
\end{equation}
where \(\Pi_i\) represents the operator that performs permutations of the elements in \(\mathbf{Z}_i^*\). \textbf{The first strategy} corresponds to the natural choice for \(g_i\), which  takes $g_i(\cdot; \mathbf{Z}^*) = -\left|\cdot\right|$. Thus the sign-flipping \(p\)-values in \eqref{equ:valid p values with g} simplify to 
\begin{equation}\label{equ:valid $p$-values}
p_i = \frac{1 + \sum_{b=1}^B \mathbb{I}\left(\left| Z_i^{*(b)}\right| \geq \left| Z^{*(0)}_i\right| \right)}{B + 1}.
\end{equation}

We investigate \textbf{a second strategy} for constructing powerful score functions by emulating the Lfdr. Our basic strategy involves specifying a working model for the \(z\)-statistics as follows: 
\[
Z^{*(0)}_i \stackrel{i.i.d.}{\sim} f(z) = (1 - \pi)f_0(z) + \pi f_1(z), \quad Z^{*(b)}_i \stackrel{i.i.d.}{\sim} f_0(z), \quad b \in [B],\; i \in [m],
\]
where \(f\) is the mixture density for \(Z^{*(0)}_i\), and \(f_0\) and \(f_1\) are the null and non-null density functions, respectively. Let $
\text{Lfdr}(z) = (1-\pi){f_0}(z)/{f}(z)$. To the best of our knowledge, the mixing strategy described in \cite{marandon2024adaptive} appears to be the only viable approach for constructing Lfdr-based scores that ensure the permutation invariance required by \eqref{equ:property of g_i} and, hence, for guaranteeing the validity of the sign-flipping \(p\)-values. This strategy involves using a new mixture density,
\begin{equation}\label{f-mix}
f_{\text{mix}}(z) = \frac{(B+1-\pi) f_0(z) + \pi f_1(z)}{B+1},
\end{equation}
to replace \(f(z)\), the mixture density for \(Z^{*(0)}_i\). Due to the monotonicity of \(f_{\text{mix}}(z)\) in \(f(z)\), \(f_0/f_{\text{mix}}\) retains the optimality of the original Lfdr-based score. 

However, estimating \(f_0/f_{\text{mix}}\) directly from data proves to be infeasible. As we will show shortly, the effective implementation of sfBH requires a large \(B\), which in turn diminishes the role of \(f_1\) in \eqref{f-mix}. For example, when \(B = 1000\) and \(\pi = 0.1\), the contribution of \(f_1\) in \(f_{\text{mix}}\) becomes almost negligible. 
This makes it extremely challenging to extract meaningful information about  \(f_1\) from the mixture samples in \(\mathbf{Z}^*\), rendering the estimates of \(f_0/f_{\text{mix}}\)  uninformative.

\begin{figure}
        \centering
        \includegraphics[scale=0.7]{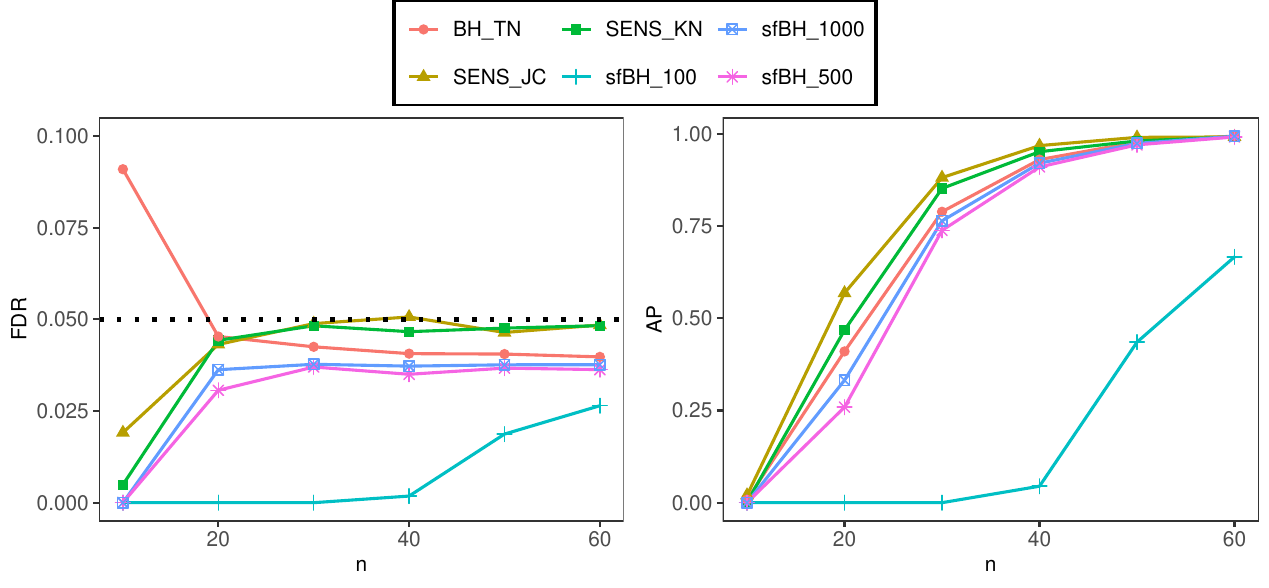}
        \caption{FDR and AP comparison: SENS vs. BH (theoretical null \(\mathcal{N}(0,1)\)) vs. sfBH.}
\label{fig:SF BH}
\end{figure}

\noindent\textbf{3. Simulation Results.} Exploring alternative score functions presents significant challenges and goes beyond the current scope of our research. Consequently, we next choose to implement sfBH using the $p$-values defined in \eqref{equ:valid $p$-values}. We set \( \alpha = 0.05 \) and use 100 replications. For each unit \( i \in [2000] \), we generate \( n \) observations as follows: \(X_{ij} = \mu_i + \epsilon_{ij}\), \( j \in [n]\), where
    \[
    \begin{aligned}
    \mu_i &\stackrel{i.i.d.}{\sim} (1 - 0.2) \delta_0 + 0.2 \delta_{\mu}, \\
    \epsilon_{ij} &\stackrel{i.i.d.}{\sim} (1 - \beta) \mathcal{N}(0, \sigma^2) + \beta \mathcal{U}(-\sqrt{3}\sigma, \sqrt{3}\sigma).
    \end{aligned}
    \]
We fix \( \mu = 0.8 \), \( \sigma = 1.2 \), \( \beta = 0.5 \), and vary \(B\) across \{100,500,1000\} and \(n\) from 10 to 60. The results are summarized in Figure \ref{fig:SF BH}. 

We can see that both SENS with sfBH control the FDR effectively, while the vanilla BH with theoretical null is invalid. Among the valid methods, their power rank as follows: 
$$\text{SENS\_JC}> \text{SENS\_KN} >\text{sfBH\_1000}>\text{sfBH\_500}> \text{sfBH\_100}.
$$
SENS outperforms sfBH due to the superior ranking of Lfdr over the $p$-values. Moreover, the power of the sfBH variants increases with $B$.

\subsubsection{Comparison with stBC}\label{subsec:stBC}
We conduct the following study to compare four methods: SENS with the ``JC'' option, stBC, which implements the BC method using symmetric t-statistics, the BH procedure, and the oracle AZ procedure, which has been proven to be optimal in \cite{sun2007oracle}. The test statistics $T_i$ and their calibration points $T^0_i$ for each unit $i \in [m]$ are generated according to the model described in Section 5.1 of \cite{TonyCaioptimal2017}. Specifically, for each unit $i \in [3000]$, we generate:
$$
T_i \stackrel{i.i.d.}{\sim} (1 - 0.1) \mathcal{N}(0, 1) + 0.1 \mathcal{N}(\mu, \sigma), \quad T^0_i \stackrel{i.i.d.}{\sim} \mathcal{N}(0, 1), \quad \text{Cor}(T_i, T^0_i) = 0.5.
$$

\begin{figure}[htbp!]
    \centering
    \includegraphics[scale=0.64]{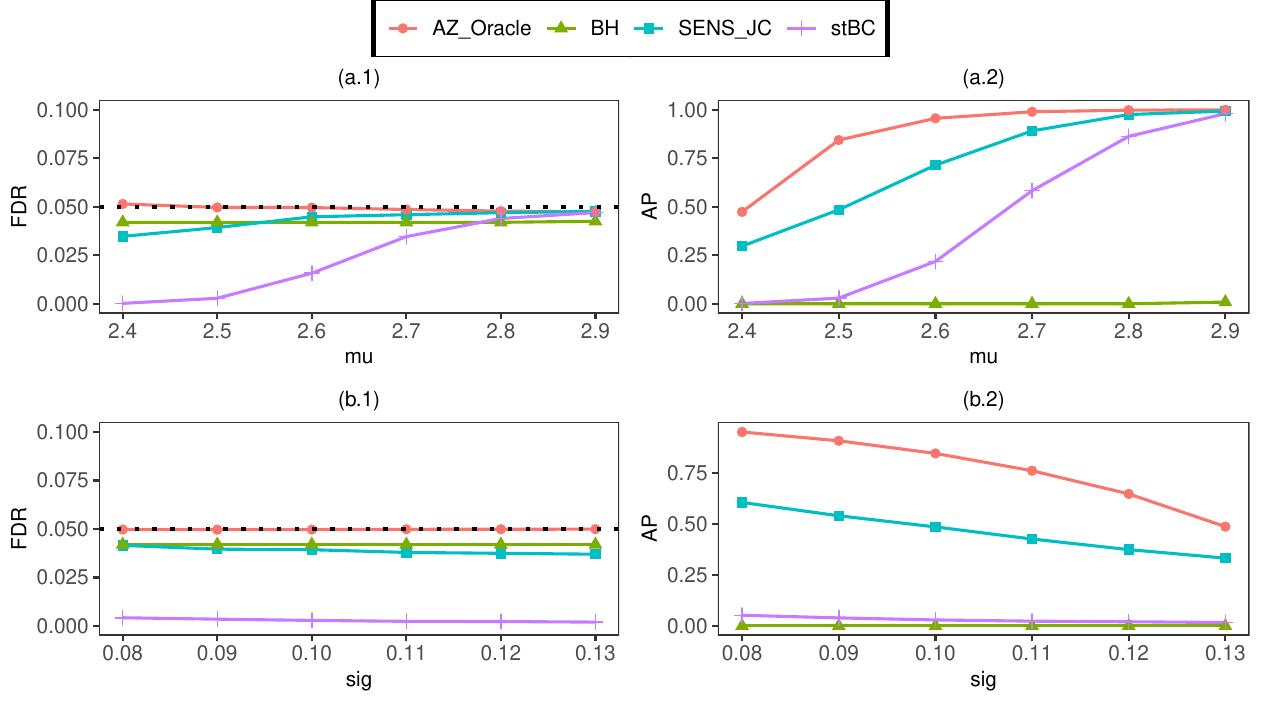}
    \caption{FDR and AP comparison: SENS\_JC vs. BH (theoretical null \(\mathcal{N}(0,1)\)) vs. AZ\_Oracle vs. stBC.}
    \label{fig:special SSMT setting}
\end{figure}

The simulation examines the following settings: (a) $\sigma = 0.1$, varying $\mu$; (b) $\mu = 2.5$, varying $\sigma$. We set $\alpha = 0.05$ and perform 100 replications. The results are presented in Figure \ref{fig:special SSMT setting}, where the power rankings from highest to lowest are as follows: AZ\_Oracle, SENS\_JC, stBC, and BH. In this setting, the significant discrepancy between the variances of $f_0$ and $f_1$ leads to a substantial decrease in power for methods that do not incorporate Lfdr, highlighting the advantage of using Lfdr in our approach.

\subsubsection{Comparison with RESS}\label{subsec:RESS}
We perform the following simulations, setting $\alpha=0.05$ and conducting 100 replications, for comparison with RESS. For each unit \(i \in [3000]\), we generate \(n\) observations:  
\(X_{ij} = \mu_i + \epsilon_{ij}\), \(j \in [n]
\), where  
$$
\begin{aligned}
\mu_i &\stackrel{i.i.d.}{\sim} (1 - 0.1) \delta_0 + 0.1 \delta_{\mu}, \\
\epsilon_{ij} &\stackrel{i.i.d.}{\sim} (1 - \beta) \mathcal{N}(0, \sigma^2) + \beta \mathcal{U}(-\sqrt{3}\sigma, \sqrt{3}\sigma).
\end{aligned}  
$$
The simulation examines the following settings:  
(a) \(\mu = 0.15\), \(\sigma = 0.15\), \(\beta = 0\), varying \(n\);
(b) \(n = 10\), \(\sigma = 0.15\), \(\beta = 0\), varying \(\mu\);  
(c) \(n = 10\), \(\mu = 0.15\), \(\beta = 0\), varying \(\sigma\);  
(d) \(n = 10\), \(\mu = 0.12\), \(\sigma = 0.15\), varying \(\beta\).  
\begin{figure}[htbp!]
    \centering
    \includegraphics[width=1\linewidth]{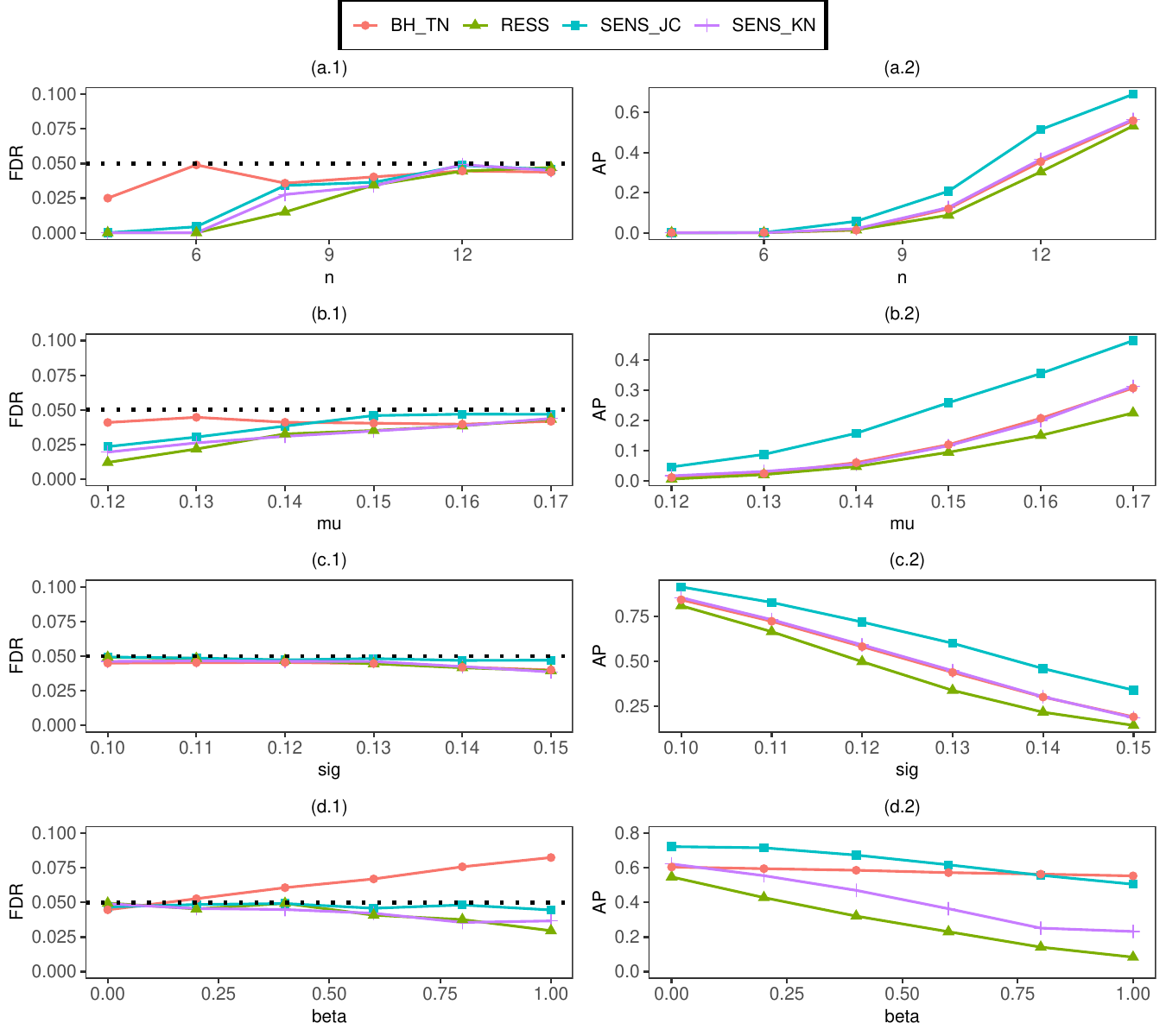}
    \caption{FDR and AP comparison: SENS vs. BH (theoretical null \(\mathcal{N}(0,1)\)) vs. RESS (\cite{zou2020new}).}
    \label{fig:compare with ress}
\end{figure}
As shown in Figure \ref{fig:compare with ress}, under normal errors in (a), (b), and (c), all methods control FDR below the pre-specified level \(\alpha\). Among them, SENS\_JC achieves the highest power, followed by BH and SENS\_KN, and finally RESS. However, as the proportion of uniform errors, i.e., \(\beta\), increases in (d), BH\_TN fails to control FDR, resulting in inflated power. In contrast, SENS\_JC, SENS\_KN, and RESS maintain valid FDR control, with their power rankings remaining consistent: SENS\_JC achieves the highest power, followed by SENS\_KN, and RESS maintains the lowest power.

\subsection{Variance of FDP and TPP}\label{subsec:variance}
In this section, we report the variances of FDP and TTP for SENS and competing methods under both one-sample and two-sample scenarios, as considered in the simulation study of Section \ref{sec:simulation}. The corresponding results are summarized in Tables \ref{tab:one-sample-EB-var}, \ref{tab:one-sample-model-free-var}, and \ref{tab:two_sample_var}.

\begin{table}[htbp!]
\centering
\caption{Variances of FDP and TPP for SENS (one-sample) and two traditional methods in Simulation 1 in Section \ref{subsec:sim-one sample}.}
\label{tab:one-sample-EB-var}
\resizebox{\textwidth}{!}{%
\begin{tabular}{lcccccccccccc}
\hline
\multirow{2}{*}{\textbf{Method}} & \multicolumn{2}{c}{$\pi$=0.01} & \multicolumn{2}{c}{$\pi$=0.05} & \multicolumn{2}{c}{$\pi$=0.09} & \multicolumn{2}{c}{$\pi$=0.13} & \multicolumn{2}{c}{$\pi$=0.17} & \multicolumn{2}{c}{$\pi$=0.21} \\
 & FDP-var & TTP-var & FDP-var & TTP-var & FDP-var & TTP-var & FDP-var & TTP-var & FDP-var & TTP-var & FDP-var & TTP-var \\
\hline
SENS\_JC & $8.77 \times 10^{-4}$ & 0.1244 & $6.67 \times 10^{-4}$ & 0.0719 & $4.88 \times 10^{-4}$ & 0.0685 & $3.94 \times 10^{-4}$ & 0.0670 & $3.98 \times 10^{-4}$ & 0.0572 & $4.08 \times 10^{-4}$ & 0.0560 \\
SENS\_KN & $1.11 \times 10^{-3}$ & 0.1083 & $6.75 \times 10^{-4}$ & 0.0623 & $5.95 \times 10^{-4}$ & 0.0528 & $4.05 \times 10^{-4}$ & 0.0517 & $3.99 \times 10^{-4}$ & 0.0482 & $3.51 \times 10^{-4}$ & 0.0481 \\
BH\_TN & $9.14 \times 10^{-4}$ & 0.0532 & $7.77 \times 10^{-4}$ & 0.0495 & $6.94 \times 10^{-4}$ & 0.0479 & $5.44 \times 10^{-4}$ & 0.0477 & $4.28 \times 10^{-4}$ & 0.0469 & $3.27 \times 10^{-4}$ & 0.0466 \\
BH\_EEN & $1.29 \times 10^{-2}$ & 0.0481 & $3.02 \times 10^{-2}$ & 0.0481 & $5.34 \times 10^{-2}$ & 0.0491 & $5.91 \times 10^{-2}$ & 0.0681 & $5.12 \times 10^{-2}$ & 0.0908 & $2.51 \times 10^{-3}$ & 0.1346 \\
\hline
\multirow{2}{*}{\textbf{Method}} & \multicolumn{2}{c}{$n$=2} & \multicolumn{2}{c}{$n$=4} & \multicolumn{2}{c}{$n$=6} & \multicolumn{2}{c}{$n$=8} & \multicolumn{2}{c}{$n$=10} & \multicolumn{2}{c}{$n$=12} \\
 & FDP-var & TTP-var & FDP-var & TTP-var & FDP-var & TTP-var & FDP-var & TTP-var & FDP-var & TTP-var & FDP-var & TTP-var \\
\hline
SENS\_JC & $6.50 \times 10^{-4}$ & 0.0459 & $5.14 \times 10^{-4}$ & 0.0895 & $4.90 \times 10^{-4}$ & 0.0360 & $4.84 \times 10^{-4}$ & 0.0245 & $5.30 \times 10^{-4}$ & 0.0199 & $4.58 \times 10^{-4}$ & 0.0144 \\
SENS\_KN & $6.02 \times 10^{-4}$ & 0.0480 & $6.49 \times 10^{-4}$ & 0.0873 & $5.24 \times 10^{-4}$ & 0.0479 & $6.61 \times 10^{-4}$ & 0.0336 & $5.93 \times 10^{-4}$ & 0.0214 & $5.43 \times 10^{-4}$ & 0.0174 \\
BH\_TN & $1.16 \times 10^{-3}$ & 0.0131 & $6.21 \times 10^{-4}$ & 0.0599 & $4.39 \times 10^{-4}$ & 0.0400 & $9.12 \times 10^{-3}$ & 0.0218 & $9.09 \times 10^{-3}$ & 0.0149 & $3.68 \times 10^{-4}$ & 0.0128 \\
BH\_EEN & $1.43 \times 10^{-1}$ & 0.2335 & $6.43 \times 10^{-2}$ & 0.0866 & $4.53 \times 10^{-2}$ & 0.0479 & $2.27 \times 10^{-2}$ & 0.0315 & $8.82 \times 10^{-3}$ & 0.0549 & $4.09 \times 10^{-3}$ & 0.0935 \\
\hline
\multirow{2}{*}{\textbf{Method}} & \multicolumn{2}{c}{$\beta$=0.0} & \multicolumn{2}{c}{$\beta$=0.2} & \multicolumn{2}{c}{$\beta$=0.4} & \multicolumn{2}{c}{$\beta$=0.6} & \multicolumn{2}{c}{$\beta$=0.8} & \multicolumn{2}{c}{$\beta$=1.0} \\
 & FDP-var & TTP-var & FDP-var & TTP-var & FDP-var & TTP-var & FDP-var & TTP-var & FDP-var & TTP-var & FDP-var & TTP-var \\
\hline
SENS\_JC & $7.05 \times 10^{-4}$ & 0.0781 & $7.99 \times 10^{-4}$ & 0.0787 & $1.05 \times 10^{-3}$ & 0.0666 & $9.24 \times 10^{-4}$ & 0.0904 & $1.19 \times 10^{-3}$ & 0.0741 & $8.05 \times 10^{-4}$ & 0.0867 \\
SENS\_KN & $1.01 \times 10^{-3}$ & 0.0819 & $6.84 \times 10^{-4}$ & 0.0555 & $8.50 \times 10^{-4}$ & 0.0753 & $9.49 \times 10^{-4}$ & 0.0673 & $1.00 \times 10^{-3}$ & 0.0778 & $9.93 \times 10^{-4}$ & 0.0626 \\
BH\_TN & $4.41 \times 10^{-4}$ & 0.0450 & $5.60 \times 10^{-4}$ & 0.0469 & $1.21 \times 10^{-3}$ & 0.0447 & $7.21 \times 10^{-4}$ & 0.0453 & $7.70 \times 10^{-4}$ & 0.0471 & $9.33 \times 10^{-4}$ & 0.0461 \\
BH\_EEN & $7.81 \times 10^{-3}$ & 0.0423 & $8.36 \times 10^{-3}$ & 0.0469 & $7.62 \times 10^{-3}$ & 0.0469 & $1.05 \times 10^{-2}$ & 0.0409 & $1.03 \times 10^{-2}$ & 0.0472 & $1.19 \times 10^{-2}$ & 0.0480 \\
\hline
\end{tabular}%
}
\end{table}

\begin{table}[htbp!]
\centering
\caption{Variances of FDP and TPP for SENS (one-sample) and three model-free methods in Simulation 2 in Section \ref{subsec:sim-one sample}.}
\label{tab:one-sample-model-free-var}
\resizebox{\textwidth}{!}{%
\begin{tabular}{lcccccccccccc}
\hline
\multirow{2}{*}{\textbf{Method}} & \multicolumn{2}{c}{$\mu=0.5$} & \multicolumn{2}{c}{$\mu=1.0$} & \multicolumn{2}{c}{$\mu=1.5$} & \multicolumn{2}{c}{$\mu=2.0$} & \multicolumn{2}{c}{$\mu=2.5$} & \multicolumn{2}{c}{$\mu=3.0$} \\
 & FDP-var & TTP-var & FDP-var & TTP-var & FDP-var & TTP-var & FDP-var & TTP-var & FDP-var & TTP-var & FDP-var & TTP-var \\
\hline
SENS\_JC & $1.14 \times 10^{-3}$ & 0.0728 & $2.02 \times 10^{-3}$ & 0.1783 & $9.36 \times 10^{-4}$ & 0.1685 & $7.82 \times 10^{-4}$ & 0.1589 & $6.85 \times 10^{-4}$ & 0.1452 & $6.06 \times 10^{-4}$ & 0.1227 \\
SENS\_KN & $1.66 \times 10^{-3}$ & 0.0539 & $2.98 \times 10^{-3}$ & 0.1644 & $1.71 \times 10^{-3}$ & 0.1752 & $1.12 \times 10^{-3}$ & 0.1692 & $9.31 \times 10^{-4}$ & 0.1628 & $8.66 \times 10^{-4}$ & 0.1484 \\
sfBH     & 0.00     & 0.0000 & 0.00     & 0.0000 & 0.00     & 0.0000 & 0.00     & 0.0000 & 0.00     & 0.0000 & 0.00     & 0.0000 \\
stBC     & $5.31 \times 10^{-5}$ & 0.0050 & $1.88 \times 10^{-4}$ & 0.0262 & $2.30 \times 10^{-4}$ & 0.0472 & $2.40 \times 10^{-4}$ & 0.0678 & $2.45 \times 10^{-4}$ & 0.0833 & $3.15 \times 10^{-4}$ & 0.0955 \\
RESS     & $3.56 \times 10^{-4}$ & 0.0055 & $1.26 \times 10^{-3}$ & 0.0656 & $1.48 \times 10^{-3}$ & 0.1325 & $1.11 \times 10^{-3}$ & 0.1541 & $9.30 \times 10^{-4}$ & 0.1501 & $1.01 \times 10^{-3}$ & 0.1509 \\
\hline
\multirow{2}{*}{\textbf{Method}} & \multicolumn{2}{c}{$\sigma_{\text{max}}=0.05$} & \multicolumn{2}{c}{$\sigma_{\text{max}}=0.15$} & \multicolumn{2}{c}{$\sigma_{\text{max}}=0.25$} & \multicolumn{2}{c}{$\sigma_{\text{max}}=0.35$} & \multicolumn{2}{c}{$\sigma_{\text{max}}=0.45$} & \multicolumn{2}{c}{$\sigma_{\text{max}}=0.55$} \\
 & FDP-var & TTP-var & FDP-var & TTP-var & FDP-var & TTP-var & FDP-var & TTP-var & FDP-var & TTP-var & FDP-var & TTP-var \\
\hline
SENS\_JC & $4.68 \times 10^{-4}$ & 0.0759 & $5.82 \times 10^{-4}$ & 0.1160 & $6.41 \times 10^{-4}$ & 0.1358 & $7.72 \times 10^{-4}$ & 0.1456 & $8.01 \times 10^{-4}$ & 0.1576 & $9.51 \times 10^{-4}$ & 0.1581 \\
SENS\_KN & $5.88 \times 10^{-4}$ & 0.0690 & $6.54 \times 10^{-4}$ & 0.1118 & $1.03 \times 10^{-3}$ & 0.1600 & $1.04 \times 10^{-3}$ & 0.1613 & $1.18 \times 10^{-3}$ & 0.1675 & $1.33 \times 10^{-3}$ & 0.1633 \\
sfBH     & 0.00     & 0.0000 & 0.00     & 0.0000 & 0.00     & 0.0000 & 0.00     & 0.0000 & 0.00     & 0.0000 & 0.00     & 0.0000 \\
stBC     & $2.82 \times 10^{-4}$ & 0.1205 & $2.97 \times 10^{-4}$ & 0.1028 & $3.38 \times 10^{-4}$ & 0.0864 & $2.52 \times 10^{-4}$ & 0.0701 & $2.42 \times 10^{-4}$ & 0.0558 & $2.42 \times 10^{-4}$ & 0.0479 \\
RESS     & $6.94 \times 10^{-4}$ & 0.1303 & $9.14 \times 10^{-4}$ & 0.1551 & $8.67 \times 10^{-4}$ & 0.1486 & $1.10 \times 10^{-3}$ & 0.1453 & $1.18 \times 10^{-3}$ & 0.1364 & $1.50 \times 10^{-3}$ & 0.1200 \\
\hline
\end{tabular}%
}
\end{table}

\newpage
\begin{table}[htbp!]
\centering
\caption{Variances of FDP and TPP for SENS (two-sample), BH (theoretical null $\mathcal{N}(0,1)$), BH (estimated empirical null via the Jin-Cai method), CLIPPER, and RESS in the simulation study described in Section \ref{subsec:sim-two sample}.}
\label{tab:two_sample_var}
\resizebox{\textwidth}{!}{%
\begin{tabular}{lcccccccccccc}
\hline
\multirow{2}{*}{\textbf{Method}} & \multicolumn{2}{c}{$\mu_y=0.6$} & \multicolumn{2}{c}{$\mu_y=0.8$} & \multicolumn{2}{c}{$\mu_y=1.0$} & \multicolumn{2}{c}{$\mu_y=1.2$} & \multicolumn{2}{c}{$\mu_y=1.4$} & \multicolumn{2}{c}{$\mu_y=1.6$} \\
 & FDP-var & TTP-var & FDP-var & TTP-var & FDP-var & TTP-var & FDP-var & TTP-var & FDP-var & TTP-var & FDP-var & TTP-var \\
\hline
SENS\_JC & $3.17 \times 10^{-4}$ & $2.10 \times 10^{-3}$ & $2.65 \times 10^{-4}$ & $2.10 \times 10^{-3}$ & $2.03 \times 10^{-4}$ & $2.09 \times 10^{-3}$ & $2.09 \times 10^{-4}$ & $1.62 \times 10^{-3}$ & $2.54 \times 10^{-4}$ & $1.41 \times 10^{-3}$ & $2.14 \times 10^{-4}$ & $9.90 \times 10^{-4}$ \\
SENS\_KN & $5.20 \times 10^{-4}$ & $3.58 \times 10^{-3}$ & $3.46 \times 10^{-4}$ & $4.58 \times 10^{-3}$ & $2.90 \times 10^{-4}$ & $6.49 \times 10^{-3}$ & $2.45 \times 10^{-4}$ & $2.72 \times 10^{-3}$ & $1.95 \times 10^{-4}$ & $1.77 \times 10^{-3}$ & $2.18 \times 10^{-4}$ & $1.50 \times 10^{-3}$ \\
CLIPPER & $1.01 \times 10^{-3}$ & $2.76 \times 10^{-4}$ & $1.24 \times 10^{-3}$ & $1.32 \times 10^{-3}$ & $8.49 \times 10^{-4}$ & $3.84 \times 10^{-3}$ & $6.40 \times 10^{-4}$ & $1.04 \times 10^{-2}$ & $3.52 \times 10^{-4}$ & $1.18 \times 10^{-2}$ & $1.99 \times 10^{-4}$ & $3.20 \times 10^{-3}$ \\
RESS & $4.50 \times 10^{-4}$ & $1.46 \times 10^{-3}$ & $3.45 \times 10^{-4}$ & $1.70 \times 10^{-3}$ & $2.75 \times 10^{-4}$ & $1.62 \times 10^{-3}$ & $2.15 \times 10^{-4}$ & $1.33 \times 10^{-3}$ & $1.95 \times 10^{-4}$ & $1.12 \times 10^{-3}$ & $1.75 \times 10^{-4}$ & $1.02 \times 10^{-3}$ \\
BH\_TN & $2.47 \times 10^{-4}$ & $4.49 \times 10^{-4}$ & $2.08 \times 10^{-4}$ & $6.56 \times 10^{-4}$ & $1.55 \times 10^{-4}$ & $6.98 \times 10^{-4}$ & $1.28 \times 10^{-4}$ & $4.63 \times 10^{-4}$ & $1.00 \times 10^{-4}$ & $3.83 \times 10^{-4}$ & $8.37 \times 10^{-5}$ & $3.72 \times 10^{-4}$ \\
BH\_EEN & $1.09 \times 10^{-4}$ & $1.79 \times 10^{-3}$ & $6.51 \times 10^{-5}$ & $2.79 \times 10^{-3}$ & $3.58 \times 10^{-5}$ & $3.62 \times 10^{-3}$ & $2.48 \times 10^{-5}$ & $4.86 \times 10^{-3}$ & $1.62 \times 10^{-5}$ & $5.61 \times 10^{-3}$ & $8.51 \times 10^{-6}$ & $5.66 \times 10^{-3}$ \\
\hline
\multirow{2}{*}{\textbf{Method}} & \multicolumn{2}{c}{$\pi_y=0.04$} & \multicolumn{2}{c}{$\pi_y=0.08$} & \multicolumn{2}{c}{$\pi_y=0.12$} & \multicolumn{2}{c}{$\pi_y=0.16$} & \multicolumn{2}{c}{$\pi_y=0.20$} & \multicolumn{2}{c}{$\pi_y=0.24$} \\
 & FDP-var & TTP-var & FDP-var & TTP-var & FDP-var & TTP-var & FDP-var & TTP-var & FDP-var & TTP-var & FDP-var & TTP-var \\
\hline
SENS\_JC & $9.80 \times 10^{-4}$ & $1.24 \times 10^{-2}$ & $5.85 \times 10^{-4}$ & $3.38 \times 10^{-3}$ & $3.79 \times 10^{-4}$ & $2.36 \times 10^{-3}$ & $2.82 \times 10^{-4}$ & $1.81 \times 10^{-3}$ & $3.02 \times 10^{-4}$ & $1.40 \times 10^{-3}$ & $2.79 \times 10^{-4}$ & $1.12 \times 10^{-3}$ \\
SENS\_KN & $1.75 \times 10^{-3}$ & $4.81 \times 10^{-2}$ & $7.58 \times 10^{-4}$ & $1.37 \times 10^{-2}$ & $4.35 \times 10^{-4}$ & $6.53 \times 10^{-3}$ & $4.07 \times 10^{-4}$ & $4.05 \times 10^{-3}$ & $3.19 \times 10^{-4}$ & $1.96 \times 10^{-3}$ & $2.35 \times 10^{-4}$ & $8.15 \times 10^{-4}$ \\
CLIPPER & $5.48 \times 10^{-3}$ & $4.62 \times 10^{-3}$ & $1.95 \times 10^{-3}$ & $1.86 \times 10^{-3}$ & $1.10 \times 10^{-3}$ & $1.06 \times 10^{-3}$ & $5.81 \times 10^{-4}$ & $5.71 \times 10^{-4}$ & $3.99 \times 10^{-4}$ & $5.11 \times 10^{-4}$ & $3.31 \times 10^{-4}$ & $2.76 \times 10^{-4}$ \\
RESS & $7.98 \times 10^{-4}$ & $5.62 \times 10^{-3}$ & $4.63 \times 10^{-4}$ & $2.68 \times 10^{-3}$ & $4.17 \times 10^{-4}$ & $1.77 \times 10^{-3}$ & $3.31 \times 10^{-4}$ & $1.74 \times 10^{-3}$ & $2.42 \times 10^{-4}$ & $1.21 \times 10^{-3}$ & $1.79 \times 10^{-4}$ & $7.62 \times 10^{-4}$ \\
BH\_TN & $9.40 \times 10^{-4}$ & $1.43 \times 10^{-3}$ & $4.34 \times 10^{-4}$ & $6.59 \times 10^{-4}$ & $3.49 \times 10^{-4}$ & $5.84 \times 10^{-4}$ & $2.01 \times 10^{-4}$ & $4.38 \times 10^{-4}$ & $2.12 \times 10^{-4}$ & $3.19 \times 10^{-4}$ & $1.48 \times 10^{-4}$ & $2.85 \times 10^{-4}$ \\
BH\_EEN & $1.83 \times 10^{-3}$ & $1.92 \times 10^{-3}$ & $2.58 \times 10^{-3}$ & $1.00 \times 10^{-3}$ & $1.12 \times 10^{-2}$ & $1.31 \times 10^{-3}$ & $2.14 \times 10^{-3}$ & $3.02 \times 10^{-3}$ & $5.71 \times 10^{-5}$ & $4.65 \times 10^{-3}$ & $2.55 \times 10^{-6}$ & $5.26 \times 10^{-2}$ \\
\hline
\end{tabular}%
}
\end{table}

\subsection{Figure in the data analysis: two-sample case}
\begin{figure}[htbp!]
    \center    \includegraphics[width=\textwidth,height=0.2\textheight]{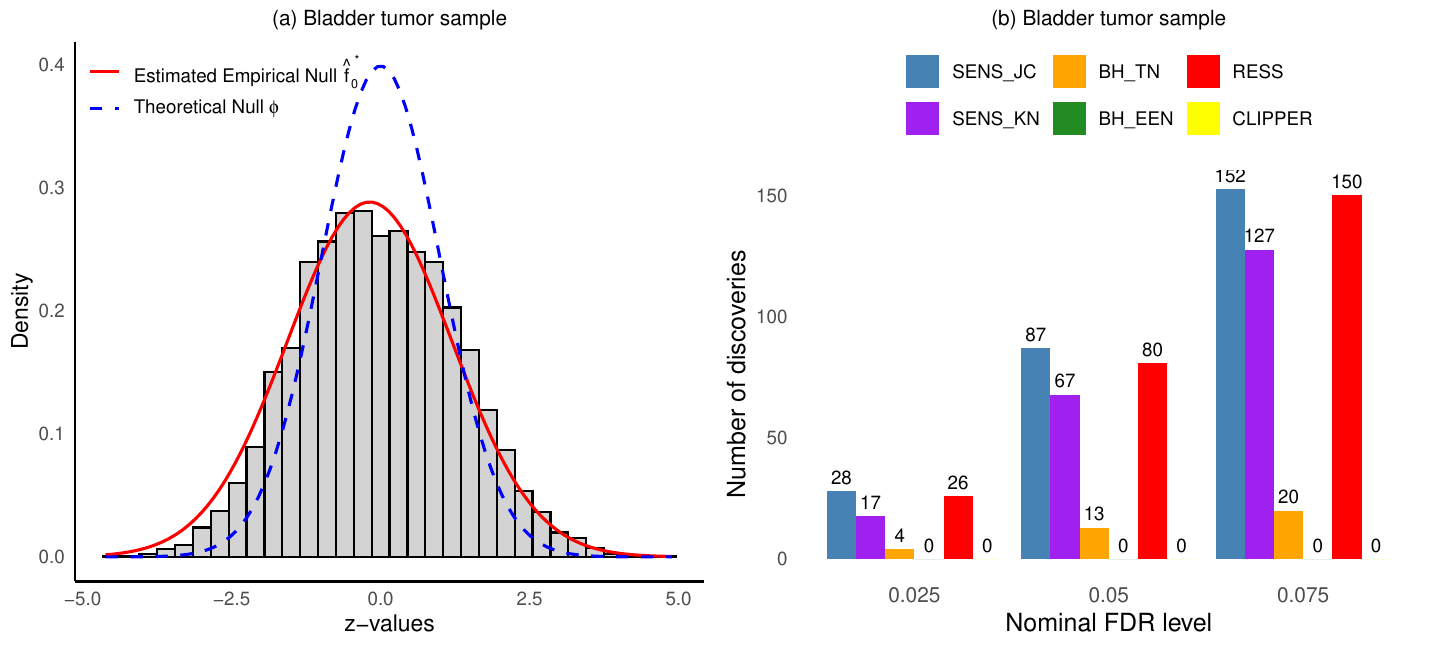}
    \caption{Analysis of the bladder tumor sample: The left plot displays the histogram of $z$-values alongside the density curves for the theoretical null $\mathcal{N}(0, 1)$ and the estimated empirical null $\mathcal{N}(-0.07, 1.34^2)$. The right plot shows the average number of discoveries for each method at various nominal FDR levels.}
    \label{fig:DE analysis_two sample}
\end{figure}

\end{document}